\documentclass[aps,prd,12pt,eqsecnum,showpacs,preprintnumbers,nofootinbib,superscriptaddress]{revtex4}

\usepackage{amssymb,amsmath,amsthm,graphicx,amscd}
\usepackage{enumerate,color,verbatim,ulem,mathrsfs,slashed,centernot,hyperref}

\begin{document}

\baselineskip=18pt
\allowdisplaybreaks

\title{Quantum thermodynamics from the nonequilibrium dynamics of open systems: energy, heat capacity and the third law}

\author{J.-T. Hsiang}
\email{cosmology@gmail.com}
\affiliation{Center for Field Theory and Particle Physics, Department of Physics, Fudan University, Shanghai 200433, China}
\author{C. H. Chou}
\email{chouch@mail.ncku.edu.tw}
\affiliation{Department of Physics, National Cheng Kung University, Tainan 70101, Taiwan}
\author{Y. Suba\c{s}\i}
\email{ysubasi@gmail.com}
\affiliation{Theoretical Division, Los Alamos National Laboratory, Los Alamos, New Mexico 87545, USA}
\author{B. L. Hu}
\email{blhu@umd.edu}
\affiliation{Center for Field Theory and Particle Physics, Department of Physics, Fudan University, Shanghai 200433, China}
\affiliation{Maryland Center for Fundamental Physics and Joint Quantum Institute, University of Maryland, College Park, Maryland 20742-4111, USA}

\date{\today}

\begin{abstract}
In a series of papers, we intend the perspective of open quantum systems and examine from their nonequilibrium dynamics the conditions when the physical quantities, their relations and the laws of thermodynamics become well defined and viable for quantum many body systems. We first describe how an open system nonequilibrium dynamics (ONEq) approach is different from the closed combined system + environment in a global thermal state (CGTs) setup. Only after the open system equilibrates will it be amenable to conventional thermodynamics descriptions, thus quantum thermodynamics (QTD) comes at the end rather than assumed in the beginning.  The linkage between the two comes from the reduced density matrix of ONEq in that stage having the same form as that of the system in the CGTs. We see the open system approach having the advantage of dealing with nonequilibrium processes as many experiments in the near future will call for. Because it spells out the conditions of QTD's existence it can also aid us in addressing the basic issues in quantum thermodynamics from first principles in a systematic way. We then study one broad class of open quantum systems where the full nonequilibrium dynamics can be solved exactly, that of the quantum Brownian motion of $N$ strongly coupled harmonic oscillators,  interacting strongly with a scalar field environment. In this paper we focus on the internal energy, heat capacity and the third law. We show for this class of physical models, amongst other findings,  the extensive property of the internal energy, the positivity of the heat capacity and the validity of the third law from the perspective of the behavior of the heat capacity toward zero temperature. These conclusions obtained from exact solutions and quantitative analysis clearly disprove claims of negative specific heat in such systems and dispel allegations that in such systems the validity of the third law of thermodynamics relies on quantum entanglement. They are conceptually and factually unrelated issues. Entropy and entanglement will be the main theme of our second paper on this subject matter.

\end{abstract}

\pacs{xyz }
\maketitle

\section{Introduction}

In this series of papers we take the perspective of open quantum systems (OQS) and examine from their nonequilibrium (NEq) dynamics the conditions when the physical quantities, concepts, constructs and the time-honored laws of thermodynamics (TD) become well defined and viable for quantum many body systems. We utilize one broad class of models where the nonequilibrium dynamics can be solved exactly -- the Brownian motion of strongly coupled (SC) harmonic oscillators,  interacting strongly with a scalar field environment --  to explore a range of basic issues in quantum thermodynamics (QTD). The exact solutions possible in these OQSs enable us to examine and define these conditions more precisely in a quantitative, systematic and transparent way. This approach hopefully compensates for the rather loose, qualitative and at times contrived way thermodynamic descriptions for quantum systems are proposed because of the need to adhere to the dictum of classical thermodynamics,  which is valid only under very special conditions.

A clarification in the meaning and contents of quantum thermodynamics (QTD) \cite{Mahler} might be useful before we proceed: to us, it is the study of the thermodynamic properties of quantum many-body systems (MBS). Quantum now refers not just to the particle spin-statistics (boson vs fermion) aspects:  the rather limited meaning of `quantum' in traditional quantum statistical mechanics (QSM), but also includes in the present era the quantum phase aspects, such as quantum coherence, quantum correlations, and quantum entanglement.  This is where quantum information has a hand in QTD \cite{ktln2}.  The thermodynamics connotation can be extended to include systems not necessarily in equilibrium at all times, thus encompassing dissipative and relaxation processes for systems deviating from equilibrium, including linear or nonlinear response theories applied to quantum MBS~\footnote{Linear response theory considers small variations in the system while staying in thermal contact with the bath. This is the underlying assumption in the use of thermal Green's functions, which is within the test-field approximation in quantum field theory terms. In a fully NEq treatment of the open system's quantum dynamics, both the system and environment variables are dynamically determined.  Thus it can cope with situations where the quantum system is small and the environment is finite.}, familiar in condensed matter or chemical physics and for classical MBS,  topics incorporated in the traditional field of NEq TD~\cite{NEqTDbooks}. These considerations can be extended to weakly nonequilibrium conditions but not for far-from-equilibrium, fully arbitrary time evolutions. That is when an open quantum system treatment becomes necessary.

The issues addressed in this first paper encompass the nature of internal energy, heat capacity and the third law for a fully nonequilibrium (NEq) system. We demonstrate what it takes for it to evolve to an equilibrium (Eq) condition, and from that point establish the connection with traditional TD theory. The conditions for traditional TD theory to be well-defined and operative for a classical or quantum system are very specific despite its wide ranging applicability: A system of relatively fewer degrees of freedom in the presence of a thermal bath of a huge number or infinite degrees of freedom (we shall consider only heat but no particle transfer here and thus the TD refers only to canonical ensembles), the coupling between the system and the bath is vanishingly small, and the system is eternally in a thermal equilibrium state by proxy with the bath which is impervious to any change in the system~\footnote{This means that action of the system on the bath is excluded from TD considerations.  In fact in TD the bath variables are not dynamical variables determined consistently by the interplay between the system and the bath through their coupled equations of motion, they only provide TD parameters such as temperature or chemical potential.}. Already for classical systems, there is a difference between equilibration and thermalization. Equilibration refers to the system evolving to a steady state after relaxation. It is broader than thermalization, which refers to the system approaching a state described by the Boltzmann distribution. When the system-bath coupling is nonvanishing, such a difference is clearly discernible. For example, the \textit{potential of mean force}~\cite{Kirkwood} is introduced to deal with such a situation. Details can be found in Appendix \ref{S:Cthrt}.

For quantum systems this difference between equilibration and thermalization certainly remains, (see, e.g., \cite{GE16}).  New challenges at zero or very low temperatures posed by non-Markovian environments and in the treatment of non-Markovian dynamics can become prominent. By virtue of its ability to provide a first principles derivation of noise from quantum fluctuations (e.g., for Gaussian noise via the Feynman-Vernon identity, instead of being put in by hand), and linking fluctuations and noise with dissipation and relaxation by dynamical relations  (such as the fluctuation-dissipation relation which can be traced to the unitarity in the original closed system before one coarse grains the environment to a description of mean field dynamics and its fluctuations), the open quantum system approach is also a natural setting for incorporating stochastic thermodynamics \cite{stochTD}, which has seen a wide range of chemical and biological science applications \footnote{Many physical systems show two intermediate stages between quantum and classical, namely, stochastic and semiclassical. Conventional stochastic thermodynamics starts from classical or macroscopic physics. Noise is added in phenomenologically for the consideration of fluctuations phenomena under different circumstances for specific purposes. Being rooted in classical physics conventional stochastic thermodynamics cannot capture the quantum features so easily.  Open quantum systems approach on the other hand starts from microphysics at the quantum level. One can derive the stochastic equations including quantum and thermal noises: Langevin, Fokker-Planck or master equations for the description of fluctuations phenomena. Thus in the quantum open system approach the pathway from the quantum regime to the stochastic regime is well laid out. Taking the distributional average of noise yields the mean field theories at the semiclassical level. To go from quantum to classical physics one needs to add decoherence considerations, but the pathway is completely accessible. The challenge is, can we come up with an appropriate quantum microphysics model for the macroscopic phenomena of interest? }.

The set-up:  In our opinion, the NEq dynamics of open quantum systems, e.g., in the tradition of Feynman-Vernon, Caldeira-Leggett \textit{et al} \cite{FeyVer,CalLeg,HPZ} even though requiring more work,  is the preferred setting for addressing  \textit{new issues in  quantum thermodynamics for future challenges}~\footnote{Similar viewpoint has been expressed by a few others, notably, Kosloff \cite{Kosloff}.}.  This is in comparison with a popular set-up which has been studied more in the literature namely, that of a global thermal state (CGTs) assumed for the combined or closed system (C)  = system (S) + bath (B)~\footnote{The CGTs set up is used by many authors, notably \cite{GelTho, KimMah10, HTL11, SIW15, Seifert16, PhiAnd, GSI88, HIT08,Srednicki94, GLTZ06, PSW06, LPSW09, SF12, Reimann08, Rigol09, PSSV11, Deutsch91, CR10,GE16}.}.    In the CGTs set-up the initial and final states of {\cal C}  are the same, namely the combined system remains in an equilibrium global thermal state, because the dynamics of the combined closed system is unitary. This is visibly closest to the setting of thermodynamics and thus naturally convenient for exploring small extensions of thermodynamics. By contrast the open system NEq (ONEq) approach deals with time evolution of the open system. It requires the specification of the initial conditions and the derivation of the late time behavior of the open system. For those systems that upon interaction with a bath equilibrate at late times, one may then connect its behavior with the descriptions of thermodynamics. For sure this is a many-to-one relation - many different initial conditions can produce the same final steady or equilibrium state, or that there is no common final steady or equilibrium state. A lot depends on the structures of the system, the properties of the bath and the way they interact. All the above mentioned factors need to be considered for interacting quantum many-body systems before we construct thermodynamical quantities, address thermodynamical issues and invoke (or hasten to claim success in revoking \cite{2LawSaga}) the well-established thermodynamical laws.  We will elaborate on their differences in the following.

\subsection{Main Contents}  There are three main components in this paper: 

1) Set-up and conditions:  The physical differences of the set-ups, comparing our open system nonequilibrium (ONEq) approach (level 2) with traditional TD (level 0) on the one hand, and with the global thermal (CGTs) state set-up (level 1) on the other. In TD, as mentioned above, the  system-bath coupling has to be vanishingly small whereas in both the Level 1 and 2 treatments the system-bath coupling can be strong.  We will mention the CGTs approach as many existing works are based on  this setup, but focus more on how to use an open system approach to define and quantify quantum thermodynamics. In a companion paper \cite{GTOS}, we will attempt to build some bridges between these two approaches, via generating functional and reduced density matrix formulations.  The hope is that from the open system perspective, one may be able to identify which entities and concepts are more suitable for treating new problems in QTD and which are residues of the old which may hinder new developments.   Other authors using an open system approach to quantum thermodynamics include Duarte and Caldeira \cite{DuaCal} who treated a coupled-oscillator system by the influence functional method, Carrega \textit{et al.} \cite{CarrWeiss15} who treated a two level system via moment-generating functionals, and Esposito \textit{et al.} \cite{EspositoPRL15} using nonequilibrium Green's functions.

2) Model with exact solutions:   We use a quantum Brownian motion (QBM) model of harmonic oscillators with strong-coupling both within the system ($\sigma$) and interacting with a scalar field bath ($\gamma$). The merit of this model, which represents a rather broad class of physical problems, is that being a Gaussian system it can yield exact solutions which enable us to cross examine the relevant issues, leaving little room for speculation. Even when familiar quantities like energy and entropy can be defined in different ways under different conditions, since we are treating NEq dynamics, if we make precise specific conditions, these quantities are defined. There is no worry about ambiguity. The results from this model study are used for addressing the following issues:

3) Issues and consequences: 
 
\paragraph{Energy extensivity.}  Thermodynamic functions are well defined under the conditions when thermodynamics theory is viable, namely, that the system is very weakly coupled to the bath, the bath being a passive source which provides a temperature parameter, not a dynamical variable which can back-react on the system. It is a meaningful question to ask if the nice properties we are accustomed to in conventional thermodynamics, e.g., the extensive property of internal energy, will still hold for strongly interacting quantum systems. In the model we studied here we answer this question in the affirmative, that the internal energy remains extensive under strong coupling.  

\paragraph{Heat capacity.}  From the internal energy we calculate the heat capacity and examine its behavior toward $T=0$. We find a power law, not an exponential decay. This has significant implications. This aids us to address a version of the third law and to resolve some puzzles raised in the literature  such as the claimed negative specific heat near absolute zero even in well behaved systems \cite{HasegawaJMP}.

\paragraph{Third law.}  There are several formulations and statements of the third law. We approach it from the behavior of the heat capacity near absolute zero, which aids us to resolve some puzzles raised in the literature such as the claimed negative specific heat near absolute zero even in well behaved systems \cite{HasegawaJMP}, and address some concerns expressed by Hanggi, Ingold, Talkner, Weiss, \textit{et al}  \cite{HanIng06,HIT08,IHT09,AIW14,SIW15,FordOC}.

\paragraph{} Vedral \textit{et al} \cite{Vedral,Vedral1} invoked heat capacity as an indicator of entanglement, and raised the issue of how the entanglement at a system's ground state bears on the third law. For the (spin) system they studied they made the claim that ``the validity of the third law of thermodynamics relies on quantum entanglement". Using the behavior of the heat capacity at $T=0$ we derived here, combined with our earlier results on the entanglement between two coupled-oscillators interacting with a zero temperature bath \cite{HHPRD}, we show that this is not the case at least for the coupled oscillator system. There is no connection between entanglement in the system and the third law.

\paragraph{} The ONEq approach we adopt for the dynamics of the system provides means to calculate entropy production, but not before the meaning and definition of entropy for interacting quantum systems can be understood and clarified. We say this because even the most commonly invoked von Neumann entropy has problems if not used and understood properly. We shall mention this issue at the end of this paper but leave a proper treatment of heat, entropy, entanglement, and from it the first and second Law,  to the second paper \cite{QTD2} in this series.

\subsection{Closed-system Global Thermal State versus Open-system Evolved Equilibrium State}

We begin by stating a few basic facts connecting the three levels of treatments:  level 0  thermodynamics (TD), level 1 closed system (system and environment combined) in a global thermal state (CGTs) and level 2 open system evolving to an equilibrium state (ONEq). 

1) Traditional statistical mechanics treats many-body systems in thermal (canonical distribution) and chemical (grand canonical) equilibrium. The starting point of quantum statistical mechanics (QSM) is probability density, no quantum phase information is invoked.  This is encoded in \textit{the two fundamental postulates of quantum statistical mechanics}: equal \textit{a priori} probability to all accessible states and random phase approximation. Thus from a quantum information viewpoint, the system of interest to QSM is already fully decohered in the energy basis and behaves classically in an effective way -- what is quantum in QSM only pertains to quantized energy levels and particle spin-statistics.    

2) Partition function is well defined only for systems in thermal equilibrium. It \textit{is ill-defined for systems under nonequilibrium conditions when the notion of temperature is lacking.  } Pathologies may ensue if it is forced upon even perfectly normal systems (in contradistinction to systems for which the canonical ensemble does not exist and the heat capacity is negative in the microcanonical ensemble, such as gravitating systems). As noted in \cite{EspositoJSM}, if one proceeds from assuming that the combined system + environment is in a thermal state the behavior of the heat capacity of the system is different when it is derived from the energy of the central system at equilibrium or  from a partition function approach \cite{HanIng06,HIT08}. By examining the open system nonequilibrium dynamics with no reference to the partition function, one could avoid these pathologies.  Likewise,  old notions such as \textit{the Hamiltonian of mean force \cite{Kirkwood} are only meaningful in the conceptual framework of equilibrium systems }\cite{HTL11} as in CGTs. 

3)  The oft-heard statement, that the generating functional (in quantum field theory) is equal to the partition function (in equilibrium statistical mechanics) is true  only for thermal fields, i.e., there exists a canonical distribution where a thermal state is well defined for all times. This statement arises from treating thermal (finite temperature) fields with imaginary (Matsubara) time quantum field theory. \textit{If one uses real time representation to describe the NEq dynamics of open systems  the generating functional remains well defined but it is not the (canonical) partition function defined in imaginary time.}

4)  If an open system upon interaction with its environment can equilibrate at late times, and if it is further thermalized, it enters a thermal state. But this equilibrium state is different from that of a system in contact with a heat bath which behaves in a totally passive and non-dynamical way, in particular, with no back-action on the system. The latter is where a theory of quantum thermodynamics is often constructed, namely, from a simple extension of conventional classical thermodynamics. The difference lies in the dynamical correlations between the system and the bath, which conventional thermodynamics ignores completely by assuming a vanishingly small coupling.

5)  There are important differences between the ONEq and the CGTs setups in their goals, approaches and consequences. There are also key differences between CGTs and thermalization in a closed quantum system in the vein of eigenstate thermalization hypothesis. Since the latter is an active topic in the last decade with many important contributions, we can only focus on the differences from the open system approach on the specific issues of interest to us here and cite some representative references and reviews for interested readers to appreciate the scope \cite{Srednicki94, GLTZ06, PSW06, LPSW09, SF12, Mahler, Reimann08, Rigol09, PSSV11, Deutsch91, CR10}. We highlight some key features below. In Appendix \ref{S:fkbser}, we will illustrate some aspects of the ONEq and the CGTs setups with a simple model calculation from the ubiquitous QBM model.
	\paragraph{} Set-up and goals.  In this work we assumed the field to be in a thermal state prior to its coupling to the system oscillators, which initially can be in an arbitrary state. Thus the system oscillators and the field are generically out of equilibrium before and after the interaction is turned on. Our focus is on the subsequent dynamics and relaxation of the system oscillators, without assuming the coupling to be weak.
	The `pure state quantum statistical mechanics' assumes the whole system is in a pure state throughout. The main goal in~\cite{GE16} is to \textit{derive} statistical mechanics and thermodynamics from quantum mechanics without resorting to the notion of ensembles. It aims to show that even pure quantum states of interacting many-body systems can display relaxation to equilibrium and, in special cases, thermalize.
	\paragraph{} The methods developed in the `pure state quantum statistical mechanics' literature are usually applied to closed systems without an intrinsic system-bath distinction. For instance, in Cramer et al.~\cite{CE09}, a one-dimensional harmonic lattice is shown to locally relax to Gaussian states for arbitrary choice of subsystem and a wide class of initial states. The authors note the following ``Every part of the system forms the environment of the other...''. Here we are only concerned with the relaxation of the system oscillators and do not require that any part the environment relaxes (In fact, in Appendix \ref{S:fkbser} we show that the bath modes never reach a steady state).
	\paragraph{} There is an important distinction in the meaning of \textit{equilibration}. In the pure state quantum statistical mechanics paradigm equilibration is used more broadly to indicate relaxation to a steady state. For instance, depending on the context, the relaxation of the expectation values of certain operators to fixed values or of the reduced density matrix is considered equilibration. In our open system approach equilibration has a very specific meaning, see Eq.~\eqref{eq:eq}. In other words, there exists an environment and an interaction Hamiltonian such that the equilibrium state is obtained by tracing out the environment in the global thermal state.
	\paragraph{} Integrability. It has been discussed in~\cite{GE16} that integrable quantum models indeed equilibrate to a suitable generalized Gibbs ensemble. Furthermore~\cite{CEF11} examine the behavior of the one- and two-point correlation functions after a quench in various models, and it is found that the relaxation dynamics and equilibrium values can be well understood by means of a generalized Gibbs ensemble.

\subsection{Key Results}  
1) Energy extensivity. In conventional thermodynamics, when the intra-system coupling is negligible, the internal energy is extensive in term of the number of the oscillators, like the case of the dilute gas. When this coupling is finite, we may instead understand the extensive property of the internal energy in terms of the normal modes of the coupled oscillators. We have shown that with this definition of extensivity the internal energy becomes extensive after the system reaches equilibrium, as implied by \eqref{E:pendj}. It is interesting to note that the degrees of freedom of the oscillators used to describe the extensive property of the internal energy are neither the original degrees of freedom associated with each oscillator, nor the modes that decouple their equations of motion. Rather they are the degrees of freedom that diagonalize the oscillation frequency matrix $\pmb{\Omega}_{p}^{2}$. In this regard, the extensive property of the internal energy in the final equilibrium state is the same as that of coupled oscillators in conventional thermodynamics, that is, in the vanishing system-bath coupling limit. This offers an explicit theoretical justification, from the open-system viewpoint, of conventional thermodynamics when applied to such a many-body system.

2) Heat capacity. When the system of $N$ coupled oscillators in a shared scalar field bath reaches equilibrium, its heat capacity is shown to be always non-negative for all nonzero bath temperatures, and it moves towards zero only if the bath temperature approaches zero. These properties are independent of the spatial arrangement of the oscillators, the inter-oscillator coupling and the system-bath interaction strength, as long as the collective non-Markovian motion of the system is stable.

3) The third law. Therefore from the viewpoint of behavior of the heat capacity at $T=0$ for this class of systems in an equilibrium state the third law is not violated.  In this connection we also addressed the issue of entanglement and the third law pertaining to heat capacity. It was stated in \cite{Vedral} the following: `` One may therefore say that in these systems the validity of the third law of thermodynamics relies on quantum entanglement ...". Our view is that The third law depends on the nondegeneracy of the ground state manifold and has nothing to do with entanglement directly.  Indeed it has been shown \cite{HHPRD} in the case of two spatially separated but coupled oscillators in a zero-temperature shared bath that the equilibrated state of this two oscillator system is not always entangled. For example, with sufficiently strong oscillator bath interaction, the reduced state of the two oscillators is separable. (See Fig.~3 in \cite{HHPRD}.) Here, we have shown that the heat capacity of the coupled harmonic oscillator system goes to  zero independent of the system-bath interaction strength. Thus it offers a counterexample to the above claim, that the validity of the third law of thermodynamics relies on quantum entanglement.

\section{Brownian motion of systems of oscillators strongly coupled to an environment }

We now begin our detailed model study for considering the viability in the establishment of a thermodynamics theory of open quantum systems.

Consider a collection of coupled quantum harmonic oscillators in a shared finite-temperature $\beta^{-1}$ bath modeled by a massless scalar field in 1+3 Minkowski spacetime.  The action of such a system is given by
\begin{align}\label{E:dbeksd}
	S&=\int\!dt\,\sum_{i}\Bigl[\frac{m}{2}\,\dot{\chi}_{i}^{2}(t)-\frac{m\omega_{b}^{2}}{2}\,\chi_{i}^{2}(t)\Bigr]-\int\!dt\,\sum_{j>i}m\sigma\,\chi_{i}(t)\chi_{j}(t)+\int\!d^{4}x\;j(x)\phi(x)\notag\\
	&\qquad\qquad\qquad\qquad+\int\!d^{4}x\;\frac{1}{2}\,\partial_{\mu}\phi(x)\partial^{\mu}\phi(x)\,,
\end{align}
with $x=(t,\mathbf{x})$. Each oscillator is located at a fixed spatial coordinate $\mathbf{z}_{i}$, and has the same mass $m$ and bare natural frequency $\omega_{b}$. The ``current'' $j(x)$ in the oscillator-bath interaction term takes the form $\displaystyle j(x)=e\sum_{i}\chi_{i}(t)\,\delta^{(3)}(\mathbf{x}-\mathbf{z}_{i})$, with $e$ the coupling strength between the oscillator and the bath. The parameter $\sigma$ is the strength of direct coupling between two oscillators and assumed to be positive for concreteness~\footnote{It can take either sign which only affects the interpretation of the normal modes. In  addition, the numerical values of $e$ and $\sigma$ are confined to ranges where instability in the dynamics is avoided. We will comment on this point later.}.

Here we suppose that the initial state of the combined system is a factorized state, given by
\begin{align}\label{E:brhtsd}
	\rho_{i}&=\rho_{i}^{(\chi)}\otimes\rho_{\beta}^{(\phi)}\,,&\rho^{(\phi)}_{\beta}&=Z_{\phi}^{-1}\,e^{-\beta H^{(\phi)}}\,,&Z_{\phi}&=\operatorname{Tr}_{\phi}\Bigl\{e^{-\beta H^{(\phi)}}\Bigr\}\,,
\end{align}
where $H^{(\phi)}$ is the free Hamiltonian of the scalar field. While the field is initially prepared in a thermal state, the initial state $\rho_{i}^{(\chi)}$ of the system can be quite arbitrary. Thus in the beginning the system and the bath are not in equilibrium, nor correlated. We will let them interact and evolve in time.  We will explore and make explicit the conditions when the system can and will reach equilibration~\footnote{The equilibration issue for classical coupled oscillator systems was studied before by, e.g., Agarwal~\cite{Agarwal}}. This equilibrium state in general will have no resemblance to the thermal state of the combined system, nor of the reduced system. Thus the setup here is in strong contrast to the closed system globally-thermal state (CGTs) often adopted in the discussions of quantum thermodynamics. There,  for  the total Hamiltonian of the combined system   
$H=H^{(\chi)}+H_{\mathrm{int}}+H^{(\phi)}$ it is assumed that 
\begin{align}
	\rho_{i}&=\rho_{\beta}\,,&\rho_{\beta}&=Z^{-1}\,e^{-\beta H}\,,&Z&=\operatorname{Tr}_{\chi,\,\phi}\,\Bigl\{e^{-\beta H}\Bigr\}\,,
\end{align}
In the global thermal state, the system has already established correlation with the bath, and the interaction between the system and the bath is such that it maintains this correlation throughout. Since they are in thermal equilibrium, the combined system will remain in the global thermal state unless an external disturbance is introduced to bring the system out of equilibrium.

The evolution of the combined system is governed by the unitary evolution operator $U$
\begin{equation}
	U(t_{f},t_{i})=\operatorname{T}\,\exp\biggl[-i\int_{t_{i}}^{t_{f}}\!ds\;H(s)\biggr]\,,
\end{equation}
where $\operatorname{T}$ denotes chronological time-ordering and $H$ is the Hamiltonian operator of the combined system that corresponds to the action \eqref{E:dbeksd}. Given the initial state \eqref{E:brhtsd} of the total system, the density matrix of the reduced system of  interest is then given by
\begin{equation}\label{E:gneuss}
	\rho^{(\chi)}(t_{f})=\operatorname{Tr}_{\phi}\Bigl\{U(t_{f},t_{i})\,\rho(t_{i})\,U^{\dagger}(t_{f},t_{i})\Bigr\}\,,
\end{equation}
after we trace out the degrees of freedom of the bath. The reduced density matrix of the system enables us to calculate the quantum expectation values of the operators, say $O^{(\chi)}$, associated with the system by
\begin{equation}
	\langle O^{(\chi)}\rangle=\operatorname{Tr}_{\chi}\Bigl\{\rho^{(\chi)}(t_{f})O^{(\chi)}\Bigr\}\,,
\end{equation}
from which we may construct the quantum thermodynamics of the system in a nonequilibrium setting.

When the initial state \eqref{E:brhtsd} is Gaussian, Eq.~\eqref{E:gneuss} can be evaluated analytically and exactly for the combined system described by \eqref{E:dbeksd}. Using a path integral representation of $U$ and $U^{\dagger}$,  the reduced density matrix elements in \eqref{E:gneuss} become
\begin{align}
	\rho^{(\chi)}(\chi_{f},\chi'_{f},t_{f})&=\int_{-\infty}^{\infty}\!d\chi_{i}\,d\chi'_{i}\;\rho^{(\chi)}(\chi_{i},\chi'_{i},t_{i})\\
	&\qquad\qquad\qquad\int_{\chi_{i}}^{\chi_{f}}\!\mathcal{D}\chi_{+}\int_{\chi'_{i}}^{\chi'_{f}}\!\mathcal{D}\chi_{-}\;\exp\Bigl\{i\,S^{(\chi)}[\chi_{+}]-i\,S^{(\chi)}[\chi_{i}]\Bigr\}\mathcal{F}[\chi_{+},\chi_{-}]\,,\notag
\end{align}
where $S^{(\chi)}$ is the action of the system alone and $\chi_{\pm}$ denotes the system variable in the respective forward and backward time branches.  This is where the `closed-time-path' integral (CTP) / `in-in'  formalism \cite{CTP} or its close kin, the Feynman-Vernon~\cite{FeyVer} influence functional (IF) $\mathcal{F}[\chi_{+},\chi_{-}]$ becomes  particularly useful.

For  a Gaussian bath, its influence  on the system  can be understood as caused by a classical noise by way of the Feynman-Vernon Gaussian identity:  the  imaginary part of the IF can be represented by a stochastic source term which inherits the quantum statistics of the bath. Using techniques from the CTP formalism, a revised imaginary part combined with the original real part of the influence functional, together with the action of the system, form a new effective action which is real, known as the stochastic effective action. Variation of this stochastic effective action produces a Langevin equation which describes the evolution of the reduced system. For details and working examples in this functional approach to open quantum system dynamics, please refer to Appendix~\ref{IF}. In what follows, we will adopt the Langevin equation approach in the discussion of quantum thermodynamics at strong coupling.

\subsection{Langevin Equation for the Reduced System}

Following this well-established procedure the Langevin equation for the stochastic dynamics of the $i^{\mathrm{th}}$ oscillator (strongly) interacting with other oscillators and their shared bath with action \eqref{E:dbeksd} is given by
\begin{align}\label{E:bejhrys}
	m\,\ddot{\chi}_{i}(t)+m\omega_{b}^{2}\,\chi_{i}(t)+\sum_{j\neq i}m\sigma\,\chi_{j}(t)-e^{2}\int^{t}_{0}\!ds\,\sum_{j}G_{R}^{(\phi)}(t-s,\mathbf{z}_{i}-\mathbf{z}_{j})\,\chi_{j}(s)=e\,\xi_{i}(t)\,.
\end{align}
In addition to the drag force and the quantum fluctuations of the bath found in a single oscillator system a new factor entering in the present coupled-oscillator shared-bath system is the induced interaction between the oscillators through their respective interaction with the scalar field bath. The field-environment mediated effect is non-Markovian in nature (see e.g., \cite{ASH06,LinHu09,HH15}), often absent~\footnote{Unless the spatial information is retained in the system-bath interaction.} in a shared bath modeled by a collections of oscillators~\footnote{The differences between a oscillator bath and a field bath will be discussed in Appendix~\ref{S:rbjdhe}.} (see, e.g., \cite{CHY,PazRec}). This feature introduces additional complications and brings forth new physics in analyzing the stochastic dynamics of the quantum many-body system, as well as its quantum thermodynamics.

The statistics of the Gaussian noise field $\xi_{i}(t)=\xi(t,\mathbf{z}_{i})$  is determined completely by the first two moments
\begin{align}
	\langle \xi_{i}(t)\rangle&=0\,,&\langle\xi_{i}(t)\xi_{j}(t')\rangle&=G_{H}^{(\phi)}(t-t',\mathbf{z}_{i}-\mathbf{z}_{j})\,.
\end{align}
All the higher even moments can be expressed by the second moment with the Wick expansion while all the odd moments vanish. The $\langle\cdots\rangle$ notation denotes either an ensemble average or expectation value, depending on whether the variable under consideration is stochastic or quantum. The two kernel functions $G_{R}^{(\phi)}(x-x')$ and $G_{H}^{(\phi)}(x-x')$ are most relevant for our present study:  They are the retarded and the Hadamard functions of the scalar field $\phi$ in its thermal state, defined by
\begin{align}
	G_{R}^{(\phi)}(x-x')&=i\,\theta(t-t')\,\bigl[\phi(x),\phi(x')\bigr]=\frac{1}{4\pi r}\,\theta(\tau)\,\Bigl[\delta(\tau-r)-\delta(\tau+r)\Bigr]\,,\label{E:ejfdhwe}\\
	G_{H}^{(\phi)}(x-x')&=\frac{1}{2}\,\langle\bigl\{\phi(x),\phi(x')\bigr\}\rangle=-\frac{1}{8\pi\beta r}\left[\coth\frac{\pi(\tau-r)}{\beta}-\coth\frac{\pi(\tau+r)}{\beta}\right]\,,\label{E:ejfdhwe1}
\end{align}
with $\tau=t-t'$ and $r=\lvert\mathbf{x}-\mathbf{x}'\rvert$. Since they are time-translation invariant, their Fourier transforms with respect to the $\tau$ variable  satisfy the well-known relation,
\begin{equation}
	\tilde{G}_{H}^{(\phi)}(\omega;r)=\coth\frac{\beta\omega}{2}\,\operatorname{Im}\tilde{G}_{R}^{(\phi)}(\omega;r)\,,
\end{equation}
where the Fourier transformation of the function $f(\tau)$ is defined by
\begin{align}
	\tilde{f}(\omega)&=\int_{-\infty}^{\infty}\!d\tau\;e^{+i\,\omega\tau}\,f(\tau)\,,&f(\tau)&=\int_{-\infty}^{\infty}\!\frac{d\omega}{2\pi}\;e^{-i\,\omega\tau}\,\tilde{f}(\omega)\,.
\end{align}

Introducing the matrix representation of the equation of motion \eqref{E:bejhrys},
\begin{align}
	\pmb{\Xi}&=\begin{pmatrix}\chi_{1}\\\chi_{2}\\\vdots\\\chi_{n}\end{pmatrix}\,,&\pmb{\Omega}_{b}^{2}&=\begin{pmatrix}
	\omega_{b}^{2} &\sigma          &\cdots &\sigma\\
	\sigma         &\omega_{b}^{2} 	&\cdots &\sigma\\
	\vdots		   &\vdots         	&\ddots &\vdots\\
	\sigma		   &\sigma 			&\cdots &\omega_{b}^{2}
	\end{pmatrix}\,,
	&\pmb{\xi}&=\begin{pmatrix}\xi_{1}\\\xi_{2}\\\vdots\\\xi_{n}\end{pmatrix}\,,
\end{align}
where $[\mathbf{G}(\tau)]_{ij}\equiv G(\tau,\mathbf{z}_{i}-\mathbf{z}_{j})$, we obtain a matrix equation 
\begin{equation}\label{E:bdeere}
	\ddot{\pmb{\Xi}}(t)+\pmb{\Omega}_{b}^{2}\cdot\pmb{\Xi}(t)-\frac{e^{2}}{m}\int_{0}^{t}\!ds\;\mathbf{G}_{R}^{(\phi)}(t-s)\cdot\pmb{\Xi}(s)=\frac{e}{m}\,\pmb{\xi}(t)\,.
\end{equation}
The solution generically takes the form
\begin{equation}\label{E:fbiiwed}
	\pmb{\Xi}(t)=\mathbf{d}_{1}(t)\cdot\pmb{\Xi}(0)+\mathbf{d}_{2}(t)\cdot\dot{\pmb{\Xi}}(0)+\frac{e}{m}\int_{0}^{t}\!ds\;\mathbf{d}_{2}(t-s)\cdot\pmb{\xi}(s)\,,
\end{equation}
where $\{\pmb{\Xi}(0),\dot{\pmb{\Xi}}(0)\}$ are the initial conditions and $\mathbf{d}_{i}(t)$ are a special set of homogeneous solutions to \eqref{E:bdeere}. The actual form of $\mathbf{d}_{1}$ is not important but the Fourier transform of $\mathbf{d}_{2}(t)$ is
\begin{equation}\label{E:efjdbjbd}
	\tilde{\mathbf{d}}_{2}(\omega)=\Bigl[\pmb{\Omega}_{b}^{2}-\omega^{2}\mathbf{I}-\frac{e^{2}}{m}\,\tilde{\mathbf{G}}_{R}^{(\phi)}(\omega)\Bigr]^{-1}\,.
\end{equation}
Later it will be shown that for certain choices of parameters, the solution to \eqref{E:bdeere} can exhibit instability and grows indefinitely when $t$ approaches infinity. In these case, the homogeneous solutions $\mathbf{d}_{i}(t)$ are not integrable, 
\begin{equation}
	\int_{-\infty}^{\infty}\!dt\;\lvert\mathbf{d}_{i}(t)\rvert\not<\infty\,,
\end{equation}
so their Fourier transforms do not exist in the usual sense. Thus when results are expressed in terms of $\tilde{\mathbf{d}}_{i}(\omega)$, it pays to be careful about their interpretations. 

Finally we note from \eqref{E:fbiiwed} that the oscillator is driven not only by the local noise at its very location, but is also affected by the quantum fluctuations of the bath at the locations of the other oscillators. This intriguing feature is essential in keeping the energy balance of the reduced system after it equilibrates.  This will become clearer when we calculate the energy balance in Sec.~\ref{S:erhbdjhfbs}.

\subsection{Covariance Matrix}
From \eqref{E:fbiiwed}, if there exists an equilibrium state~\footnote{The existence of the equilibrium state is related to the fact that the complex poles of $\tilde{\mathbf{d}}_{2}(\omega)$ lie on the upper half of the complex $\omega$ plane. See Sec.~\ref{S:jkjdf}. This is also the very basis on which we can discuss the fluctuation-dissipation relation of the reduced system and the energy balance among the dissipative, retarded and noise force terms.} for the reduced system, then the moment
\begin{align}
	\pmb{\sigma}_{\chi\chi}(t)=\frac{1}{2}\,\langle\{\pmb{\Xi}(t),\pmb{\Xi}^{T}(t)\}\rangle
\end{align}
at late time,  after the reduced system completely relaxes,  is well defined and given by
\begin{align}\label{E:uirysbf}
	\lim_{t\to\infty}\pmb{\sigma}_{\chi\chi}(t)&=\frac{e^{2}}{m^{2}}\int_{-\infty}^{\infty}\!\frac{d\omega}{2\pi}\;\tilde{\mathbf{d}}_{2}(\omega)\cdot\tilde{\mathbf{G}}_{H}^{(\phi)}(\omega)\cdot\tilde{\mathbf{d}}_{2}^{\dagger}(\omega)\,,
\end{align}
where the superscripts $T$ and $\dagger$ denote the transposition and Hermitian conjugate of the matrix, respectively.

Since $\mathbf{d}_{2}$ is a symmetric matrix, we observe that 
\begin{align}
	\tilde{\mathbf{d}}_{2}(\omega)-\tilde{\mathbf{d}}_{2}^{\dagger}(\omega)=2i\,\operatorname{Im}\tilde{\mathbf{d}}_{2}(\omega)&=2i\,\frac{e^{2}}{m}\,\tilde{\mathbf{d}}_{2}(\omega)\cdot\operatorname{Im}\tilde{\mathbf{G}}_{R}^{(\phi)}(\omega)\cdot\tilde{\mathbf{d}}_{2}^{\dagger}(\omega)\,,
\end{align}
with the help of the matrix identity
\begin{equation}\label{E:eridsd}
	\mathbf{A}^{-1}-\mathbf{B}^{-1}=\mathbf{A}^{-1}\cdot\bigl(\mathbf{B}-\mathbf{A}\bigr)\cdot\mathbf{B}^{-1}\,,
\end{equation}
for two nonsingular matrices $\mathbf{A}$, $\mathbf{B}$. Thus we can write $\pmb{\sigma}_{\chi\chi}(\infty)$ in \eqref{E:uirysbf} as
\begin{align}\label{E:fbkfq1}
	\pmb{\sigma}_{\chi\chi}(\infty)=\frac{1}{m}\int_{-\infty}^{\infty}\!\frac{d\omega}{2\pi}\;\coth\frac{\beta\omega}{2}\,\operatorname{Im}\tilde{\mathbf{d}}_{2}(\omega)=\operatorname{Im}\int_{-\infty}^{\infty}\!\frac{d\omega}{2\pi}\;\coth\frac{\beta\omega}{2}\,\tilde{\mathbf{G}}_{R}^{(\chi)}(\omega)\,,
\end{align}
where the retarded Green's function $\tilde{\mathbf{G}}_{R}^{(\chi)}(\omega)$ of the reduced system is in fact
\begin{equation}
	\tilde{\mathbf{G}}_{R}^{(\chi)}(\omega)=\frac{1}{m}\,\tilde{\mathbf{d}}_{2}(\omega)\,.
\end{equation}
Similarly we introduce
\begin{align}
	\pmb{\sigma}_{\upsilon\upsilon}(t)=\frac{1}{2}\,\langle\{\dot{\pmb{\Xi}}(t),\dot{\pmb{\Xi}}^{T}(t)\}\rangle\,,
\end{align}
and at  late times it becomes
\begin{align}\label{E:fbkfq2}
	\pmb{\sigma}_{\upsilon\upsilon}(\infty)=\operatorname{Im}\int_{-\infty}^{\infty}\!\frac{d\omega}{2\pi}\;\omega^{2}\coth\frac{\beta\omega}{2}\,\tilde{\mathbf{G}}_{R}^{(\chi)}(\omega)\,.
\end{align}
This integral in general is not well-defined due to the presence of ultraviolet (UV) divergence so regularization is needed.

\subsection{Internal Energy}

We define the internal energy of the system as its total mechanical energy. The total mechanical energy $E$ of the coupled oscillators is
\begin{align}
	E(t)&=\sum_{i}\Bigl[\frac{m}{2}\,\langle\dot{\chi}_{i}^{2}(t)\rangle+\frac{m\omega_{p}^{2}}{2}\,\langle\chi_{i}^{2}(t)\rangle\Bigr]+\sum_{j>i}m\sigma\,\langle\chi_{i}(t)\chi_{j}(t)\rangle\notag\\
	&=\frac{m}{2}\operatorname{Tr}\,\Bigl\{\pmb{\sigma}_{\upsilon\upsilon}(t)+\pmb{\Omega}_{p}^{2}\cdot\pmb{\sigma}_{\chi\chi}(t)\Bigr\}\,.
\end{align}
Here $\operatorname{Tr}$ is the matrix trace, and the matrix $\pmb{\Omega}_{p}^{2}$ is defined in a way similar to $\pmb{\Omega}_{b}^{2}$ except that the elements $\omega_{b}^{2}$ in $\pmb{\Omega}_{b}^{2}$ are replaced $\omega_{p}^{2}$, where $\omega_{p}$ is the renormalized or physical frequency, which will be determined by the system preparation at the experimental energy scale. The difference between them is not necessarily large and depends on the choice of the cutoff frequency $\Lambda$, such that
\begin{equation}\label{E:bsrthbf}
	\omega^{2}_{p}-\omega_{b}^{2}=-\frac{4\gamma\Lambda}{\pi}\,,
\end{equation}
where the damping constant $\gamma$ is equal to $\gamma=e^{2}/(8\pi m)$. In the equilibrium state (note it is not necessarily the Gibbs state \cite{SFTH}), the total mechanical energy becomes
\begin{align}\label{E:efkbsdf}
	E(\infty)=\frac{m}{2}\operatorname{Tr}\,\Bigl\{\pmb{\sigma}_{\upsilon\upsilon}(\infty)+\pmb{\Omega}_{p}^{2}\cdot\pmb{\sigma}_{\chi\chi}(\infty)\Bigr\}&=\frac{1}{2}\operatorname{Im}\int_{-\infty}^{\infty}\!\frac{d\kappa}{2\pi}\;\coth\frac{\beta\kappa}{2}\,\operatorname{Tr}\,\Bigl\{\Bigl [\kappa^{2}\mathbf{I}+\pmb{\Omega}_{p}^{2}\Bigr]\cdot\tilde{\mathbf{d}}_{2}(\kappa)\Bigr\}\,.
\end{align}
The heat capacity $C$ is then given by 
\begin{equation}\label{E:dnerjdgd} 
	C=\frac{\partial E}{\partial T}=-\beta^{2}\,\frac{\partial E}{\partial\beta}\,.
\end{equation}
The evaluation of $E(\infty)$ can be trickier than expected if regularization is not properly introduced.

At this point it may be desirable to get some physical feel of the dynamics and thermodynamics of the system. In Appendix~\ref{S:one} we treat a simpler system of one and two oscillators so that we can see the subtleties involved in the non-Markovian dynamics and thermodynamics of a strongly interacting open quantum system. Otherwise, we may proceed to the formal development for the $N$ oscillator system. 
\begin{figure}
\centering
    \scalebox{0.7}{\includegraphics{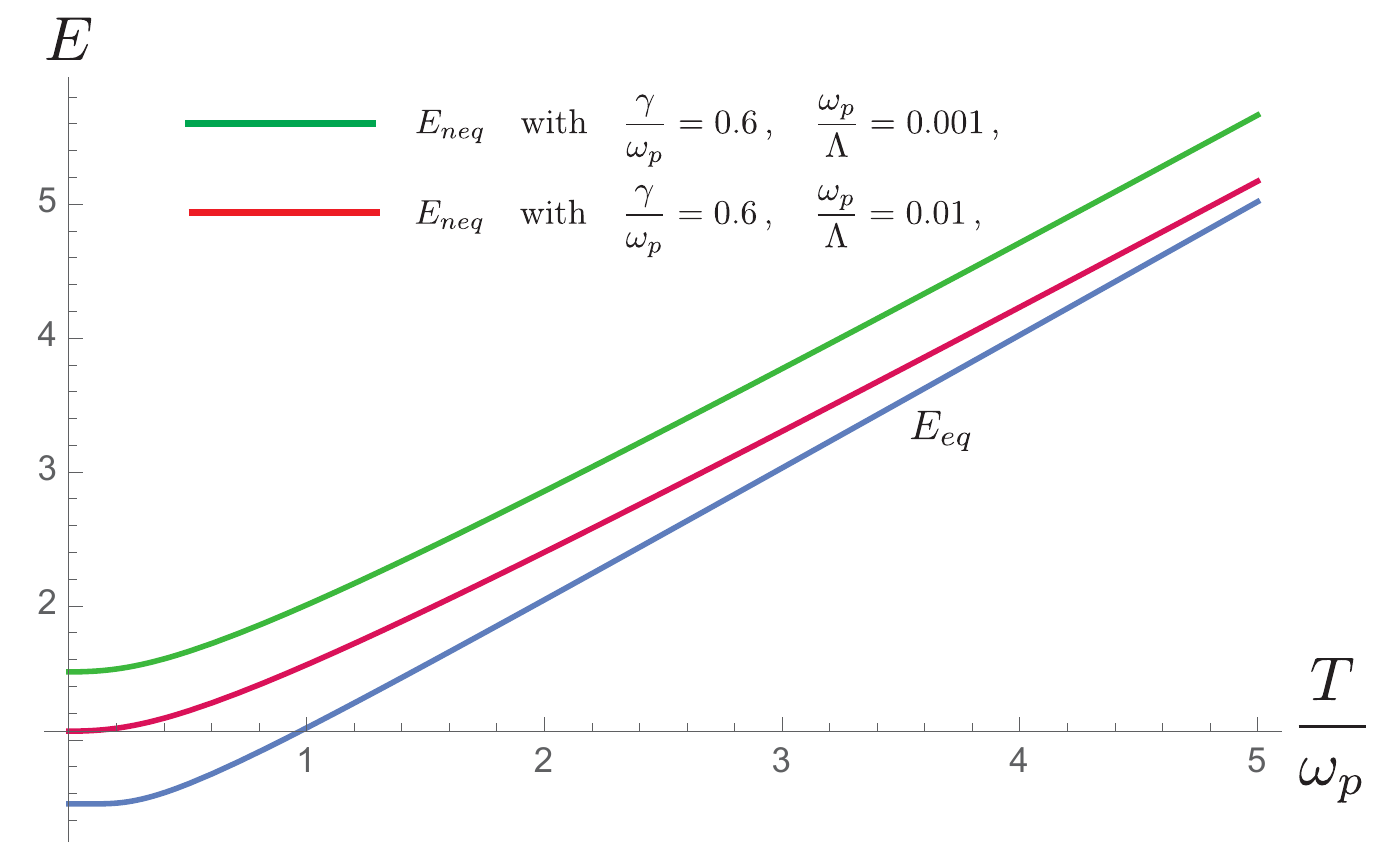}}
    \caption{We compare the difference in predictions from the equilibrium and nonequilibrium approaches. The blue (solid) curve describes the temperature dependence of the mechanical energy of the oscillator in the equilibrium thermal state. The other two curves give the temperature variations of the mechanical energy of the oscillator which interacts with a bath that is initially in the thermal state. As an illustration of the nontrivial effect in the choice of the cutoff scale, these two curves have different cutoff scales.}\label{Fi:energy}
\end{figure}
\section{Thermodynamics of Open Quantum Systems}\label{S:betiyer}

As mentioned in the beginning, in this paper we use the model of an $N$ coupled-oscillator system interacting with a scalar field bath  to address the energy and heat capacity issues and discuss the third law of thermodynamics.

Before we launch our studies of the $N$ coupled-oscillator model in full rigor, it would be useful to gain some feeling of the anticipated physical results for simpler cases. Hence, we summarize what we have learned from the one and two coupled harmonic oscillators examples below. Details for these two cases are placed in Appendix \ref{S:one}.
\begin{figure}
\centering
    \scalebox{0.7}{\includegraphics{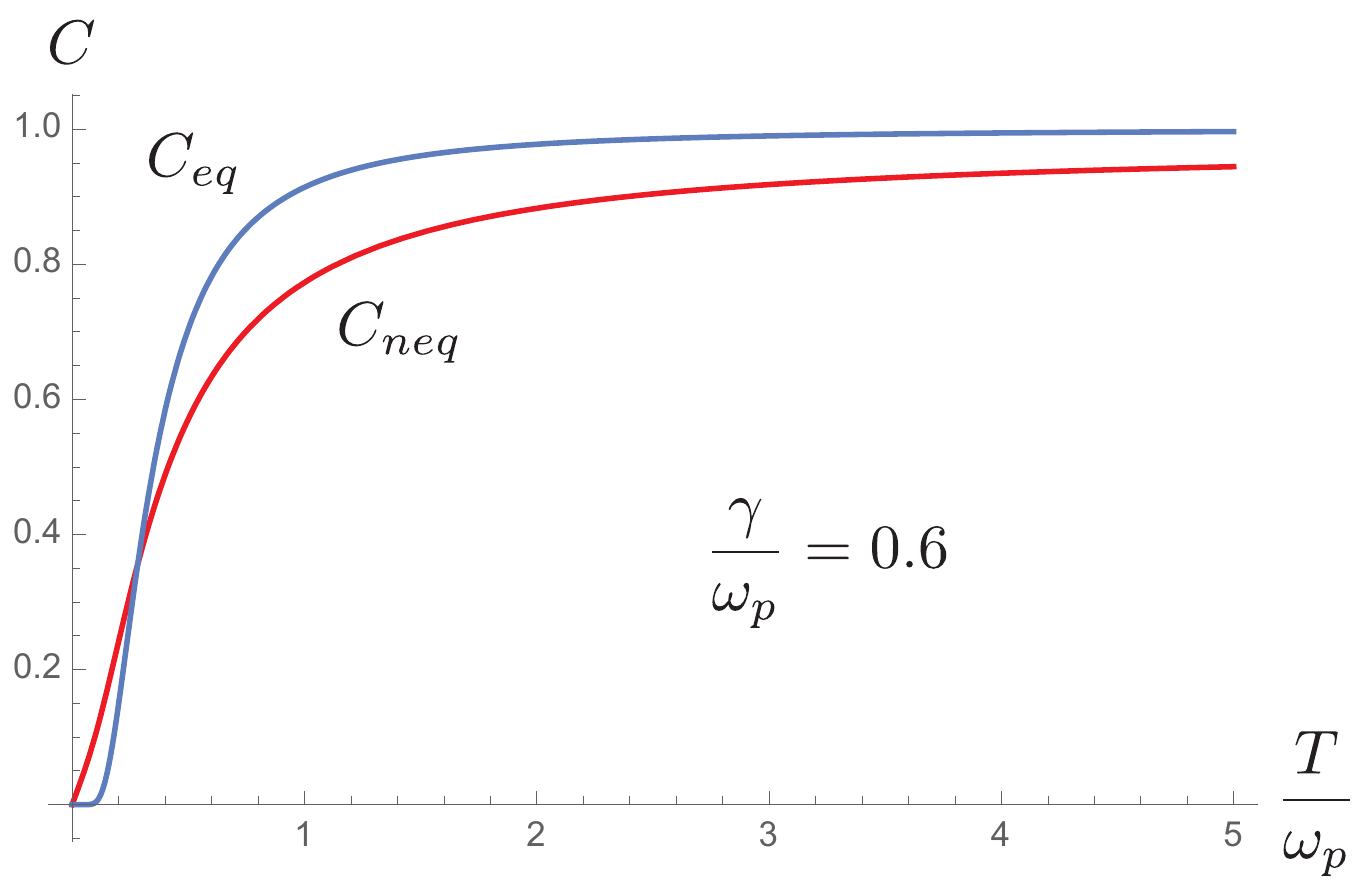}}
    \caption{We show the difference of the heat capacity of the single oscillator obtained from equilibrium $C_{eq}$ and nonequilibrium approaches $C_{neq}$. The difference is more significant for stronger coupling between the oscillator and the bath. At low temperature, 
	$C_{eq}$ approaches zero exponentially, while $C_{neq}$  has an algebraic fall-off.}\label{Fi:heatC}
\end{figure}
For a system containing just one harmonic oscillator coupled to  a thermal bath with finite coupling strength the heat capacity behaves qualitatively different at low temperatures from traditional thermodynamics, which assumes that the system-bath coupling is vanishingly small.  
A new scale associated with the coupling strength  $\gamma$ appears. As shown in \eqref{E:btweyw}, the heat capacity approaches zero following a  \textit{power law} when the bath temperature is lowered to zero. In contrast, quantum statistical mechanics calculations assuming  vanishing system-bath coupling predict in \eqref{E:beersd} that the heat capacity approaches zero \textit{exponentially} as the bath temperature is lowered to zero.  This is mostly transparently seen in Fig.~\ref{Fi:energy} and~\ref{Fi:heatC}.

With a mere increase of the number of system oscillators from one to two, the physics of the reduced system becomes more intricate because the two oscillators will have, on top of their direct coupling, also an indirect coupling  mediated by the ambient scalar field bath, which introduces non-Markovian effects in the reduced system dynamics. As for quantum entanglement,  in addition to the system-bath entanglement in the one-oscillator case, one needs to consider also entanglement between the constituent oscillators.  Noteworthy on this issue is, as shown in Eq.~\eqref{E:gfkgbrsda} and Fig. \ref{Fi:twosc}-(c) of Appendix \ref{S:one}, the behavior of the heat capacity for the two-oscillator system near absolute zero temperature does not depend on the presence or the absence of  quantum entanglement between the two system oscillators. The heat capacity still approaches zero no matter what, and has a qualitatively similar behavior as the one-system-oscillator case.

\subsection{System of $N$ Coupled Oscillators in a Common Bath}

Now we consider a system that contains $N$ coupled harmonic oscillators in a shared thermal bath. Their spatial locations, specified by $\mathbf{z}_{i}$ with $i=1$, $\cdots$, $N$, is arbitrary, and their initial states can be far from equilibrium. From the previous discussions, we have learned that their motion is highly non-Markovian and intertwined, so it is not obvious whether systems that contain a large number of constituents always equilibrate. This would be the most important issue to address, namely, identify the conditions, or lack thereof, for an $N$ coupled harmonic oscillators in a shared thermal bath to reach in time an \textit{equilibrium} (note different from \textit{thermal}) state. We will show that indeed it exists. Then in this equilibrium state, we can discuss for this non-Markovian system the fluctuation-dissipation relation and the energy balance. We then advance towards the thermodynamics issues, beginning with a proof of the extensivity of the internal energy, the positivity of the heat capacity and finally, the behavior of the heat capacity as the temperature approaches to zero, pertaining to the issues of the third law.

The number $N$ of the system constituents can be arbitrary but cannot be infinite,  because when it is comparable with the number of degrees of freedom of the bath, 1) it may lose the character of a system in contradistinction to its environment, as the basic definition of open systems calls for. 2) The system and environment should in this situation be considered as two equal subsystems interacting with each other which have a very different dynamics from open systems, e.g., recurrence; More seriously 3) the system may never equilibrate because any oscillator will be continually perturbed by the non-Markovian influences from its faraway counterparts all the time. This will make the motion of the system difficult to settle down.  

For a finite $N$, following our earlier analysis outlined in the two-oscillator case, we note there are exceptional cases that equilibration may not be always possible. For example, we exclude those arrangements where some of the oscillators are placed remotely from all others because such a setup can render the relaxation time unusually long. From these considerations we assume the number $N$ is much smaller than the number of degrees of freedom of the bath, and that the oscillators are all localized within a finite region. As a reminder, this still does not exclude the possibility that when the non-Markovian effects are sufficiently strong, albeit not enough to induce instability, the system might still have an extraordinarily long relaxation time so as to behave almost like an undamped one.

Since analytical results for the $N$-oscillator system are unavailable, we will provide a qualitative but general analysis based on the mathematical properties of positive matrices. We start with two simpler topics by first examining the fluctuation-dissipation relation of the reduced system in the final equilibrium state, and then the energy balance between the reduced system and the bath. They provide the basis for extensivity of internal energy and positivity of heat capacity. We will save the discussion on the existence of this equilibrium state for the end.

\subsection{Fluctuation-Dissipation Relation, Stationarity}

We direct our attention now to the correlation function of $\chi_{i}(t)$ and derive the corresponding fluctuation-dissipation relation when the reduced system reaches equilibrium. From \eqref{E:fbiiwed}, we find the correlation function, namely, the Hadamard function of $\pmb{\Xi}(t)$ given by
\begin{align}
	\mathbf{G}_{H}^{(\chi)}(t,t')&=\mathbf{d}_{1}(t)\cdot\mathbf{d}_{1}^{T}(t')\,\langle\chi_{i}^{2}(0)\rangle+\mathbf{d}_{2}(t)\cdot\mathbf{d}_{2}^{T}(t')\,\langle\dot{\chi}_{i}^{2}(0)\rangle\notag\\
	&\qquad\qquad+\mathbf{d}_{1}(t)\cdot\mathbf{d}_{2}^{T}(t')\,\langle\chi_{i}(0)\dot{\chi}_{i}(0)\rangle+\mathbf{d}_{2}(t)\cdot\mathbf{d}_{1}^{T}(t')\,\langle\dot{\chi}_{i}(0)\chi_{i}(0)\rangle\notag\\
	&\qquad\qquad\qquad\qquad+\frac{e^{2}}{m^{2}}\int_{0}^{t}\!ds\int_{0}^{t'}\!ds'\;\mathbf{d}_{2}(t-s)\cdot\mathbf{G}_{H}^{(\phi)}(s-s')\cdot\mathbf{d}_{2}^{T}(t'-s')\,.\label{E:dneurs}
\end{align}
Again $[\mathbf{G}_{H}^{(\phi)}(s-s')]_{ij}=G_{H}^{(\phi)}(s-s',\mathbf{z}_{i}-\mathbf{z}_{j})$. It is not invariant in time translation so the intermediate state is not an equilibrium state. If we choose the parameters of the configuration in such a way that no runaway solution is allowed, then $\mathbf{d}_{i}(t)$ exponentially decays with time. Thus in \eqref{E:dneurs}, those terms that are not inside integrals will be exponentially small at late times. The double integrals in \eqref{E:dneurs} can be written as
\begin{align}
	\int_{0}^{t}\!ds\!\int_{0}^{t'}\!ds'\;\mathbf{d}_{2}(t-s)\cdot\mathbf{G}_{H}^{(\phi)}(s-s')\cdot\mathbf{d}_{2}^{T}(t'-s')&\approx\int\!\frac{d\kappa}{2\pi}\;\tilde{\mathbf{d}}_{2}^{*}(\kappa)\cdot\tilde{\mathbf{G}}_{H}^{(\phi)}(\kappa)\cdot\tilde{\mathbf{d}}_{2}^{T}(\kappa)\,e^{-i\kappa(t-t')}\,,
\end{align}
among which we have ignored terms that are exponentially small at late times and have used the approximation that when $t$ is sufficiently large,
\begin{align}
	\int_{0}^{t}\!ds\;\mathbf{d}_{2}(t-s)\,e^{-i\,\kappa s}&=e^{-i\,\kappa t}\,\tilde{\mathbf{d}}_{2}^{*}(\kappa)+\mathcal{O}(e^{-\alpha t})\,,
\end{align}
with $\alpha$ being some positive number to describe the generic decaying behavior of $\mathbf{d}_{2}$ with time. Thus we see the nonstationary components in $\mathbf{G}_{H}^{(\chi)}(t,t')$ becomes negligibly small as $t$, $t'\to\infty$. We can then focus on the stationary component,
\begin{align}
	\lim_{\substack{t\to\infty \\ t'\to\infty}}\mathbf{G}_{H}^{(\chi)}(t,t')=\mathbf{G}_{H}^{(\chi)}(t-t')&=\int\!\frac{d\kappa}{2\pi}\;\coth\frac{\beta\kappa}{2}\,\operatorname{Im}\Bigl\{\tilde{\mathbf{G}}_{R}^{(\chi)}(\kappa)\Bigr\}\,e^{-i\,\kappa(t-t')}\,,\label{E:euhfshf}
\end{align}
where we have invoked the fluctuations-dissipation relation of the free (stand-alone) scalar field
\begin{equation}\label{E:dbjerssd}
	\tilde{\mathbf{G}}_{H}^{(\phi)}(\kappa)=\coth\frac{\beta\kappa}{2}\,\operatorname{Im}\tilde{\mathbf{G}}_{R}^{(\phi)}(\kappa)\,.
\end{equation}
In \eqref{E:euhfshf}, we notice that the integrand in fact is $\tilde{\mathbf{G}}_{H}^{(\chi)}(\kappa)$ by the definition of the Fourier integral, and thus we arrive at
\begin{equation}\label{E:dbhejrs}
	\tilde{\mathbf{G}}_{H}^{(\chi)}(\kappa)=\coth\frac{\beta\kappa}{2}\,\operatorname{Im}\tilde{\mathbf{G}}_{R}^{(\chi)}(\kappa)\,,
\end{equation}
when the reduced system reaches equilibrium. Thus, from the derivation we see that the correlation function of the reduced system is not stationary in time during the nonequilibrium evolution, but dissipation causes the nonstationary component of the correlation to decay with time such that when the dynamics of the reduced system is relaxed, the correlation becomes stationary. This reflects the presence of a final equilibrium state. 

Stationarity enables us to express the fluctuation-dissipation relation of the reduced system in the frequency domain, similar to that of the bath. However, even though they appear deceptively similar in structure, they are utterly different in physical contents. Essentially \eqref{E:dbjerssd} is based on the initial thermal state of the bath, while \eqref{E:dbhejrs} is established only because there exists a final equilibrium state, which by no means is necessarily a Gibbs thermal state; however it still inherits the information of the initial thermal state of the bath. This is related to the fact the late-time statistics of the reduced system is governed by the bath. A similar behavior is also observed for the case when a charged oscillator interacts with a quantized electromagnetic field, initially prepared in a squeezed vacuum~\cite{WHL12}. The only difference is that the proportionality constant in the fluctuation-dissipation relation like \eqref{E:dbhejrs} takes a different form and depends on the squeeze parameters of the bath's initial squeezed vacuum state. 

\subsection{Energy Balance in the Equilibrium State}\label{S:erhbdjhfbs}

Now we turn to the energy balance of the reduced system described by \eqref{E:bejhrys} in the equilibrium state,
\begin{align}\label{E:dkfeo}
	&m\,\ddot{\chi}_{i}(t)+m\bigl(\pmb{\Omega}_{b}^{2}\bigr)_{ij}\,\chi_{j}(t)-e^{2}\int^{t}_{0}\!ds\;G_{R}^{(\phi)}(t-s,\mathbf{0})\,\chi_{i}(s)\notag\\
	&\qquad\qquad\qquad\qquad-e^{2}\int^{t}_{0}\!ds\;\sum_{j\neq i}G_{R}^{(\phi)}(t-s,\mathbf{z}_{i}-\mathbf{z}_{j})\,\chi_{j}(s)=e\,\xi_{i}(t)\,,
\end{align}
where complexity arises from the frequency renormalization and the nonlocal causal influence among oscillators. In the single oscillator case,  when equilibrium is reached, the net energy flow between the oscillator and the bath stops. The energy flowing in from the noise force of the bath is counterbalanced by the energy flowing out of the oscillator due to the frictional force, as captured by the fluctuation-dissipation relation. In the multi-oscillator case, it is then interesting to ask whether only the same two factors are needed to balance the energy flow in the course of equilibration, or other mechanisms are also involved? If so, what are their roles in the fluctuation-dissipation relation?

We will show that 
\begin{equation}\label{E:pendj}
	\lim_{t\to\infty}\sum_{i}\frac{m}{2}\,\langle\dot{\chi}_{i}^{2}(t)\rangle+\frac{m}{2}\sum_{i,j}\bigl(\pmb{\Omega}_{p}^{2}\bigr)_{ij}\langle\chi_{i}(t)\chi_{j}(t)\rangle=\text{const.}\,,
\end{equation}
that is, the energy transfer mediated by the shared bath ceases after the motion of the reduced system reaches equilibrium.

We first rewrite the third term in \eqref{E:dkfeo} as,
\begin{align}\label{E:dbfehrsd}
	-e^{2}\int^{t}_{0}\!ds\;G_{R}^{(\phi)}(t-s,\mathbf{0})\,\chi_{i}(s)&=-e^{2}\Gamma^{(\phi)}(0)\,\chi_{i}(t)+e^{2}\int^{t}_{0}\!ds\;\Gamma^{(\phi)}(t-s)\,\dot{\chi}_{i}(s)\,,
\end{align}
where $\Gamma^{(\phi)}(t)$ vanishes for a scalar-field bath at late times, and we have introduced a new kernel function $\Gamma^{(\phi)}(\tau)$,
\begin{align}
	G_{R}^{(\phi)}(\tau,\mathbf{0})&=\bigl[\mathbf{G}_{R}^{(\phi)}(\tau)\bigr]_{ii}=-\frac{d}{d\tau}\Gamma^{(\phi)}(\tau)\,,&&\Leftrightarrow&\tilde{G}_{R}^{(\phi)}(\omega,\mathbf{0})&=i\,\omega\,\tilde{\Gamma}^{(\phi)}(\omega)\,.
\end{align}
In what follows we will calculate the power delivered to the $i^{th}$ oscillator in the equilibrium state.

Each of the terms in \eqref{E:dbfehrsd} gives a contribution with a distinct physical interpretation. The first term on the righthand side of \eqref{E:dbfehrsd} will be absorbed into the bare frequency $\omega_{b}$ to form the physical frequency $\omega_{p}$
\begin{equation}
	\omega_{p}^{2}=\omega_{b}^{2}-\frac{e^{2}}{m}\,\Gamma^{(\phi)}(0)\,.
\end{equation}
The second term on the righthand side of \eqref{E:dbfehrsd} thus represents the dissipative force whose mean power delivered to the $i^{\mathrm{th}}$ oscillator is
\begin{equation}\label{E:kendkfs1}
	P_{\gamma}^{(i)}(t)=-e^{2}\int^{t}_{0}\!ds\;\Gamma^{(\phi)}(t-s)\,\langle\dot{\chi}_{i}(s)\dot{\chi}_{i}(t)\rangle\,.
\end{equation}
The mean power exerted by the noise force on the $i^{th}$ oscillator is
\begin{equation}\label{E:kendkfs2}
	P_{\xi}^{(i)}(t)=e\,\langle\xi_{i}(t)\dot{\chi}_{i}(t)\rangle\,. 
\end{equation}
Finally the net power delivered by the other oscillators to the $i^{th}$ oscillator via the nonlocal causal influence transmitted by the field is given by
\begin{equation}\label{E:kendkfs3}
	P_{c}^{(i)}(t)=e^{2}\int^{t}_{0}\!ds\;\sum_{j\neq i}G_{R}^{(\phi)}(t-s,\mathbf{z}_{i}-\mathbf{z}_{j})\,\langle\chi_{j}(s)\dot{\chi}_{i}(t)\rangle\,.
\end{equation}
These three contributions look very distinct in nature, but we will show that at late times after the system of oscillators relaxes, their sum vanishes. Let us rewrite \eqref{E:kendkfs1}--\eqref{E:kendkfs3} in the limit $t\to\infty$,
\begin{itemize}
	\item $P_{\gamma}^{(i)}(\infty)$ : It is given by
		\begin{align}
			P_{\gamma}^{(i)}(\infty)&=i\,e^{2}\int^{\infty}_{-\infty}\!\frac{d\kappa}{2\pi}\;\kappa\,\operatorname{Im}\bigl[\tilde{\mathbf{G}}_{R}^{(\phi)}(\kappa)\bigr]_{ii}\bigl[\tilde{\mathbf{G}}_{H}^{(\chi)}(\kappa)\bigr]_{ii}\,,
		\end{align}
		where we have used several facts
			\begin{enumerate}[(a)]
				\item In general $\mathbf{G}_{H}^{(\chi)}(t,s)=\langle\{\chi_{i}(t),\,\chi_{i}(s)\}\rangle/2$ is not invariant with time translation unless $t$, $s$ are sufficiently large. That is, the non-stationary components will decay with time, so when $t$, $s\to\infty$, we can write $\mathbf{G}_{H}^{(\chi)}(t,s)$ into $\mathbf{G}_{H}^{(\chi)}(t-s)$. \label{E:kbfheer}
				\item The real part of the Fourier transform of a retarded Green's function $\tilde{\mathbf{G}}_{R}^{(\phi)}(\kappa)$ is an even function in $\kappa\in\mathbb{R}$ , but the imaginary part is an odd function.
				\item $\operatorname{Im}\bigl[\tilde{G}_{R}^{(\phi)*}(\kappa,\mathbf{0})\bigr]_{ii}=-\operatorname{Im}\bigl[\tilde{G}_{R}^{(\phi)}(\kappa,\mathbf{0})\bigr]_{ii}$.
			\end{enumerate}
	\item $P_{\xi}^{(i)}(\infty)$ : It is given by
		\begin{align}
			P_{\xi}^{(i)}(\infty)&=-i\,e^{2}\sum_{j}\int_{-\infty}^{\infty}\!\frac{d\kappa}{2\pi}\;\kappa\,\operatorname{Im}\bigl[\tilde{\mathbf{G}}_{R}^{(\chi)}(\kappa)\bigr]_{ij}\bigl[\tilde{\mathbf{G}}_{H}^{(\phi)}(\kappa)\bigr]_{ij}\,,
		\end{align}
		where we have made use of the fluctuation-dissipation relation \eqref{E:dbhejrs} for the reduced system.
	\item $P_{c}^{(i)}(\infty)$ : It is given by
		\begin{align}
			P_{c}^{(i)}(\infty)&=i\,e^{2}\sum_{j\neq i}\int^{\infty}_{-\infty}\!\frac{d\kappa}{2\pi}\;\kappa\,\operatorname{Im}\bigl[\tilde{\mathbf{G}}_{R}^{(\phi)}(\kappa)\bigr]_{ij}\bigl[\tilde{\mathbf{G}}_{H}^{(\chi)}(\kappa)\bigr]_{ij}\,.
		\end{align}
\end{itemize}
We observe that unlike the one-oscillator case,
\begin{equation}
	P_{\gamma}^{(i)}(\infty)+P_{\xi}^{(i)}(\infty)\neq0\,,
\end{equation}
so in the multi-oscillator case, the energy balance is more delicate. On the other hand, the contribution $P_{\gamma}^{(i)}(\infty)$ can be combined with $P_{c}^{(i)}(\infty)$ to form
		\begin{equation}\label{E:berhegh}
			P_{\gamma}^{(i)}(\infty)+P_{c}^{(i)}(\infty)=i\,e^{2}\sum_{j}\int^{\infty}_{-\infty}\!\frac{d\kappa}{2\pi}\;\kappa\,\operatorname{Im}\bigl[\tilde{\mathbf{G}}_{R}^{(\phi)}(\kappa)\bigr]_{ij}\bigl[\tilde{\mathbf{G}}_{H}^{(\chi)}(\kappa)\bigr]_{ij}\,,
		\end{equation}
which turns out to be the negative of $P_{\xi}^{(i)}(\infty)$. We thus see in fact we should have
\begin{equation}\label{E:ebdkjfd}
	P_{\gamma}^{(i)}(\infty)+P_{c}^{(i)}(\infty)+P_{\xi}^{(i)}(\infty)=0\,,
\end{equation}
if both of the fluctuation-dissipation relations 
\begin{align}\label{E:fskfsdgs}
	\tilde{\mathbf{G}}_{R}^{(\phi)}(\kappa)&=\coth\frac{\beta\kappa}{2}\,\operatorname{Im}\tilde{\mathbf{G}}_{R}^{(\phi)}(\kappa)\,,&\tilde{\mathbf{G}}_{R}^{(\chi)}(\kappa)&=\coth\frac{\beta\kappa}{2}\,\operatorname{Im}\tilde{\mathbf{G}}_{R}^{(\chi)}(\kappa)\,,
\end{align}
hold.

Eq.~\eqref{E:ebdkjfd} immediately implies \eqref{E:pendj}. Here we see additional mechanisms are at play in the energy transfer between coupled oscillators. The motion of any oscillator is, apart from direct coupling, causally affected by all the other oscillators via the shared bath. These coherent and correlated contributions from the other oscillators, depending on their individual evolution history, do not necessarily induce a drag nor a push force on that very oscillator. The net effects of the retarded influence are thus highly complicated, hinging on the distance between any two oscillators and their states of motion. It is not obvious how they participate in balancing the energy flow between each system oscillator and the bath. However, we have mentioned earlier that each oscillator, in addition to experiencing the disturbance from the noise of the bath locally, is also affected by the bath fluctuations at the locations of the other oscillators. We can see that the correlations of the bath fluctuations will be passed on to the oscillators such that their motions are also correlated. These correlated noises can be the counterparts of the causal influences, both of which are the off-diagonal elements in the fluctuation-dissipation relation of the bath \eqref{E:fskfsdgs}, in the roles of either the fluctuation-dissipation relation or the energy balance of the system.  Moreover once we observe that the retarded influence is in fact related to the Li\'enard-Wiechert-type radiation of the scalar field as a consequence of the oscillators' motion, it is clear that the damping force is of the same physical origin as the non-Markovian causal influences. Thus grouping $P_{\gamma}^{(i)}(\infty)$ and $P_{c}^{(i)}(\infty)$ together in \eqref{E:berhegh} becomes natural, and from that \eqref{E:ebdkjfd} follows.

Eq.~\eqref{E:pendj} says that even with the presence of non-Markovian influences in the motion of the reduced system, the interaction of the system with the bath is such that when the system settles down in its equilibrium state, its total mechanical energy becomes constant in time. Hereafter the reduced system acts as a collection of coupled \textit{undamped} oscillators, oscillating at the physical frequency $\omega_{p}$, and can be completely described by the final equilibrium density matrix. That is, the reduced system becomes self-contained and free from any further intervention from the bath. This motivates us to assign the total mechanical energy as the internal energy of the system.

Next we will discuss the extensive property of the internal energy of a system of $N$ coupled oscillators in a shared bath.

\subsection{Extensivity of Internal Energy}

Before proceeding to the coupled system in a nonequilibrium configuration, we first delineate the extensivity of the internal energy in the simpler equilibrium case.

Formally, equilibrium thermodynamics is realized in the limit $\gamma\to0$, whereby the matrix $\tilde{\mathbf{G}}_{R}^{(\chi)}(\omega)$ reduces to
\begin{equation}\label{E:iwugd}
	\lim_{\gamma\to0}\tilde{\mathbf{G}}_{R}^{(\chi)}(\omega)=\frac{1}{m}\Bigl [\pmb{\Omega}^{2}_{p}-\omega^{2}\,\mathbf{I}\Bigr]^{-1}\to\frac{1}{m}\Bigl [\pmb{\Omega}^{2}_{p}-\bigl(\omega+i\,\epsilon\bigr)^{2}\,\mathbf{I}\Bigr]^{-1}\,,
\end{equation}
where in order to preserve the retarded property of $\mathbf{G}_{R}^{(\chi)}(\tau)$, we have let $\omega\to\omega+i\,\epsilon$ with $\epsilon>0$. Since the matrix $\pmb{\Omega}^{2}_{p}$ is real and symmetrical, we may find a real orthogonal constant matrix $\mathbf{U}$, independent of $\omega_{p}$ and $\sigma$, to diagonalize it, that is,
\begin{align}
	\pmb{\Omega}^{2}_{p}&=\mathbf{U}\cdot\mathbf{W}_{p}^{2}\cdot\mathbf{U}^{T}\,,&\mathbf{U}\cdot\mathbf{U}^{T}&=\mathbf{I}\,.
\end{align}
The matrix $\mathbf{W}_{p}$ is real and diagonal and we have assumed that its diagonal elements remain positive definite, with the appropriate choice of $\omega_{p}$ and $\sigma$ to avoid instability in motion. Thus we write \eqref{E:iwugd} as
\begin{align}
	\lim_{\gamma\to0}\tilde{\mathbf{G}}_{R}^{(\chi)}(\omega)=\frac{1}{m}\,\mathbf{U}\cdot\Bigl[\mathbf{W}_{p}^{2}-\bigl(\omega+i\,\epsilon\bigr)^{2}\,\mathbf{I}\Bigr]^{-1}\cdot\mathbf{U}^{T}=\mathbf{U}\cdot\tilde{\pmb{\mathfrak{G}}}_{R,\,0}^{(\chi)}(\omega)\cdot\mathbf{U}^{T}\,,
\end{align}
with the diagonalized $\pmb{\mathfrak{G}}_{R,\,0}^{(\chi)}(\omega)$ matrix given by
\begin{equation}
	\tilde{\pmb{\mathfrak{G}}}_{R,\,0}^{(\chi)}(\omega)=\frac{1}{m}\Bigl [\mathbf{W}_{p}^{2}-\bigl(\omega+i\,\epsilon\bigr)^{2}\,\mathbf{I}\Bigr]^{-1}\,.
\end{equation}
Now since the symmetric matrices $\pmb{\sigma}_{\chi\chi}(\infty)$ and $\pmb{\sigma}_{\upsilon\upsilon}(\infty)$ are related to $\tilde{\mathbf{G}}_{R}^{(\chi)}(\omega)$, according to \eqref{E:fbkfq1} and \eqref{E:fbkfq2}, we can write them into the diagonal forms as well with the help of $\mathbf{U}$,
\begin{align}
	\pmb{\sigma}_{\chi\chi}(\infty)&=\mathbf{U}\cdot\mathbf{n}(\infty)\cdot\mathbf{U}^{T}\,,&&\text{and}&\pmb{\sigma}_{\upsilon\upsilon}(\infty)&=\mathbf{U}\cdot\mathbf{m}(\infty)\cdot\mathbf{U}^{T}\,,
\end{align}
in which the diagonal matrices $\mathbf{n}$, $\mathbf{m}$ are
\begin{align}
	\mathbf{n}(\infty)&=\operatorname{Im}\int_{-\infty}^{\infty}\!\frac{d\omega}{2\pi}\;\coth\frac{\beta\omega}{2}\,\tilde{\pmb{\mathfrak{G}}}_{R,\,0}^{(\chi)}(\omega)\,,\\
	\mathbf{m}(\infty)&=\operatorname{Im}\int_{-\infty}^{\infty}\!\frac{d\omega}{2\pi}\;\omega^{2}\coth\frac{\beta\omega}{2}\,\tilde{\pmb{\mathfrak{G}}}_{R,\,0}^{(\chi)}(\omega)\,.
\end{align}
So far what we have done is equivalent to expressing the results in terms of the normal modes of the coupled oscillators when their interaction with the shared bath is almost nonexistent. The matrices $\mathbf{n}$, $\mathbf{m}$ are nothing but the position and velocity uncertainties of the normal-mode coordinates. This decomposition implies that the mean mechanical energy in the equilibrium thermal state, due to the presence of the trace, is invariant under the orthogonal transformation acted by $\mathbf{U}$,
\begin{align}
	E(\infty)=\frac{m}{2}\operatorname{Tr}\,\Bigl\{\pmb{\sigma}_{\upsilon\upsilon}(\infty)+\pmb{\Omega}_{p}^{2}\cdot\pmb{\sigma}_{\chi\chi}(\infty)\Bigr\}&=\frac{m}{2}\operatorname{Tr}\,\Bigl\{\mathbf{m}(\infty)+\mathbf{W}_{p}^{2}\cdot\mathbf{n}(\infty)\Bigr\}\,.\label{E:bdfehrf}
\end{align}
The advantage of the form \eqref{E:bdfehrf} is that since every matrix in it is diagonal, \eqref{E:bdfehrf} can be literally and formally written as
\begin{equation}\label{E:bsjdferys}
	E(\infty)=\sum_{i}E_{i}(\infty)\,,
\end{equation}
where
\begin{equation}
	E_{i}(\infty)=\frac{m}{2}\Bigl\{\mathbf{m}_{ii}(\infty)+\bigl(\mathbf{W}^{2}_{p}\bigr){}_{ii}\,\mathbf{n}_{ii}(\infty)\Bigr\}
\end{equation}
is essentially the mechanical energy associated with each normal mode. That is, the total mechanical energy is the sum of the mechanical energy of each normal mode. Thus when the interaction between the coupled oscillators and the shared bath is negligible, the mechanical or internal energy is extensive, at least with respect to the normal modes.  This is the limiting condition underlying conventional thermodynamics.

When the oscillator-bath interaction is not negligible, the Green's function matrix $\tilde{\mathbf{G}}_{R}^{(\chi)}(\omega)$ of the oscillators contains the contribution from the retarded Green's function matrix $\tilde{\mathbf{G}}_{R}^{(\phi)}(\omega)$ of the free scalar field,
\begin{equation}\label{E:urbfd}
	\tilde{\mathbf{G}}_{R}^{(\chi)}(\omega)=\frac{1}{m}\,\Bigl[\pmb{\Omega}_{b}^{2}-\omega^{2}\mathbf{I}-\frac{e^{2}}{m}\,\tilde{\mathbf{G}}_{R}^{(\phi)}(\omega)\Bigr]^{-1}\,.
\end{equation}
Since the values of the elements of the matrix $\tilde{\mathbf{G}}_{R}^{(\phi)}(\omega)$ depend on the locations of the coupled oscillators,
\begin{equation}
	\bigl[\tilde{\mathbf{G}}_{R}^{(\phi)}\bigr]_{ij}(\omega)=\tilde{G}_{R}^{(\phi)}(\omega,\mathbf{z}_{i}-\mathbf{z}_{j})\,,
\end{equation}
the orthogonal matrix $\mathbf{U}$ that can diagonalize $\pmb{\Omega}_{b}^{2}$ in general cannot diagonalize $\tilde{\mathbf{G}}_{R}^{(\phi)}(\omega)$, because the latter two matrices do not commute in general unless the locations of the oscillators are especially arranged. That is, in general the matrices
\begin{align}
	\tilde{\pmb{\mathfrak{G}}}_{R}^{(\phi)}(\omega)&=\mathbf{U}^{T}\cdot\tilde{\mathbf{G}}_{R}^{(\phi)}(\omega)\cdot\mathbf{U}\,,&&\Rightarrow&\tilde{\pmb{\mathfrak{G}}}_{R}^{(\chi)}(\omega)&=\mathbf{U}^{T}\cdot\tilde{\mathbf{G}}_{R}^{(\chi)}(\omega)\cdot\mathbf{U}
\end{align}
are not diagonal, neither are $\mathbf{n}$ and $\mathbf{m}$ in this case. Even though \eqref{E:bdfehrf} always holds, and it will give an impression that the total mechanical energy can still be expressed as a sum like \eqref{E:bsjdferys}, here $E_{i}$ does not enjoy the special significance of the mechanical energy of each normal mode for the full equation of motion \eqref{E:bdeere}
\begin{equation}\label{E:bdeerke}
	\ddot{\pmb{\Xi}}(t)+\pmb{\Omega}_{b}^{2}\cdot\pmb{\Xi}(t)-\frac{e^{2}}{m}\int_{0}^{t}\!ds\;\mathbf{G}_{R}^{(\phi)}(t-s)\cdot\pmb{\Xi}(s)=\frac{e}{m}\,\pmb{\xi}(t)\,.
\end{equation}
This may be most easily understood if we apply the transformation $\mathbf{U}$ to the coupled equations of motion \eqref{E:bdeerke}, and it becomes
\begin{equation}\label{E:btryere}
	\ddot{\mathbf{X}}(t)+\mathbf{W}_{b}^{2}\cdot\mathbf{X}(t)-\frac{e^{2}}{m}\int_{0}^{t}\!ds\;\pmb{\mathfrak{G}}_{R}^{(\phi)}(t-s)\cdot\mathbf{X}(s)=\frac{e}{m}\,\mathbf{U}^{T}\cdot\pmb{\xi}(t)\,.
\end{equation}
where $\mathbf{X}=\mathbf{U}^{T}\cdot\pmb{\Xi}$ represents the coordinates of the normal modes of the coupled oscillators in the absence of the bath as is discussed earlier, but is not the normal modes of the coupled oscillators in the presence of the shared bath. The off-diagonal elements $\pmb{\mathfrak{G}}_{R}^{(\phi)}(t-s)$ will link up any given element in $\mathbf{X}$ with all other elements.

Therefore, from \eqref{E:pendj} we arrive at some interesting conclusions. When the coupled oscillators interact with a shared bath, \textit{after the coupled system reaches equilibrium, the internal energy of the system oscillators becomes extensive}; however this extensivity is expressed by neither its original degrees of freedom nor the decoupled degrees of freedom. Instead, the extensive property of the system's internal energy is only manifested by a specified set of modes obtained from the orthogonal transformation that diagonalizes $\pmb{\Omega}^{2}_{p}$, as can be seen from Eq.~\eqref{E:pendj}. Moreover, before the motion of the reduced system equilibrates, this special extensiveness property does not hold, as is implied by \eqref{E:pendj}. Thus we are not able to discuss the extensive property of the system's internal energy during the nonequilibrium evolution of the reduced system, until the final equilibrium state of the reduced system is attained.

In particular, this seemingly mundane conclusion, together with \eqref{E:pendj}, justifies or explicitly demonstrates, in the weak oscillator-bath coupling, why conventional thermodynamics (at least for the system that constitutes coupled oscillators) works, why we need only the density matrix of the system to describe the behaviors of the system, and why we need not be concerned with renormalization, relaxation, damping, bath noise.

\subsection{Positivity of Heat Capacity and Existence of the Equilibrium State}\label{S:jkjdf}
Now we would like to discuss the positivity of the heat capacity for a system of $N$ coupled oscillators in a shared bath in the context of nonequilibrium thermodynamics. The positivity of heat capacity, the decaying behavior and the retarded nature of $\mathbf{d}_{2}(t)$ all hinge on the existence of the equilibrium state. Thus in this section, we will also address the conditions that a nonequilibrium system settles into an equilibrium state at late times.

Given the internal energy \eqref{E:efkbsdf} when the system reaches equilibrium, we proceed to examine the positivity property of the heat capacity $C(\infty)$, given by
\begin{align}\label{E:dfbjhre}
	C(\infty)&=\frac{1}{2}\operatorname{Im}\int_{-\infty}^{\infty}\!\frac{d\kappa}{2\pi}\;\biggl(\frac{\frac{\beta\kappa}{2}}{\sinh\frac{\beta\kappa}{2}}\biggr)^{2}\frac{1}{\kappa}\,\operatorname{Tr}\,\Bigl\{\Bigl[\kappa^{2}\mathbf{I}+\pmb{\Omega}_{p}^{2}\Bigr]\cdot\tilde{\mathbf{d}}_{2}(\kappa)\Bigr\}\,e^{-\frac{\lvert\kappa\rvert}{\Lambda}}\,,
\end{align}
from \eqref{E:efkbsdf} and \eqref{E:dnerjdgd}. The damping factor $e^{-\frac{\lvert\kappa\rvert}{\Lambda}}$, with $\Lambda\to+\infty$, is necessary to regularize the integral.

If we consider only the case that there is no runaway solution in the motion of the system, such as with the inverted oscillator, then this requires that the matrix $\pmb{\Omega}_{p}^{2}$ should be at least positive definite~\footnote{Positive semi-definiteness can be too weak because the non-Markovian contributions can easily induce instability in the strong system-bath coupling regime or in the limit of extremely short separations among the oscillators.}. This allows us later to define a matrix that would be the square root of $\kappa^{2}\mathbf{I}+\pmb{\Omega}_{p}^{2}$. The imaginary part of $\tilde{\mathbf{d}}_{2}(\kappa)$ can be written as
\begin{equation}\label{E:ahfruw}
	\operatorname{Im}\tilde{\mathbf{d}}_{2}(\kappa)=\tilde{\mathbf{d}}_{2}^{\vphantom{\dagger}}(\kappa)\cdot\Bigl[2\gamma\kappa\,\mathbf{I}+\frac{e^{2}}{m}\,\operatorname{Im}\tilde{\slashed{\mathbf{G}}}_{R}^{(\phi)}(\kappa)\Bigr]\cdot\tilde{\mathbf{d}}_{2}^{\dagger}(\kappa)\,.
\end{equation}
Using \eqref{E:efjdbjbd} we know that $\tilde{\mathbf{d}}_{2}(\kappa)$ takes the form
\begin{equation*}
	\tilde{\mathbf{d}}_{2}(\kappa)=\Bigl[\pmb{\Omega}_{b}^{2}-\kappa^{2}\mathbf{I}-\frac{e^{2}}{m}\,\tilde{\mathbf{G}}_{R}^{(\phi)}(\kappa)\Bigr]^{-1}=\Bigl[\pmb{\Omega}_{p}^{2}-\kappa^{2}\mathbf{I}-i\,2\gamma\kappa\,\mathbf{I}-\frac{e^{2}}{m}\,\tilde{\slashed{\mathbf{G}}}_{R}^{(\phi)}(\kappa)\Bigr]^{-1}\,,
\end{equation*}
with the help of \eqref{E:eridsd}. The matrix $\tilde{\slashed{\mathbf{G}}}_{R}^{(\phi)}(\kappa)$ is $\tilde{\mathbf{G}}_{R}^{(\phi)}(\kappa)$ with its diagonal elements  removed,
\begin{equation}
	\bigl[\tilde{\slashed{\mathbf{G}}}_{R}^{(\phi)}(\kappa)\bigr]_{ij}=\begin{cases}
	0\,,&i=j\,,\vspace{6pt}\\
	\bigl[\tilde{\mathbf{G}}_{R}^{(\phi)}(\kappa)\bigr]_{ij}\,,&i\neq j\,.
	\end{cases}
\end{equation}
The diagonal elements of $\tilde{\mathbf{G}}_{R}^{(\phi)}(\kappa)$ account for the usual damping term and the frequency renormalization. Now we introduce the matrix $\tilde{\mathbf{D}}_{2}(\kappa)$ by
\begin{equation}
	\Bigl[\kappa^{2}\mathbf{I}+\pmb{\Omega}_{p}^{2}\Bigr]^{\frac{1}{2}}\cdot\tilde{\mathbf{d}}_{2}(\kappa)\equiv\tilde{\mathbf{D}}_{2}(\kappa)\,,
\end{equation}
such that
\begin{align}
	\operatorname{Im}\operatorname{Tr}\,\Bigl\{\Bigl[\kappa^{2}\mathbf{I}+\pmb{\Omega}_{p}^{2}\Bigr]\cdot\tilde{\mathbf{d}}_{2}(\kappa)\Bigr\}&=\operatorname{Tr}\,\Bigl\{\Bigl[2\gamma\kappa\,\mathbf{I}+\frac{e^{2}}{m}\,\operatorname{Im}\tilde{\slashed{\mathbf{G}}}_{R}^{(\phi)}(\kappa)\Bigr]\cdot\tilde{\mathbf{D}}_{2}^{\dagger}(\kappa)\cdot\tilde{\mathbf{D}}_{2}^{\vphantom{\dagger}}(\kappa)\Bigr\}\,,
\end{align}
where we have used the cyclic property of the matrix trace. The product of the last pair of matrices is positive, so we would like to examine whether the matrix sandwiched by $\tilde{\mathbf{d}}_{2}^{\vphantom{\dagger}}(\kappa)$ and $\tilde{\mathbf{d}}_{2}^{\dagger}(\kappa)$ in \eqref{E:ahfruw} is positive as well. In general, it takes the form
\begin{align}
	2\kappa\,\pmb{\Gamma}&=2\gamma\kappa\,\mathbf{I}+\frac{e^{2}}{m}\,\operatorname{Im}\tilde{\slashed{\mathbf{G}}}_{R}^{(\phi)}(\kappa)\notag\\
	&=\begin{pmatrix}
		2\gamma\kappa	&\dfrac{2\gamma}{\ell_{12}}\,\sin\kappa\ell_{12} &\dfrac{2\gamma}{\ell_{13}}\,\sin\kappa\ell_{13} &\cdots &\dfrac{2\gamma}{\ell_{12}}\,\sin\kappa\ell_{1N}\vspace{6pt}\\
		\dfrac{2\gamma}{\ell_{21}}\,\sin\kappa\ell_{21}&2\gamma\kappa&\dfrac{2\gamma}{\ell_{23}}\,\sin\kappa\ell_{23}&\cdots &\dfrac{2\gamma}{\ell_{2N}}\,\sin\kappa\ell_{2N}\vspace{6pt}\\
		\dfrac{2\gamma}{\ell_{31}}\,\sin\kappa\ell_{31}&\dfrac{2\gamma}{\ell_{32}}\,\sin\kappa\ell_{32}&\ddots&&\vdots\\
		\vdots&\vdots&&\ddots&\vdots\vspace{6pt}\\
		\dfrac{2\gamma}{\ell_{N1}}\,\sin\kappa\ell_{N1}&\dfrac{2\gamma}{\ell_{N2}}\,\sin\kappa\ell_{N2}&\cdots&\cdots&2\gamma\kappa\vspace{6pt}
	\end{pmatrix}
\end{align}
where $\ell_{ij}=\ell_{ji}=\lvert\mathbf{z}_{i}-\mathbf{z}_{j}\rvert$. Here we assume that the choices of the parameters $\omega_{p}$, $\sigma$ and $\ell_{ij}$ are such that the matrix $\pmb{\Gamma}$ is strictly diagonally dominant, that is, its elements satisfying 
\begin{equation}\label{E:rvuedfs}
	\lvert\Gamma_{ii}\rvert>\sum_{j\neq i}\lvert\Gamma_{ij}\rvert\,.
\end{equation}
At the first sight, this assumption looks pretentious; however we observe that the strictly diagonally dominant matrix has a nice property of being positive definite~\cite{HJ85}. That is, its eigenvalues are all positive. Thus this assumption, together with positive definiteness of $\pmb{\Omega}_{p}^{2}$, implies that the integrand in \eqref{E:dfbjhre} is always positive. Since the integral is well defined, we conclude the value of the heat capacity $C(\infty)$ remains positive for all temperatures $\beta^{-1}$ with one exception that $\beta\to\infty$. In that limit the factor  
\begin{equation*}
	\left(\frac{\frac{\beta\kappa}{2}}{\sinh\frac{\beta\kappa}{2}}\right)^{2}
\end{equation*}
goes to zero~\footnote{We can generalize the more sophisticated arguments in the context of the two-oscillator system to the current case.}, so it indicates that the heat capacity for the system of coupled oscillators will be zero at zero temperature; otherwise it is always positive.

Physically the assumption \eqref{E:rvuedfs} amounts to the existence of the effective damping constants for all modes of motion, and thus the motion of the system, described by $\tilde{\mathbf{d}}_{2}$ reduces to that of a collection of coupled damped oscillators. This can be read off from the denominator of $\tilde{\mathbf{d}}_{2}(\kappa)$. Suppose the real matrix $\pmb{\Gamma}$ can be diagonalized by the orthogonal matrix $\mathbf{V}$
\begin{equation}
	\mathbf{V}\cdot\pmb{\Gamma}\cdot\mathbf{V}^{T}=\pmb{\Gamma}'=\operatorname{diag}(\gamma_{1},\gamma_{2},\dots,\gamma_{N })\,.
\end{equation}
The matrix $\mathbf{V}$ in general cannot diagonalize $\pmb{\Omega}_{p}^{2}$ unless $\pmb{\Omega}_{p}^{2}$ commutes with $\pmb{\Gamma}$, but it will transform $\pmb{\Omega}_{p}^{2}$ to another symmetric, positive matrix, which we denote by $\pmb{\mathcal{W}}^{2}_{p}$. The diagonal elements of $\pmb{\mathcal{W}}^{2}_{p}$ describe the same physical frequencies $\omega_{p}$ of the transformed modes and the off-diagonal ones account for the coupling among them. The mode-mode couplings are usually different among pairs of modes. Explicitly the denominator of $\tilde{\mathbf{d}}_{2}(\kappa)$ is transformed to
\begin{align}\label{E:kbfkarfs}
	&\kappa^{2}\mathbf{I}+i\,2\kappa\,\pmb{\Gamma}'-\pmb{\mathcal{W}}^{2}_{p}\notag\\
	=&\begin{pmatrix}
	\kappa^{2} 	&0          &\cdots	&0\\
	0		 	&\kappa^{2}	&		&\vdots\\
	\vdots		&			&\ddots	&0\\
	0			&\cdots		&0		&\kappa^{2}	
 \end{pmatrix}+i\,2\kappa
 \begin{pmatrix}
	\gamma_{1} 	&0          &\cdots	&0\\
	0		 	&\gamma_{2}	&		&\vdots\\
	\vdots		&			&\ddots	&0\\
	0			&\cdots		&0		&\gamma_{N}	
 \end{pmatrix}-
 \begin{pmatrix}
	\omega_{p}^{2} 	&\sigma_{12}    &\cdots			&\sigma_{1N}\\
	\sigma_{21}		&\omega_{p}^{2}	&				&\vdots\\
	\vdots			&				&\ddots			&\sigma_{N-1,N}\\
	\sigma_{N1}		&\cdots			&\sigma_{N,N-1}	&\omega_{p}^{2}	
 \end{pmatrix}\,,
\end{align}
and the zeros of its determinant 
\begin{equation}\label{E:dfbsjher}
	\det\bigl(\kappa^{2}\mathbf{I}+i\,2\kappa\,\pmb{\Gamma}'-\pmb{\mathcal{W}}_{p}^{2}\bigr)=0
\end{equation}
identify the eigen-modes of the motion of the system. The signs of the imaginary part of the solutions to \eqref{E:dfbsjher} provide information about the stability of the motion. If there exists a solution whose imaginary part is positive, then instability of the collective motion~\footnote{Unless the unstable mode is not excited, and that is highly unlikely for a generic initial state.} will occur. Eq.~\eqref{E:kbfkarfs} in fact corresponds to a simultaneous set of equations of motion that describes a system of coupled damped oscillators,
\begin{equation}\label{E:ernfdg}
	\ddot{\pmb{\Xi}}(t)+2\,\pmb{\Gamma}'\cdot\dot{\pmb{\Xi}}(t)+\pmb{\mathcal{W}}_{p}^{2}\cdot\pmb{\Xi}(t)=0\,.
\end{equation}
Thus the stability condition associated with \eqref{E:ernfdg} is equivalent to whether the characteristic polynomial \eqref{E:dfbsjher}, when $\kappa=-i\,s$, is a (strict) Hurwitz polynomial~\cite{AH95}, whose zeros are all located on the left half of the complex $s$ plane. In other words, the motion described by \eqref{E:ernfdg} is stable if the characteristic polynomial associated with the Laplace transformation of the lefthand side of \eqref{E:ernfdg}
\begin{equation}\label{E:ubeiuhsd}
	p(s)=\det\bigl(s^{2}\mathbf{I}+2s\,\pmb{\Gamma}'+\pmb{\mathcal{W}}^{2}_{p}\bigr)
\end{equation}
is Hurwitz. In general, a sufficient and necessary condition is provided by the Routh-Hurwitz stability criterion~\cite{PD85}, which states that all principal minors of the Hurwitz matrix associated with  $p(s)$ are positive.

This criterion becomes computationally cumbersome as $n$ grows, and it is very hard to establish an apparent connection between this criterion and the physically meaningful matrices $\pmb{\Gamma}'$ and $\pmb{\mathcal{W}}^{2}_{p}$. For this reason we turn to finding arguments to directly relate the properties of the matrices $\pmb{\Gamma}'$ and $\pmb{\mathcal{W}}^{2}_{p}$ with the stability condition of motion described by \eqref{E:ernfdg}. These arguments, although mathematically less rigorous, are physically more transparent. The idea is that solving the polynomial $p(s)=0$ is equivalent to finding the eigenvalue $s$ of the system~\cite{VE11}
\begin{align}\label{E:bruwerw}
	\bigl(s^{2}\mathbf{I}+2s\,\pmb{\Gamma}'+\pmb{\mathcal{W}}^{2}\bigr)\cdot \mathbf{x}=0\,,
\end{align}
with the normalized column eigenvector $\mathbf{x}$, with $\mathbf{x}^{T}\cdot\mathbf{x}=1$. We multiply \eqref{E:bruwerw} from the left with $\mathbf{x}^{T}$, transforming the matrix expression \eqref{E:bruwerw} to an ordinary quadratic equation of $s$,
\begin{align}\label{E:dbhfsj}
	s^{2}+2bs+c&=0\,,&b&=\mathbf{x}^{T}\cdot\pmb{\Gamma}'\cdot\mathbf{x}\,,&c&=\mathbf{x}^{T}\cdot\pmb{\mathcal{W}}^{2}_{p}\cdot\mathbf{x}\,.
\end{align}
so that
\begin{equation}
	s=-b\pm i\sqrt{c-b^{2}}\,.
\end{equation}
Since we have required that $\pmb{\Gamma}'$ and $\pmb{\mathcal{W}}^{2}_{p}$ are (strictly) positive definite, the variables $b$ and $c$ are also positive by construction. This implies that 
\begin{align}
	b^{2}-c&<b^{2}\,,&&\Rightarrow&\operatorname{Re}s&<0\,.
\end{align}
Thus the positive definiteness of $\pmb{\Gamma}'$ and $\pmb{\mathcal{W}}^{2}_{p}$ is sufficient to ensure the stability of the motion \eqref{E:ernfdg} which in turn signals the existence of an equilibrium state. In addition, the expressions in \eqref{E:dbhfsj} resemble  those we have seen for the case of one oscillator interacting with a bath, where $\sqrt{c-b^{2}}\propto\sqrt{\smash[b]{\omega_{p}^{2}-\gamma^{2}_{\vphantom{p}}}}$ is related to the resonance frequency.

In summary, the requirement that $\pmb{\Gamma}'$ and $\pmb{\mathcal{W}}^{2}_{p}$ are positive matrices implies that $\mathbf{d}_{2}(\tau)$ is indeed a retarded Green's function and $\tilde{\mathbf{d}}_{2}^{\vphantom{\dagger}}(\kappa)$ does not have any pole along the real axis of $\kappa$ and on the upper half of the complex $\kappa$ plane. Therefore the integrand in \eqref{E:dfbjhre} is positive and bounded, so the heat capacity \eqref{E:dfbjhre} is positive and approaches zero as $\beta\to\infty$. In addition it ensures the existence of the equilibrium state, which is needed a) for the reduced system to have a meaningful  fluctuation-dissipation relation, b) to show the energy balance between the system and the bath, c) to ensure the extensive nature of the internal energy of the system and finally d) for the associated heat capacity to be positive definite in our framework of open system nonequilibrium dynamics approach to quantum thermodynamics.

\section{Summary and Discussions}

\subsection{Summary of major results}

As a preamble we bring up the rather special conditions whereupon the foundation of thermodynamics is laid, from an open system perspective:  A small open system interacting with a vast environment (whose thermal properties can be captured by a few physical parameters, its temperature, chemical potential), it is in the limit of vanishing coupling between them, only when the system can equilibrate and thermalize at late times, that thermodynamics makes sense. These considerations can be extended to nonequilibrium conditions but not for far from equilibrium, fully arbitrary time evolutions.  We mentioned the important differences in the setups for treating quantum thermodynamics (QTD), namely, between level 1 Assuming the closed system (comprising the system and its environment) remains in a global thermal state (which we call CGTs)  and level 2 an open system approaching equilibrium at late times (we call it ONEq), which is the preferred approach we adopt for the discussion of level 0.  The centroid of this paper is a detailed model study, that of a system of $N$ coupled, spatially separated quantum oscillators interacting with a common scalar quantum field bath at finite temperature,  where the existence of exact solutions can provide unambiguous quantification of physical variables, thermodynamic relations and help to clarify many basic issues in QTD. The set of issues we addressed include the following.

\subsubsection{Gateway to thermodynamics: The existence of an equilibrium state}

1) Equilibrium state at late times:  Let the system initially be prepared in a state that is not in thermal equilibrium with the shared bath, it has been known that if the coupling between the system and the bath is vanishingly weak, the reduced system will equilibrate at late times. This is the pre-condition for talking about its thermodynamic behavior. The new challenge is whether the system will equilibrate for strong coupling.  This point has been emphasized in e.g.~\cite{SFTH} who used the quantum Brownian motion model where the system consists of $N$ quantum harmonic oscillators and the environment is an infinite-oscillator bath.

2) Equilibration, not thermalization:  The strong coupling regime poses new challenges: Allowing the coupling between the system oscillators and the interaction between the system and the bath to be strong, and assuming that the dynamics of the system remains stable, the first and foremost statement is that due to non-weak system-bath interaction, this final state (of the system) is not described by a density matrix of the Gibbs form with respect to the system Hamiltonian. Therefore one should refrain from using the word thermalization to describe the end result,  and note that conventional thermodynamics need not apply. The tough question is, \textit{when will TD remain a viable theory for this equilibrated strongly coupled system}.

3) Environment-induced non-Markovian inter-oscillator interaction:  The newer challenge which we need to take on here is to show equilibration for a system of strongly coupled $N$ quantum oscillators \textit{at finite spatial separation} and strongly interacting with an environment composed of a quantum scalar field. The case of $N$ oscillators in the same spatial location is easier to prove because one needs not worry about the field-induced non-Markovian effects.  However, beware of the pathology of even two oscillators stacked up at the same spatial location, as described in Appedix.~\ref{S:rbjdhe}. The added complication is due to the non-Markovian nature of the induced interaction discovered in \cite{ASH06,LinHu09,HHPRD} amongst the system oscillators (or qubits) mediated by the field environment. This issue has not been dealt with in this context before, as far as we know.

4) The existence of an equilibrated state for the case of two coupled oscillators has been demonstrated. The conditions for $N$ oscillators are discussed in Sec.~\ref{S:jkjdf}. We argue that certain positive-definiteness requirements must be satisfied to the effect that the effective damping constants of the oscillators stay positive and the effective oscillator frequencies remain real.

5) With the assurance of an equilibrated state, many nice properties follow. Specifically, the extensivity of the internal energy and the positivity of the heat capacity. The absence of such a state for open quantum systems severs the linkage to thermodynamics. QTD in the form described here does not exist for these systems.

\subsubsection{Internal energy, heat capacity and the third law}

1) The internal energy for certain strongly bounded systems may not be straighforward to define (e.g., the presence of self energy as when gravity is involved) but fortunately not so in the model we studied: it is the sum of the kinetic energy of each oscillator, the harmonic potential energy, and their coupling energy. Heat capacity is the derivative of the internal energy with respect to the bath temperature. 

2) We examine the third law from the behavior of the heat capacity at low and zero temperatures. We are concerned with a) low temperature behavior, b) the positivity and c) the extensivity of heat capacity.

3) The internal energy and the heat capacity for a system consisting of only one harmonic oscillator have been derived  before in the CGTs setup~\cite{HanIng06}. They are derived here in an open quantum system ONEq setup,  which in the epoch after equilibration,  can be compared, in the weak oscillator-bath coupling limit, with the quantities derived in the CGTs and in the conventional thermodynamics. They all agree with each other.
  
4) Complexity arises when the system has more than one constituent. The bath-induced non-Markovian effects cannot be properly described in conventional thermodynamics.  This also brings in question the validity of energy extensivity because the system constituents are not only directly coupled (which is easy to deal with by normal mode separation) but are indirectly coupled or intertwined in a non-Markovian way by the induced interaction through their common environment. 

5) With the proven existence of an equilibrated state for $N$ spatially separated but mutually coupled system oscillators we have shown that even a strongly coupled system can still have asymptotic extensivity of the internal energy, and the heat capacity remains positive as long as the motion is stable. 

6) To compare results calculated in the three different levels, CGTs, ONEq and conventional thermodynamics, we need to beware of their respective regimes of validity and identify their common denominators, e.g.,  the common physical quantities and the states they are in.  We work here with an open system ONEq set up, namely, we allow the system to evolve from a nonequilibrium initial state to a final equilibrium state. If the system thermalizes then the results obtained in the ONEq setup can be compared to results in conventional thermodynamics, as we did. Since the reduced density matrix of the ONEq after equilibration and that of the system in the  CGTs setups are the same~\cite{SFTH}, the results from these two setups can be compared using this quantity.

\subsection{Heat capacity and third law}
For the same system we have studied here there are claims of negative heat capacity in variance to our findings. For example, the author of \cite{HasegawaJMP} working with a global thermal state CGTs setup  claims that the heat capacity of the system  of multiple quantum harmonic oscillators can be negative at low temperatures for  1) stronger system-bath interaction, or 2) smaller number of system constituents. However, we have demonstrated under rather general conditions that after the system reaches equilibration, the heat capacity of the system is always positive and approaches zero, for the full range of system-bath interaction strength. Thus the third law, from the aspect of low-temperature behavior of the heat capacity of the system, is not violated. 
Our findings of the  extensivity of the energy and the positivity of heat capacity for $N$ coupled oscillator system should hold in all three levels of inquiry.

\subsection{On entanglement witnesses and heat capacity}

In Ref.~\cite{Vedral} Wie{\'s}niak {\it et al} have ``shown that the low-temperature behavior of the specific heat can reveal the presence of entanglement in bulk bodies in the thermodynamical equilibrium". They drew this conclusion by showing that the heat capacity is an entanglement witness for some spin models. This involves finding a lower bound on heat capacity that can be achieved by separable states of the system. The third law requires the heat capacity to approach zero asymptotically with the temperature. This means that at low enough temperatures the behavior of heat capacity is not compatible with separable states, and is thus an indicator of entanglement.

Here we find that specific heat is not a reliable indicator of entanglement for our model.\footnote{It is worth mentioning that the models studied in Ref.~\cite{Vedral} differ from the ones studied in this paper in two important aspects. First, we are studying harmonic oscillators with infinite dimensional Hilbert spaces as opposed to the finite dimensional spin systems. Second, we are taking an open system approach, allowing the system-bath coupling to be finite, and we are dealing with equilibrium states as opposed to thermal states.} We see no requirement~\cite{HHPRD} that the zero temperature state is always entangled. For example, when the coupling between the constituents in the system is sufficiently strong, the system tends to relax to an entangled state at zero temperature. However, for sufficiently strong system-bath coupling, the system can relax to a separable state even at zero temperature.  In this equilibrium state, the system oscillators are disentangled among themselves, but can get entangled with the bath oscillators. This can be understood as a consequence of the monogamy of entanglement.

At the same time, we have shown that the third law is valid for our model in all parameter regimes. There is no connection between the third law and entanglement. Moreover, even in the models or parameter regimes in which this connection exist, we do not interpret this observation as the third law relying on quantum entanglement. The third law stands. The claim that the third law implies the existence of entanglement could well be affected by the use of entanglement witness as a criterion.

\subsection{Relation with global thermal state formulation, sequel on heat, energy and entropy}

\subsubsection{Relation to global thermal state formulation: Seifert's systematics of energy heat and entropy for quantum systems}

Much work on QTD has been done under the closed system in a global thermal state (CGTs) set up. It would be useful to find a link between it and our open system nonequilibrium (ONEq) approach.  We have carried out a first step towards this goal. We focused on Seifert's rendition of energy and entropy for classical thermodynamics \cite{Seifert16} in a closed system global thermal state (CGTs) set up. In a companion paper~\cite{GTOS} we have generalized his results for the thermodynamics of classical systems to quantum systems.  This may enable us to use his systematics for discussing the First and Second Laws for quantum systems.

The diversity of how thermodynamic functions are defined is both a resource of adaptivity and at times a source of confusion. To show how the thermodynamic functions are used and how they enter into the TD relations, in this same companion paper we studied the approach of Gelin and Thoss \cite{GelTho} who also work in the CGTs setup but adopt a different set of  thermodynamics functions from Seifert's.  Consulting Seifert's systematics and allowing for varying thermodynamic functions we hope to construct more links between our open systems ONEq formulation and the prevailing CGTs formulations of quantum thermodynamics.

\subsubsection{Sequel: On heat, entropy, entanglement and the second law}

The issues of heat, entropy and entanglement in strongly coupled open quantum systems will be the center of attention in our second paper.  Notice the subtle yet important difference between energy and heat.  In this paper we have focused on the internal energy and heat capacity of the system, but not heat, which is the energy transfer between the system and the environment.  This is because heat transfer as energy change contains ambiguities in an open system context. For example Esposito \textit{et al} \cite{EspositoPRB15} found that any heat definition expressed as an energy change in the reservoir energy plus any fraction of the system-reservoir interaction is not an exact differential when evaluated along reversible isothermal transformations, except when that fraction is zero. Even in that latter case the reversible heat divided by temperature, namely entropy, does not satisfy the third law of thermodynamics and diverges in the low temperature limit. 

We also have reservations in some claims of violation of the Second Law for quantum  systems \cite{2LawSaga}. For example if one uses the Clausius inequality representation for the Second Law, we know it is only valid for classical systems at high temperatures. One needs to scrutinize the different definitions of entropy for strongly coupled quantum systems to make sure they are physically sound in the quantum regimes (such as at low temperatures), including possible non-Markovian behaviors, before adopting them to address foundational issues.

The relevant issues ranging from quantum correlations, entanglement, information to entropy production and heat can be sampled in these references of the last 15 years \cite{EisPle,HB05,HBJSP08,HiltLutz,ELvdB,DeffLutz,Kim12,EspositoJSM,AurEic15,AnkPek14,EspositoPRB15,CarrWeiss16}, which span the scope of our sequel studies using the paradigm established in this paper.

\section*{Acknowledgement}
B.L.H enjoyed the hospitality of the Physics Theory Center of Fudan University when this work was conceived while J.T.H enjoyed the hospitality of the  Physics Theory Center at the University of Maryland when it was set in its final form. B.L.H thanks E. Bianchi for discussions related to his recent work on eigenvalue thermalization. C.H.C acknowledges the support from Ministry of Science and Technology, Taiwan under Grant No. MOST 105-2112-M-006-011 and National Center for Theoretical Sciences (NCTS). Y.S. acknowledges that this work was performed under the auspices of the U.S. DOE Contract No. DE-AC52-06NA25396 through the LDRD program at LANL.

\appendix

\section{Influence Functional Formalism for Quantum Thermodynamics}
\label{IF}

For the study of the thermodynamical properties of a system one mainly focuses on the dynamics and thermal properties of the system under the influence of a thermal bath it interacts with, not particularly about the bath itself. This is the arena where the open system conceptual framework is most suited. When the system of interest interacts with the bath strongly, one needs to take into account not only the influence of the bath on the system but also the backreaction of the system on the bath.  The influence functional formalism is particularly adept  for this description because it respects the self-consistency of the system-bath evolutionary dynamics (an example being the fluctuation-dissipation relation) which has increased significance for treating systems with non-vanishing coupling with its environment.

In this appendix we give a self-contained description of the influence functional formalism used in this series of papers for obtaining the thermodynamic properties of physical quantities of a system interacting strongly with a thermal bath.  Over and beyond the standard classic sources of Feynman-Vernon \cite{FeyVer} and Caldeira-Leggett \cite{CalLeg}, we also provide additional materials in the use of the coarse-grained effective action \cite{CGEA} and stochastic effective action \cite{JH1} developed in the 1990s and 2000s. And, from the stochastic effective action, as a further development, we follow a recent work \cite{HH15} to expound the different advantages of using the Langevin equation route which is more intuitive versus the more formal route via the \textit{reduced density operator}, which can account for the full quantum dynamics of the reduced system and enforce the operator ordering. This system of tools was used to describe the thermodynamics of quantum many-body systems in a nonequilibrium steady state.  The reader can find more details of this formalism and examine its application to a more complex problem in \cite{HH15}.  In this paper we shall use it for the development of quantum thermodynamics at strong coupling, starting with the third law.

\subsection{Influence functional for open quantum systems}\label{S:erbdjh}

Consider for our system a quantum harmonic oscillator (called an Unruh-DeWitt detector in relativistic quantum information) moving along a prescribed spatial trajectory $\mathbf{z}$ in $1+3$ Minkowski spacetime. We can call $\mathbf{z}$ its external degree of freedom, while its internal degree of freedom is the oscillator's displacement $\chi$. The bath is represented by a massless scalar field $\phi$. The action of the combined system is 
\begin{align}\label{E:cnmxrere}
	 S[\chi,\phi]&=S_{\chi}[\chi]+S_{I}[\chi,\phi]+S_{\phi}[\phi]\,,
\end{align}
where $S_{\chi}$, $S_{\phi}$,  $S_{I}$ are the actions which describe the free quantum oscillators, the bath field and their interaction, given respectively by
\begin{align*}
	S_{\chi}[\chi]&=\int_{0}^{t}\!ds\;\frac{m}{2}\Bigl[\dot{\chi}^{2}(s)-\omega_{b}^{2}\chi^{2}(s)\Bigr]\,,&S_{\phi}[\phi]&=\int_{0}^{t}\!d^{4}x\;\frac{1}{2}\,\partial_{\mu}\phi(x)\,\partial^{\mu}\phi(x)\,,\\
	S_{I}[\chi,\phi]&=\int_{0}^{t}\!d^{4}x\;e\,\chi(s)\,\delta^{3}[\mathbf{x}-\mathbf{z}(s)]\phi(x)\,,
\end{align*}
where $x=(t,\mathbf{x})$, $\omega_{b}$ is the bare natural frequency of the oscillator, and an overdot denotes taking the time derivative of a variable. The internal degree of freedom of the quantum oscillator is assumed to be linearly coupled to its bath with coupling strength $e$, which can take on a finite (non-vanishing) value. 

Assume that the combined system at time $t=0$ is in a product state \footnote{For a discussion of the physical consequences of factorizable initial conditions and generalizations, see e.g., \cite{HPZ,JH1,PazRom,FRH}.}.
\begin{equation}
	\rho(0)=\rho^{(\chi)}\otimes\rho_{\beta}^{(\phi)}\,,
\end{equation}
where $\rho^{(\chi)}$ is the initial density operator for the internal degree of freedom, and has a Gaussian form
\begin{equation}
	 \rho^{(\chi)}(\chi_{i},\chi'_{i};0)=\left(\frac{1}{\pi\varsigma^{2}}\right)^{1/2}\exp\left[-\frac{1}{2\varsigma^{2}}\bigl(\chi^{2}_{i}+\chi'^{2}_{i}\bigr)\right]\,.
\end{equation}
The parameter $\varsigma$ is the width of the wavepacket, and $\chi_{i}$, $\chi_{f}$ are shorthands for $\chi$ at the initial time $t=0$ and the final time $t$ respectively, that is, $\chi_{i}=\chi(0)$ and $\chi_{f}=\chi(t)$. This subscript convention will be adopted for the other variables. The bath is initially in its own thermal state at temperature $\beta^{-1}$, with the density matrix
\begin{align}
	\rho_{\beta}^{(\phi)}(0)&=\frac{e^{-\beta H^{(\phi)}}}{Z_{\phi}}\,,&Z_{\phi}&=\operatorname{Tr}_{\phi}e^{-\beta H^{(\phi)}}
\end{align}
Here $H^{\phi}[\phi]$ is the free scalar field Hamiltonian associated with the action $S_{\phi}[\phi]$.

The time evolution of the density matrix operator of the combined system is then described by the unitary evolution operator $U(t,0)$ associated with the action \eqref{E:cnmxrere},
\begin{equation}\label{E:woejs}
	\rho(t)=\Bigl\{U(t,0)\,\rho(0)\,U^{-1}(t,0)\Bigr\}\,.
\end{equation}
In the path-integral representation the total density matrix at time $t$ is related to its values at an earlier moment $t=0$ by
\begin{align}
	\rho(\chi_{f},\chi'_{f};\phi_{f},\phi'_{f};t)&=\int_{-\infty}^{\infty}\!d\chi_{i}d\chi'_{i}\!\int_{-\infty}^{\infty}\!d\phi_{i}d\phi'_{i}\!\int_{\chi_{i}}^{\chi_{f}}\!\mathcal{D}\chi_{+}\!\int_{\chi'_{i}}^{\chi'_{f}}\!\mathcal{D}\chi_{-}\!\int_{\phi_{i}}^{\phi_{f}}\!\mathcal{D}\phi_{+}\!\int_{\phi'_{i}}^{\phi'_{f}}\!\mathcal{D}\phi_{-}\notag\\
	 &\qquad\quad\exp\Bigl\{i\,S[\chi_{+},\phi_{+}]-i\,S[\chi_{-},\phi_{-}]\Bigr\}\,\rho^{(\chi)}(\chi_{i},\chi'_{i};0)\rho_{\beta}^{(\phi)}(\phi_{i},\phi'_{i};0)\,.
\end{align}
The variables evaluated along the forward and backward time paths, respectively corresponding to $U$ and $U^{-1}$ in \eqref{E:woejs}, will be distinguished by the subscripts $+$, $-$.

\subsection{Reduced Density Operator and Coarse-Grained Effective Action}

When we are interested only in  the dynamics of the system as influenced by the bath, we can work with the reduced density matrix of the system, obtained by tracing out the microscopic degrees of freedom of the bath in the total density matrix, namely,
\begin{align}
	\rho^{(\chi)}(\chi_{f},\chi'_{f};t)&=\operatorname{Tr}_{\phi}\rho(\chi_{f},\chi'_{f};\phi_{f},\phi'_{f};t)\notag\\
	 &=\int_{-\infty}^{\infty}\!d\chi_{i}d\chi'_{i}\;\rho^{(\chi)}(\chi_{i},\chi'_{i},0)\int_{\chi_{i}}^{\chi_{f}}\!\mathcal{D}\chi_{+}\!\int_{\chi'_{i}}^{\chi'_{f}}\!\mathcal{D}\chi_{-}\;\exp\Bigl\{i\,S_{\chi}[\chi_{+}]-i\,S_{\chi}[\chi_{-}]\Bigr\}\notag\\
	 &\qquad\times\exp\biggl\{\frac{i}{2}\,e^{2}\int_{0}^{t}\!ds\,ds'\biggl(\Bigl[\chi_{+}(s)-\chi_{-}(s)\Bigr]G^{(\phi)}_{R}(s,s')\Bigl[\chi_{+}(s')+\chi_{-}(s')\Bigr]\biggr.\biggr.\notag\\
	 &\qquad\qquad\qquad+\biggl.\biggl.i\,\Bigl[\chi_{+}(s)-\chi_{-}(s)\Bigr]G_{H}^{(\phi)}(s,s')\Bigl[\chi_{+}(s')-\chi_{-}(s')\Bigr]\biggr)\biggr\}\,,\label{E:nwjdjaa}
\end{align}
where the retarded Green's function $G_{R}^{(\phi)}$ of the scalar field $\phi$ is defined by
\begin{align}
	G_{R}^{(\phi)}(s,s')&=i\,\theta(s-s')\operatorname{Tr}\Bigl(\rho_{\beta}\Bigl[\phi(z(s),s),\phi(z(s'),s')\Bigr]\Bigr)\,,
\intertext{and the Hadamard function $G_{H}^{(\phi)}$ by}
	 G_{H}^{(\phi)}(s,s')&=\frac{1}{2}\operatorname{Tr}\Bigl(\rho_{\beta}\Bigl\{\phi(z(s),s),\phi(z(s'),s')\Bigr\}\Bigr)\,.
\end{align}
The Hadamard function is the expectation value of the anti-commutator of the quantum field $\phi$, and is hence temperature dependent. The retarded Green's function, on the other hand, does not have any temperature dependence. The exponential containing $G_{R}^{(\phi)}$ and $G_{H}^{(\phi)}$ in \eqref{E:nwjdjaa} is called  the Feynman-Vernon influence functional $\mathcal{F}$, and we may write it as
\begin{align}
	\mathcal{F}[\chi_{+},\chi_{-}]&=e^{i\,S_{IF}[\chi_{+},\chi_{-}]}\,,
\end{align}
where $S_{IF}$ is called the influence action. It consistently incorporates all the influences of the bath on the system of our interest.

From \eqref{E:nwjdjaa}, we may view the time evolution of the reduced density matrix as a map from its initial value $\rho_{\chi}(0)$ to its final value $\rho_{\chi}(t)$ by a superoperator $J$,
\begin{align}
	 \rho^{(\chi)}(\chi_{f},\chi'_{f};t)&=\int_{-\infty}^{\infty}\!dq_{i}dr_{i}\;J(q_{f},r_{f},t;q_{i},r_{i},0)\;\rho^{(\chi)}(q_{i},r_{i};0)\,.\label{E:fnjdsa}
\end{align}
The matrix elements of the superoperator $J$ are expressed by
\begin{align}
	 J(q_{f},r_{f},t;q_{i},r_{i},0)&=\int_{q_{i}}^{q_{f}}\!\mathcal{D}q\!\int_{r_{i}}^{r_{f}}\!\mathcal{D}r\;\exp\Bigl\{i\,S_{CG}[q,r]\Bigr\}\,,
\end{align}
where $q$, $r$ are respectively the relative coordinate and the center-of-mass coordinate,
\begin{align}
	q&=\chi_{+}-\chi_{-}\,,&r=\frac{1}{2}\bigl(\chi_{+}+\chi_{-}\bigr)\,.
\end{align}
Here, $S_{CG}$, called \textit{the `coarse-grained' effective action},  governs the dynamics of the reduced system under the influence of the bath. It contains the actions of the system plus the influence of the bath on the system described by the influence action $S_{IF}$,
\begin{align}\label{E:oweioiana}
	S_{CG}[q,r]&=S_{\chi}[\chi_{+}]-S_{\chi}[\chi_{-}]+S_{IF}[\chi_{+},\chi_{-}]\\
	 &=\int_{0}^{t}\!ds\;\biggl\{m\dot{q}(s)\dot{r}(s)-m\omega^{2}q(s)r(s)\biggr\}\notag\\
	 &\qquad\qquad\qquad+e^{2}\int_{0}^{t}\!ds\,ds'\biggl[q(s)G_{R}^{(\phi)}(s,s')r(s')+\frac{i}{2}\,q(s)G_{H}^{(\phi)}(s,s')q(s')\biggr]\,,\notag
\end{align}
The path integral in the evolutionary operator $J$ can be evaluated exactly because the coarse-grained effective action \eqref{E:oweioiana} is quadratic in $q$ and $r$.

\subsection{Stochastic Effective Action and Langevin Equations}\label{S:reeih3}
With the help of the Feynman-Vernon identity, we can express the imaginary part of the coarse-grained effective action $S_{CG}$ in \eqref{E:oweioiana} in terms of a Gaussian noise $\xi$,
\begin{align}
	 \exp\left[-\frac{e^{2}}{2}\int_{0}^{t}\!ds\!\int_{0}^{t}\!ds'\;q(s)G_{H}^{(\phi)}(s,s')q(s')\right]=\int\mathcal{D}\xi\;\mathcal{P}[\xi]\,\exp\left[i\,e\int_{0}^{t}\!ds\;q(s)\xi(s)\right]\,,
\end{align}
with
\begin{align}
	 \langle\xi(s)\rangle&=0\,,&\langle\xi(s)\xi(s')\rangle&=G_{H}^{(\phi)}(s,s')\,.
\end{align}
Here the angular brackets represent the ensemble average over the probability distribution functional $\mathcal{P}[\xi]$. Thus the exponential of the coarse-grained effective action $S_{CG}$ can be expressed as a distributional integral
\begin{align}
	 e^{i\,S_{CG}[q,r]}&=\int\mathcal{D}\xi\;\mathcal{P}[\xi]\,e^{i\,S_{SE}[q,r;\xi]}\,,
\end{align}
where $S_{SE}$ is the \textit{stochastic effective action} 
\begin{align}\label{E:hunxms}
	 S_{SE}[q,r;\xi]&=\int_{0}^{t}\!ds\;\biggl\{m\,\dot{q}(s)\dot{r}(s)-m\omega_{b}^{2}\,q(s)r(s)+q(s)\xi(s)+e^{2}\int_{0}^{s}\!ds'\;q(s)G_{R}^{(\phi)}(s,s')r(s')\biggr\}\,.
\end{align}
At this point, we may use the stochastic effective action to either derive the Langevin equation, or to construct the stochastic reduced density matrix. We first take the former route.

\subsubsection{Langevin Equations}
Taking the variation of $S_{SE}$ with respect to $q$ and then letting $q=0$, we arrive at the Langevin equation,
\begin{align}\label{E:euwbnss}
	 m\,\ddot{\chi}(s)+m\omega_{b}^{2}\,\chi(s)-e^{2}\int_{0}^{s}\!ds'\;G^{(\phi)}_{R}(s,s')\chi(s')&=e\,\xi(s)\,.
\end{align}
It describes the time evolution of the reduced system under the non-Markovian influence of the bath. In this case, the influence is manifested in the form of the local stochastic driving noise $\xi$ and the nonlocal dissipative force,
\begin{equation*}
	 e^{2}\int_{0}^{s}\!ds'\;G^{(\phi)}_{R}(s,s')\chi(s')\,.
\end{equation*}
In general, this nonlocal expression implies the evolution of the reduced system is history-dependent. However, in this one-oscillator example, the retarded Green's functions matrix has a simple form
\begin{equation}
	G_{R}^{(\phi)}(s,s')=-\frac{1}{2\pi}\,\theta(s-s')\,\delta'(s-s')\,,
\end{equation}
so the Langevin equation reduces to a purely local expression
\begin{align}
	 m\,\ddot{\chi}(s)+2m\gamma\,\dot{\chi}(s)+m\omega^{2}\,\chi(s)&=e\,\xi(s)\,,\label{E:knkdnfks}
\end{align}
where the renormalized frequency $\omega$ is obtained by lumping the divergence term of $G_{R}^{(\phi)}(s,s')$ with the original bare frequency $\omega_{b}$, and $\gamma=e^{2}/8\pi m>0$ is the damping constant, which serves as a convenient measure for the system-bath coupling strength. Eq.~\eqref{E:knkdnfks} is seen to be the Langevin equation for a driven, damped oscillator, as anticipated.

The reduced system is superficially non-conservative with the presence of friction and noise forces, which originate from the interaction between the system and its environment. These two processes are, however, connected by the fluctuation-dissipation relation. This relation plays a fundamental role in the energy flow balance between the system and the bath: fluctuations in the bath show up as noise and its backaction on the system gives rise to dissipative dynamics.

The general solution to \eqref{E:euwbnss} or \eqref{E:knkdnfks} 
\begin{equation}
	 \chi(s)=d_{1}(s)\chi(0)+d_{2}(s)\dot{\chi}(0)+\frac{e}{m}\int_{0}^{s}\!ds'\;d_{2}(s-s')\xi(s')
\end{equation}
are expanded in terms of fundamental solution matrices $d_{1}$ and $d_{2}$. They are simply the homogeneous solutions of the corresponding equation of motion but satisfy a particular set of initial conditions,
\begin{align}
	d_{1}(0)&=1\,,&\dot{d}_{1}(0)&=0\,,\\
	d_{2}(0)&=0\,,&\dot{d}_{2}(0)&=1\,.
\end{align}	
This can be the starting point for computing the physical observables of the reduced system. For example, the power delivered by the noise $\xi$ at any given time $t>0$ is given by
\begin{align}
	P_{\xi}(t)=e\,\langle\xi(t)\dot{\chi}(t)\rangle&=\frac{e^{2}}{m}\int_{0}^{t}\!ds\;\dot{d}_{2}(t-s)\,\langle\xi(t)\xi(s)\rangle\\
	&=\frac{e^{2}}{m}\int_{0}^{t}\!ds'\;\dot{d}_{2}(t-s')\,G_{H}^{(\phi)}(t-s')\,,\label{E:rnddsa}
\end{align}
since initially the oscillator's displacement $\chi$ is not correlated with the noise force $\xi$.  Likewise the power output at time $t$ due to the dissipative self-force is given by
\begin{align}
	P_{\gamma}(t)&=-2m\gamma\,\langle\dot{\chi}^{2}(t)\rangle\,,
\end{align}
where the velocity uncertainty $\langle\dot{\chi}^{2}(t)\rangle$ takes the form 
\begin{align}
	 \langle\dot{\chi}^{2}(t)\rangle&=\dot{d}_{1}^{2}(t)\,\langle\chi^{2}(0)\rangle+\dot{d}_{2}^{2}(t)\,\langle\dot{\chi}^{2}(0)\rangle\notag\\
	 &\qquad\qquad\qquad\qquad+\frac{e^{2}}{m^{2}}\int_{0}^{t}\!ds\!\int_{0}^{t}\!ds'\;\dot{d}_{2}(t-s)\dot{d}_{2}(t-s')\,\langle\xi(s)\xi(s')\rangle\,,\label{E:mnxeruea}
\end{align}
if initially $\chi(0)$ and $\dot{\chi}(0)$ are not correlated. The first two terms in \eqref{E:mnxeruea} exponentially decay with time, so at late times the third term dominates. This reflects the fact that at late times the dynamics of the reduced system is governed by the bath.

\subsubsection{Stochastic Reduced Density Matrix}
The Langevin equation approach illustrated above to obtain the desired physical quantities associated with the dynamics of the reduced system is less formal, but more flexible and physically intuitive. It is particularly convenient if the quantities at hand involve noise either from the bath or externally introduced. Alternatively, a more systematic and complete approach is by means of the reduced density operator, which accounts for the full quantum dynamics of the reduced system and enforces the operator ordering. The drawback is that since the influence functional does not have explicit dependence on the noise, it is not straightforward to implement it for the cases that explicitly depend on the bath noise. An example is the average power input by the noise shown in \eqref{E:rnddsa}. Only after invoking the Feynman-Vernon identity could the bath noise be made explicit.  Next we will show a way to combine the advantages of these two approaches, by incorporating the noise from the bath in the reduced density matrix.

Let us rewrite the reduced density matrix \eqref{E:fnjdsa} in terms of the stochastic effective action $S_{SE}$ in \eqref{E:hunxms},
\begin{align}
	 \rho^{(\chi)}(q_{f},r_{f};t)&=\int_{-\infty}^{\infty}\!dq_{i}dr_{i}\;\rho^{(\chi)}(q_{i},r_{i};0)\int_{q_{i}}^{q_{f}}\!\mathcal{D}q\!\int_{r_{i}}^{r_{f}}\!\mathcal{D}r\;\exp\Bigl\{i\,S_{CG}[q,r]\Bigr\}\notag\\
	&=\int\mathcal{D}\xi\;\mathcal{P}[\xi]\,\rho^{(\chi)}(q_{f},r_{f},t;\xi]\,,
\end{align}
The $\rho^{(\chi)}(q_{f},r_{f},t;\xi]$ is called the stochastic reduced density matrix, which has explicit dependence on the noise $\xi$ of the bath
\begin{equation}\label{E:eucbma}
	 \rho_{\chi}^{(\chi)}(q_{f},r_{f},t;\xi]=\int_{-\infty}^{\infty}\!dq_{i}dr_{i}\;\rho^{(\chi)}(q_{i},r_{i};0)\int_{q_{i}}^{q_{f}}\!\mathcal{D}q\!\int_{r_{i}}^{r_{f}}\!\mathcal{D}r\;\,e^{i\,S_{SE}[q,r,\xi]}\,,
\end{equation}
with the stochastic effective action given by \eqref{E:hunxms}. In this representation, we see that the reduced system, now driven by a classical stochastic force of the bath, is described by the stochastic density matrix. For each realization of the bath noise, the reduced system evolves to a state described by the density matrix \eqref{E:eucbma}. Different realizations make the system end up at different final states with probability given by $\mathcal{P}[\xi]$.

To compute the quantum and stochastic average of a dynamical variable, say,  $f(\chi;\xi]$ at time $t$, which depends on both the stochastic variable $\xi$ and the quantum operator $\chi$ of the reduced system, we simply evaluate the trace associated with the system variables and the ensemble average associated with the bath noise,
\begin{equation}\label{E:mcbers}
	\langle f(\chi;\xi]\rangle=\int\mathcal{D}\xi\;\mathcal{P}[\xi]\,\operatorname{Tr}_{\chi}\Bigl\{\rho^{(\chi)}(t;\xi]\,f(\chi;\xi]\Bigr\}\,.
\end{equation}
The procedure in \eqref{E:mcbers} is understood as follows: for each specific realization of the stochastic noise $\xi$, we first calculate the expectation value of the quantum operator $f(\chi;\xi]$ for the state described by the reduced density operator $\rho^{(\chi)}(t;\xi]$. The obtained result, still dependent on the stochastic variable, will then be averaged over according to the probability distribution $\mathcal{P}[\xi]$ of the noise.

As an example, we will compute the same average power $P_{\xi}$ delivered by the stochastic force $\xi$ from bath as in \eqref{E:rnddsa}. Once we note that $p=m\dot{\chi}$, the power $P_{\xi}$ is then give by
\begin{align}
	P_{\xi}(t)&=\frac{e}{m}\,\langle\,\xi(t)\,p(t)\,\rangle\notag\\
			 &=-i\,\frac{e}{m}\int\mathcal{D}\xi\;\mathcal{P}[\xi]\,\int_{-\infty}^{\infty}\!dq_{f}dr_{f}\;\delta(q_{f})\,\xi\;\frac{\partial}{\partial\chi_{f}}\,\rho^{(\chi)}(q_{f},r_{f},t;\xi)\,,\label{E:tnbrex}
\end{align}
where the momentum $p$ canonical to the coordinate $\chi$ is given by
\begin{equation}
	p=-i\,\frac{\partial}{\partial\chi}\,,
\end{equation}
and the trace over the dynamical variables of the reduced system is defined as
\begin{equation}
	\operatorname{Tr}_{\chi}=\int_{-\infty}^{\infty}\!dq_{f}dr_{f}\;\delta(q_{f})\,.
\end{equation}
Since the initial state of the reduced system is a Gaussian state and the stochastic effective action is quadratic in the system's variables, the final state will remain Gaussian and the corresponding reduced density operator thus can be evaluated exactly. To derive the explicit form of the reduced density matrix, we first evaluate the path integrals in \eqref{E:eucbma},
\begin{align}
	 &\quad\int_{q_{i}}^{q_{f}}\!\mathcal{D}q\!\int_{r_{i}}^{r_{f}}\!\mathcal{D}r\;\exp\left\{i\int_{0}^{t}\!ds\;\biggl[m\,\dot{q}(s)\dot{r}(s)-m\omega_{b}^{2}(s)\,q(s)r(s)+q(s)\xi(s)\biggr.\right.\notag\\
	 &\qquad\qquad\qquad\qquad\qquad\qquad\qquad\qquad\qquad\qquad+\left.\biggl.e^{2}\int_{0}^{s}\!ds'\;q(s)G_{R}^{(\phi)}(s,s')r(s')\biggr]\right\}\notag\\
	 &=\mathcal{N}\,\exp\biggl[i\,m\,q_{f}\dot{\overline{r}}_{f}-i\,m\,q_{i}\dot{\overline{r}}_{i}\biggr]\,,
\end{align}
where $\mathcal{N}$ is the normalization constant, and can be determined by the unitarity requirement. It is given by
\begin{equation}
	\mathcal{N}=\left(\frac{m}{2\pi}\right)^{2}\det\dot{\mu}(0)\,.
\end{equation}
Note that the mean trajectories $\overline{q}$, $\overline{r}$ are solutions to the Langevin equation \eqref{E:knkdnfks} with the boundary conditions $\overline{q}(t)=q_{i}$, $\overline{q}(0)=q_{i}$ and $\overline{r}(t)=r_{b}$, $\overline{r}(0)=r_{i}$. Thus they and their time derivatives are functionals of the stochastic noise $\xi$. Explicitly, in terms of the boundary values, we can write $\overline{r}(s)$ as
\begin{equation}
	 \overline{r}(s)=\nu(s)\,r_{i}+\mu(s)\,r_{f}+\mathcal{J}_{r}(s)\,,
\end{equation}
for $0\leq s\leq t$. The functions $\mu(s)$, $\nu(s)$ are 
\begin{align}
	\mu(s)&=\frac{d_{2}(s)}{d_{2}(t)}\,,&\nu(s)&=d_{1}(s)-d_{2}(s)\,\frac{d_{1}(t)}{d_{2}(t)}\,,\label{E:zzz2}
\end{align}
and the current $\mathcal{J}_{r}(s)$ is given by
\begin{equation}
	 \mathcal{J}_{r}(s)=\frac{e}{m}\int_{0}^{s}\!ds'\;d_{2}(s-s')\xi(s')-\frac{e}{m}\int_{0}^{t}\!ds'\;\frac{d_{2}(s)}{d_{2}(t)}\,d_{2}(t-s')\,\xi(s')\,.
\end{equation}
Moreover, we can write the partial derivative $\partial/\partial\chi$ as
\begin{equation}
	\frac{\partial}{\partial\chi}=\frac{\partial}{\partial q}+\frac{1}{2}\frac{\partial}{\partial r}\,.
\end{equation}
Now we are ready to evaluate the power delivered by the stochastic force $\xi$, Eq.~\eqref{E:tnbrex} becomes
\begin{align}
	 P_{\xi}(t)&=\frac{\mathcal{N}}{\det\dot{\mu}(0)}\left(\frac{2\pi}{m}\right)^{2}\int\mathcal{D}\xi\;\mathcal{P}[\xi]\;\xi\,\Bigl[\dot{\mathcal{J}}_{r}(t)-\frac{\dot{\mu}(t)}{\dot{\mu}(0)}\,\dot{\mathcal{J}}_{r}(0)\Bigr]\,.
\end{align}
The expressions in the square brackets can be reduced to
\begin{align}
	 \dot{\mathcal{J}}_{r}(t)-\frac{\dot{\mu}(t)}{\dot{\mu}(0)}\,\dot{\mathcal{J}}_{r}(0)&=\frac{e}{m}\int_{0}^{t}\!ds'\;\dot{d}_{2}(t-s')\,\xi(s')\,.
\end{align}
Thus the power delivered to the system from the bath is equal to
\begin{align}
	 P_{\xi}(t)&=\frac{e^{2}}{m}\int\mathcal{D}\xi\;\mathcal{P}[\xi]\;\xi(t)\int_{0}^{t}\!ds'\;\dot{d}_{2}(t-s')\,\xi(s')=\frac{e^{2}}{m}\int_{0}^{t}\!ds'\;\dot{d}_{2}(t-s')\,G_{H}^{(\phi)}(t-s')\,.\label{E:bvnorue}
\end{align}
This is exactly the same as \eqref{E:rnddsa}.

\section{Thermodynamics of simple systems in a common bath}\label{S:one}

\subsection{System of One Harmonic Oscillator}

We first examine the one-oscillator system. The absence of mutual interaction between the constituents of the system renders it exactly and  fully solvable without approximation. It can readily be compared with the corresponding case in conventional thermodynamics, where the system is extremely weakly coupled to and assumed to be always in thermal equilibrium with a heat bath, whose dynamics is of no concern to the system beyond its being in a thermal state with a temperature parameter.   In this case the mechanical energy is given by
\begin{equation}\label{E:btwiaf}
	E(\infty)=\frac{1}{2}\operatorname{Im}\int_{-\infty}^{\infty}\!\frac{d\kappa}{2\pi}\;\coth\frac{\beta\kappa}{2}\,\bigl(\kappa^{2}+\omega_{p}^{2}\bigr)\,\tilde{d}_{2}(\kappa)\,,
\end{equation}
where $\tilde{d}_{2}(\kappa)$ is given by
\begin{equation}
	\tilde{d}_{2}(\kappa)=\frac{1}{\omega_{b}^{2}-\kappa^{2}-\dfrac{e^{2}}{m}\,G_{R}^{(\phi)}(\kappa)}=\frac{1}{\omega_{p}^{2}-\kappa^{2}-i\,2\gamma\kappa}\,.
\end{equation}
The integral in \eqref{E:btwiaf} is logarithmically divergent, so we will introduce a regularization scheme in due course. In addition since it is much more difficult to evaluate the integral of the hyper-trigonometric function, we write the factor $\coth\dfrac{\beta\kappa}{2}$ as the summation of the algebraic function of the Matsubara frequency $\nu_{n}$, 
\begin{align}
	\coth\frac{\beta\kappa}{2}&=\frac{2}{\beta}\sum_{n=-\infty}^{\infty}\frac{\kappa}{\kappa^{2}+\nu_{n}^{2}}\,,&\nu_{n}&=\frac{2n\pi}{\beta}\,.
\end{align}
in hope that the resulting integral contains only the algebraic function.

On account of regularization, we may assume it valid to exchange the order of integration and summation, so that \eqref{E:btwiaf} becomes 
\begin{align}\label{E:dfebdss}
	E(\infty)=\operatorname{Im}\frac{2}{\beta}\sum_{n=-\infty}^{\infty}\int_{-\infty}^{\infty}\!\frac{d\kappa}{2\pi}\;\frac{\kappa}{\kappa^{2}+\nu_{n}^{2}}\frac{i\,\gamma\kappa-\omega_{p}^{2}}{\kappa^{2}-\omega_{p}^{2}+i\,2\gamma\kappa}\,.
\end{align}
The evaluation of the integral in \eqref{E:dfebdss} is straightforward except for the contribution of the zero mode $n=0$, which needs a separate treatment from the $n\neq 0$ case. The contribution from the $(n=0)$ zero mode has an infrared (IR) divergence. We introduce an IR cutoff $\delta$ and obtain
\begin{align}
	\int_{-\infty}^{\infty}\!\frac{d\kappa}{2\pi}\;\frac{1}{\kappa}\frac{i\,\gamma\kappa-\omega_{p}^{2}}{\kappa^{2}-\omega_{p}^{2}+i\,2\gamma\kappa}=\lim_{\delta\to0^{+}}\int_{-\infty}^{\infty}\!\frac{d\kappa}{2\pi}\;\frac{\kappa}{\kappa^{2}+\delta^{2}}\frac{i\,\gamma\kappa-\omega_{p}^{2}}{\kappa^{2}-\omega_{p}^{2}+i\,2\gamma\kappa}&=\frac{i}{2}\,.
\end{align}
For $n\neq0$, the integral in the summation is well defined and it gives
\begin{equation}
	\int_{-\infty}^{\infty}\!\frac{d\kappa}{2\pi}\;\frac{\kappa}{\kappa^{2}+\nu_{n}^{2}}\frac{i\,\gamma\kappa-\omega_{p}^{2}}{\kappa^{2}-\omega_{p}^{2}+i\,2\gamma\kappa}=\frac{i}{2}\left(\frac{\gamma\nu_{n}+\omega_{p}^2}{\omega_{p}^2+2\gamma\nu_{n}+\nu_{n}^2}\right)\,.
\end{equation}
Now Eq.~\eqref{E:dfebdss} becomes
\begin{align}
	E(\infty)=\frac{1}{\beta}\sum_{n=-\infty}^{\infty}\frac{\gamma\lvert\nu_{n}\rvert+\omega_{p}^2}{\omega_{p}^2+2\gamma\lvert\nu_{n}\rvert+\nu_{n}^2}\,.
\end{align}
The summation including infinitely high Matsubara frequencies will give an inevitable UV logarithmic divergence, as expected from \eqref{E:btwiaf}. We insert a damping factor $e^{-\nu_{n}/\Lambda}$, with $\Lambda>0$, to regularize the summation and arrive at
\begin{align}
	&\quad\frac{2}{\beta}\sum_{n=1}^{\infty}\frac{\gamma\nu_{n}+\omega_{p}^2}{\omega_{p}^2+2\gamma\nu_{n}+\nu_{n}^2}\,e^{-\nu_{n}/\Lambda}\notag\\
	&=2e^{-\frac{2\pi}{\beta\Lambda}}\,\operatorname{Im}\,\biggl\{\frac{w_{+}}{1-i\,\beta w_{+}}\,~_{2}F_{1}(1,\,1-i\,\beta w_{+},\,2-i\,\beta w_{+};\,e^{-\frac{2\pi}{\beta\Lambda}})\biggr\}\,,
\end{align}
where $w_{\pm}=(W\pm i\,\gamma)/2\pi$ and $W$ is the resonance frequency, given by $W=\sqrt{\omega_{p}^{2}-\gamma^{2}}$. The hypergeometric function ${}_{2}F_{1}(a,\,b,\,c;\,z)$ is defined by
\begin{align}
	~_{2}F_{1}(a,\,b,\,c;\,z)&=\sum_{k=0}^{\infty}\frac{(a)_{k}(b)_{k}}{(c)_{k}}\frac{z^{k}}{k!}\,,&(a)_{k}&=\prod_{n=0}^{k-1}(a+n)=\frac{(a+k-1)!}{(a-1)!}\,,
\end{align}
and has a branch-cut on the complex $z$ plane along the real axis from 1 to $\infty$.

In the limit $\Lambda\to+\infty$, the mechanical energy of the oscillator becomes
\begin{align}\label{E:dferbes}
	E(\infty)=\frac{1}{\beta}-\frac{\gamma}{\pi}\,\ln\frac{2\pi}{\beta\Lambda}-2\operatorname{Im}\,\biggl\{w_{+}\,\mathbb{H}(-i\,\beta w_{+})\biggr\}\,,
\end{align}
where $\mathbb{H}(n)$ is the $n^{\mathrm{th}}$ harmonic number. The cutoff parameter $\Lambda$ defines the highest energy scale in the problem, Its presence can be understood as the consequence that the oscillator couples with a bath that contains a huge number of degrees of freedom. It results from the bath contribution on very short length scales. Since $\ln\Lambda$ is accompanied by the damping constant $\gamma$, the cutoff-dependent term in \eqref{E:dferbes} is negligible for weak oscillator-bath interaction while it can have a significant contribution in the strong interaction limit. Thus the internal energy $E(\infty)$ in principle can depend on the cutoff scale. Note that since the cutoff-dependent term does not depend on temperature, it will not appear in the heat capacity.

This yields for the heat capacity
\begin{align}\label{E:btweyw}
	C=1-\frac{\gamma\beta}{\pi}-2\operatorname{Re}\biggl\{\beta^{2}w_{+}^{2}\;\psi^{(1)}(1-i\,\beta w_{+})\biggr\}=\begin{cases}
										1-\dfrac{\gamma\beta}{\pi}+\mathcal{O}(\beta^{2})\,,&\beta\omega\ll1\,,\vspace{3pt}\\
										\dfrac{2\pi\gamma}{3\beta\omega_{p}^{2}}+\mathcal{O}(\beta^{-2})\,,&\beta\gamma\gg1\,,
									 \end{cases}
\end{align}
when $\gamma<\omega_{p}$. We thus see the heat capacity grows algebraically from zero at low temperature and then saturates to unity at high temperature. The function $\psi^{(n)}(z)$ is the $n^{\mathrm{th}}$ derivative of the digamma function.

Next we examine the weak oscillator-bath coupling limit $\gamma\to0$. In this limit the mean mechanical energy \eqref{E:dferbes} becomes
\begin{equation}\label{E:ojernd}
	\lim_{\gamma\to0}E(\infty)=\frac{\omega_{p}}{2}\,\coth\frac{\beta\omega_{p}}{2}-\frac{\gamma}{\pi}\,\ln\frac{2\pi}{\beta\Lambda}+\mathcal{O}(\gamma)\,,
\end{equation}
where $\mathcal{O}(\gamma)$ contains the finite contribution of the order $\gamma$.~\footnote{It also contains the cutoff-dependent contributions but they are of the order $\Lambda^{-1}$ and higher. contribution of the order $\gamma$. {In principle} as $\gamma\to0$, the physical frequency $\omega_{p}$ will approach to the bare value $\omega_{b}$. This seems innocuous at first sight for the weak oscillator-bath interaction regime, but their values can be drastically different in the strong interaction and the large cutoff scale limit. In particular since $\omega_{p}$ and $\omega_{b}$ are related by \eqref{E:bsrthbf}, the choice of $\gamma$ and $\Lambda$ must be restricted so that $\omega_{p}^{2}$ remains positive definite to prevent unstable dynamics. In fact the physical frequency $\omega_{p}$  is determined by experimental preparation; thus from the operational viewpoint we can let the physical frequency be fixed at the energy scale of measurement. Alternatively, in the action~\eqref{E:dbeksd}, we can assume the system parameters take the physical values, and we introduce counter terms to cancel contributions due to interactions with the bath~\cite{DC06}. Both approaches eventually produce equivalent results.} We can compare results for different values of the system-bath interaction strength $\gamma$.

Following this protocol, the corresponding heat capacity is then given by
\begin{equation}\label{E:berwaa}
	\lim_{\gamma\to0}C=\frac{\left(\dfrac{\beta\omega_{p}}{2}\right)^{2}}{\sinh^{2}\dfrac{\beta\omega_{p}}{2}}+\mathcal{O}(\gamma)\,.
\end{equation}
It is divergence-free due to the fact that the logarithmic divergence does not depend on temperature.

Let us now compare this expression with the corresponding heat capacity in conventional thermodynamics. Consider a harmonic oscillator in its thermal state
\begin{align}
	\rho_{\beta}^{(\chi)}&=Z_{\chi}^{-1}\,e^{-\beta H^{(\chi)}}\,,&&\text{with}&H^{(\chi)}&=\frac{m}{2}\,\dot{\chi}^{2}+\frac{m\omega^{2}_{p}}{2}\,\chi^{2}\,.
\end{align}
Its mean mechanical energy is given by
\begin{equation}
	E=\operatorname{Tr}\Bigl\{\rho_{\beta}^{(\chi)}\,H^{(\chi)}\Bigr\}=\frac{\omega_{p}}{2}\,\coth\frac{\beta\omega_{p}}{2}\,.
\end{equation}
It does not depend on the coupling between the oscillator and the bath (because of this, one may not realize that conventional thermodynamics is an open-system theory) and in fact this expression is the same as the dominant term of \eqref{E:ojernd} in the weak coupling limit. For this system the heat capacity in conventional thermodynamics is exactly given by
\begin{equation}\label{E:beersd}
	C=-\beta^{2}\,\frac{\partial E}{\partial\beta}=\frac{\left(\dfrac{\beta\omega_{p}}{2}\right)^{2}}{\sinh^{2}\dfrac{\beta\omega_{p}}{2}}=\begin{cases}
			1-\dfrac{\beta^{2}\omega_{p}^{2}}{12}+\cdots\,,&\beta\omega_{p}\ll1\,,\vspace{3pt}\\
			\beta^{2}\omega_{p}^{2}\,e^{-\beta\omega_{p}}+\cdots\,,&\beta\omega_{p}\gg1\,,
	\end{cases}
\end{equation}
and is equal to the leading term of \eqref{E:berwaa} in the weak coupling limit. Note that it has a different low temperature asymptote from that in \eqref{E:btweyw}. This is central to the consideration of the third law. Similar results have been obtained for the same configuration, which is nonetheless initially prepared in the equilibrium global thermal state~\cite{HanIng06}, but the physical contents are different. The similarity in outcomes based on the nonequilibrium initial state and equilibrium global thermal state is not a coincidence, as has been discussed in~\cite{SFTH}. Essentially it is the consequence of the damped, stable motion of the reduced system due to the interaction with the bath.

For nonvanishing oscillator-bath coupling, there is a stark difference between the conventional thermodynamical equilibrium and our open-system nonequilibrium approaches. In general, the results in the open-system framework may have cutoff-dependent contributions, as a result of the huge number of degrees of freedom in the bath. Since this cutoff-dependent term is proportional to the oscillator-bath interaction strength, it tends to be ignored in the weak coupling approximation. The second distinction is related to the observation that there is one more scale $\gamma$ in the open-systems nonequilibrium framework, in addition to $\omega$ and $\beta^{-1}$ already existent in the conventional equilibrium thermodynamics framework. This introduces an additional subtlety in defining the low temperature limit $\beta\omega\gg1$. In the open-systems nonequilibrium framework, there may be a more stringent criterion such as $\beta\gamma\gg1$ when $\gamma/\omega<1$, or $\beta f(\gamma,\beta)\gg1$, where $f(\gamma,\beta)$ is a dimensional function of $\gamma$, $\omega$ and $[f]=L^{-1}$, that is, inversely proportional to the length scale $L$. The presence of this additional scale contributes to different predictions of the heat capacity between the two frameworks in the low temperature regime, as can be seen from \eqref{E:btweyw} and \eqref{E:beersd}.

The low-temperature behavior of the heat capacity in \eqref{E:btweyw} has been argued~\cite{HZ95}, for the global thermal state case, to be related to the density of the state of the harmonic oscillator in the thermal bath. There it has been shown that if the combined system is initially in the equilibrium global thermal state, then the original discrete energy spectrum of the undamped oscillator will become a continuous one with a unique ground level. That is, the oscillator-bath interaction renders the oscillator a gapless system. This interesting observation has not been proven for a nonequilibrium initial state. It may still apply because the spectrum depends on the effective Hamiltonian (Lagrangian) instead of the prepared initial state.

\subsection{System of two coupled oscillators in a common bath}\label{S:ebrej}

When the system has two or more oscillators, the dynamics of the reduced system becomes much more intricate. Other than their direct coupling, the oscillators also interact with each other indirectly through their shared bath. This indirect influence by one oscillator will propagate in the form of the bath or field disturbance exerted onto all the other oscillators. In turn, more and more subsequent repercussions will be proliferated among the perturbed oscillators. Thus the total effect on the system as a whole depends on the history of each oscillator, leading to very complex evolution. {In addition to this, further complication in the interpretation of the results arises} from the reduced (environment-influenced) system's parameter space containing regions where its motion is unstable.  We need to identify and exclude this case, then expound the results of the reduced system in an equilibrium state after its motion is fully relaxed.

\subsubsection{Dynamics}

From \eqref{E:bejhrys}, the equations of motion for two coupled oscillators are
\begin{align}
	\ddot{\chi}_{1}(t)+\omega_{p}^{2}\,\chi_{1}(t)+\sigma\,\chi_{2}(t)+2\gamma\,\dot{\chi}_{1}(t)-\frac{2\gamma}{\ell}\,\theta(t-\ell)\,\chi_{2}(t-\ell)&=\frac{1}{m}\,\xi_{1}(t)\,,\label{E:dfkdsdaq}\\
	\ddot{\chi}_{2}(t)+\omega_{p}^{2}\,\chi_{2}(t)+\sigma\,\chi_{1}(t)+2\gamma\,\dot{\chi}_{2}(t)-\frac{2\gamma}{\ell}\,\theta(t-\ell)\,\chi_{1}(t-\ell)&=\frac{1}{m}\,\xi_{2}(t)\,,\label{E:dfkdsdaq1}
\end{align}
where the oscillators 1, 2 are respectively located at $\mathbf{z}_{1}$, $\mathbf{z}_{2}$ so that $\ell=\lvert\mathbf{z}_{1}-\mathbf{z}_{2}\rvert$. It is convenient to reorganize the coupled motion of these two oscillators to an uncoupled motion of a fast mode $\Sigma=(\chi_{1}+\chi_{2})/2$ and a slow mode $\Delta=\chi_{1}-\chi_{2}$,
\begin{align}
	\ddot{\Sigma}(t)+\omega_{+}^{2}\,\Sigma(t)+2\gamma\,\dot{\Sigma}(t)-\frac{2\gamma}{\ell}\,\theta(t-\ell)\,\Sigma(t-\ell)&=\frac{1}{2m}\bigl [\xi_{1}(t)+\xi_{2}(t)\bigr]\,,\label{E:dkeje1}\\
	\ddot{\Delta}(t)+\omega_{-}^{2}\,\Delta(t)+2\gamma\,\dot{\Delta}(t)+\frac{2\gamma}{\ell}\,\theta(t-\ell)\,\Delta(t-\ell)&=\frac{1}{m}\bigl [\xi_{1}(t)-\xi_{2}(t)\bigr]\,,\label{E:dkeje2}
\end{align}
with $\omega^{2}_{\pm}=\omega^{2}_{p}\pm\sigma$. Although in appearance the variables $\Sigma$ and $\Delta$ satisfy separate equations \eqref{E:dkeje1} and \eqref{E:dkeje2}, it does not mean that their motions are  decoupled. This is because the righthand sides of \eqref{E:dkeje1} and \eqref{E:dkeje2} indicate that the noises from two locations get mingled together. This clearly shows that the two oscillators are correlated due to arbitration of the ambient quantum field bath.

Their solutions are most easily found if we perform the Laplace transformation over this set of equations of motion and turn them into a simultaneous set of algebraic equations
\begin{align}
	\Bigl[z^{2}+2\gamma\,z+\omega_{+}^{2}-\frac{2\gamma}{\ell}\,e^{-z\ell}\Bigr]\tilde{\Sigma}(z)&=\Bigl(z+2\gamma\Bigr)\Sigma(0)+\dot{\Sigma}(0)+\frac{1}{m}\,\tilde{\xi}_{+}(z)\,,\\
	\Bigl[z^{2}+2\gamma\,z+\omega_{-}^{2}+\frac{2\gamma}{\ell}\,e^{-z\ell}\Bigr]\tilde{\Delta}(z)&=\Bigl(z+2\gamma\Bigr)\Delta(0)+\dot{\Delta}(0)+\frac{1}{m}\,\tilde{\xi}_{-}(z)\,,
\end{align} 
where $\xi_{+}=(\xi_{1}+\xi_{2})/2$ and $\xi_{-}=\xi_{1}-\xi_{2}$. Unstable motion occurs when the solutions to
\begin{equation}\label{E:iernsdw}
	z^{2}+2\gamma\,z+\omega_{\pm}^{2}\mp\frac{2\gamma}{\ell}\,e^{-z\ell}=0\,,
\end{equation}
have the positive real parts. The solution that corresponds to unstable (runaway) motion can be shown to be always located on the positive real axis of the complex $z$ plane. Since \eqref{E:iernsdw} is not a simple algebraic equation, it will have an infinite number of solutions, symmetrically distributed on both sides of the real axis, except for the previously mentioned runaway solution. It reflects the mutual undulant disturbance mediated by the field from each oscillator. Thus we expect the spatial non-Markovianity renders the motion of two oscillators much more intricate than that of one oscillator. Finding the solutions to \eqref{E:iernsdw} is then nontrivial but since the details about locating the perturbative or the asymptotic solutions have been discussed in~\cite{HHPRD, HWL08,BC63,LinHu09}, they will not be repeated here.

However, a word of caution about the choice of $\ell$: When $\ell$ is extremely small such that $2\gamma/\ell\gg1$ the contribution from the retardation term is comparable with the frequency renormalization due to the interaction of the oscillator with the bath. Therefore the expression for the equation of motion like \eqref{E:iernsdw} becomes dubious in the sense that 1) a point particle model is not always feasible in the context of the self-force, as was long pointed out by Rohrlich and others~\cite{roh}, 2) the equation of motion has a different damping term, proportional to the third-order time derivative, instead of the first-order one~\cite{HHPRD}. Thus the effect of finite size of the oscillator must be taken into consideration.

In summary, instability of motion occurs when the formal effective oscillating frequencies of at least one of the two modes
\begin{align}
	\omega_{eff}^{(\pm)}&=\omega^{2}\pm\sigma\mp\frac{2\gamma}{\ell}\,\cos z\ell\,,
\end{align}
become negative. It is likely to happen when 1) the oscillating frequencies of the normal mode $\omega_{\pm}$ become imaginary, and/or 2) the non-Markovian field-induced effect becomes too extreme. Finally we remark that the ratio $\varsigma=\sigma\ell/(2\gamma)$ measuring the relative strength between the direct inter-oscillator coupling and the indirect environment-induced non-Markovian effect is a useful quantity for this consideration, as was introduced in~\cite{HHPRD} for expounding the competing physical mechanisms determining the quantum entanglement between two coupled oscillators in a shared bath.

\subsubsection{Internal energy, Heat capacity}

In the matrix notations \eqref{E:bdeere} for this two-oscillator system, we have
\begin{align}
	\pmb{\Omega}_{b}^{2}&=\begin{pmatrix}
	\omega_{b}^{2} &\sigma\\
	\sigma     &\omega_{b}^{2}
	\end{pmatrix}\,,&\mathbf{G}_{R}^{(\phi)}(\kappa)=\begin{pmatrix}
	\dfrac{\delta(0)}{2\pi}+i\,\dfrac{\kappa}{4\pi} &\dfrac{e^{i\,\kappa\ell}}{4\pi\ell}\\
	\dfrac{e^{i\,\kappa\ell}}{4\pi\ell}     &\dfrac{\delta(0)}{2\pi}+i\,\dfrac{\kappa}{4\pi}
	\end{pmatrix}\,,
\end{align}
where $\ell=\lvert\mathbf{z}_{1}-\mathbf{z}_{2}\rvert$, and from \eqref{E:ejfdhwe}, we can find $\tilde{G}_{R}^{(\phi)}(\omega;\,r)$ given by
\begin{align}
	\tilde{G}_{R}^{(\phi)}(\kappa;\,r)=\int_{-\infty}^{\infty}\!d\tau\;G_{R}^{(\phi)}(\tau,\,r)\,e^{i\,\kappa\tau}&=\begin{cases}
		\dfrac{e^{i\,\kappa r}}{4\pi r}\,,&r\neq0\,,\vspace{6pt}\\
		\dfrac{\delta(0)}{2\pi}+i\,\dfrac{\kappa}{4\pi}\,,&r=0\,.
	  \end{cases}
\end{align}
In the diagonal elements of $\dfrac{e^{2}}{m}\,\mathbf{G}_{R}^{(\phi)}(\kappa)$, the divergent or cutoff-dependent term will be absorbed with the diagonal elements $\omega_{b}^{2}$ in $\pmb{\Omega}_{b}^{2}$ to form the physical frequency $\omega^{2}_{p}$. The remaining term in the diagonal elements of $\dfrac{e^{2}}{m}\,\mathbf{G}_{R}^{(\phi)}(\kappa)$ will then give $i\,2\gamma\kappa$. Thus $\tilde{\mathbf{d}}_{2}(\kappa)$ in \eqref{E:efjdbjbd} becomes
\begin{align}
	\tilde{\mathbf{d}}_{2}(\kappa)&=\begin{pmatrix}
	\omega^{2}_{p}-\kappa^{2}-i\,2\gamma\kappa 			&\sigma-\dfrac{2\gamma}{\ell}\,e^{i\,\kappa\ell}\\
	\sigma-\dfrac{2\gamma}{\ell}\,e^{i\,\kappa\ell} &\omega^{2}_{p}-\kappa^{2}-i\,2\gamma\kappa
	\end{pmatrix}^{-1}\,.
\end{align}
This is essentially what we need to compute the total mechanical energy \eqref{E:efkbsdf} of two oscillators when their motion reaches equilibrium after relaxation.

We have assumed somewhat artificially that $\sigma$ is a constant independent of the separation between the two oscillators. We may relax this restriction to allow it to be a function of $\ell_{ij}=\lvert\mathbf{z}_{i}-\mathbf{z}_{j}\rvert$, namely, 
\begin{equation}
	\sigma=f(\ell_{ij})\,.
\end{equation}
Here $f(z)$ is a monotonically decreasing function of $z$, except for the case $z=0$, where we require $f(0)=0$, i.e, no self-interaction.

\begin{figure}
\centering
    \scalebox{0.62}{\includegraphics{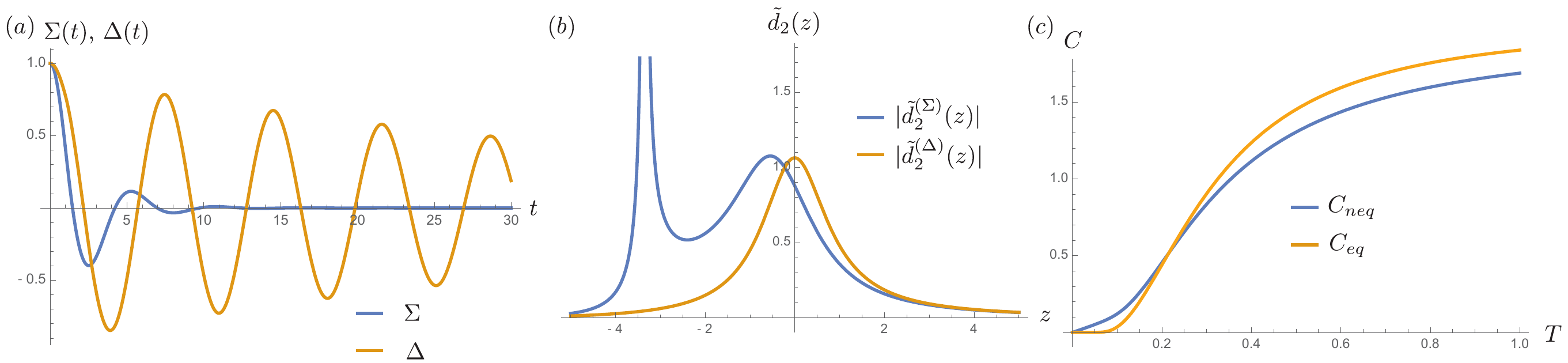}}
    \caption{The case of two coupled quantum oscillators in a common thermal bath. (a) The general behaviors of the motions of the fast mode $\Sigma(t)$ and the slow mode $\Delta(t)$. (b) The behaviors of $\lvert\tilde{d}_{2}^{(\Sigma)}(z)\rvert$ and $\lvert\tilde{d}_{2}^{(\Delta)}(z)\rvert$:  the divergence signifies the presence of the pole, and the positive pole implies instability in motion. (c) The temperature dependence of the heat capacity in the equilibrium vs nonequilibrium approaches. The parameters are chosen such that the resonance frequency $\omega_{r}=\sqrt{\omega^{2}_{p}-\gamma^{2}}=1$, $\gamma=0.2$, $\sigma=0.5$, $\ell=1$, and the cutoff scale $\Lambda^{-1}=0.01$.}\label{Fi:twosc}
\end{figure}

The total mechanical energy for the two coupled oscillator system is then given by
\begin{align}
	E(\infty)&=\frac{1}{2}\operatorname{Im}\int_{-\infty}^{\infty}\!\frac{d\kappa}{2\pi}\;\coth\frac{\beta\kappa}{2}\biggl\{\frac{\omega_{+}^{2}+\kappa^{2}}{\omega_{+}^{2}-\kappa^{2}-i\,2\gamma\kappa-\dfrac{2\gamma}{\ell}\,e^{i\,\kappa\ell}}+\frac{\omega_{-}^{2}+\kappa^{2}}{\omega_{-}^{2}-\kappa^{2}-i\,2\gamma\kappa+\dfrac{2\gamma}{\ell}\,e^{i\,\kappa\ell}}\biggr\}\,e^{-\frac{\lvert\kappa\rvert}{\Lambda}}\,,\notag
\end{align}
and the corresponding heat capacity is
\begin{align}\label{E:dkbdkjse}
	C(\infty)&=\operatorname{Im}\int_{-\infty}^{\infty}\!\frac{d\kappa}{2\pi}\;\left(\frac{\dfrac{\beta\kappa}{2}}{\sinh\dfrac{\beta\kappa}{2}}\right)^{2}\notag\\
	&\qquad\times\frac{1}{\kappa}\biggl\{\frac{\omega_{+}^{2}+\kappa^{2}}{\omega_{+}^{2}-\kappa^{2}-i\,2\gamma\kappa-\dfrac{2\gamma}{\ell}\,e^{i\,\kappa\ell}}+\frac{\omega_{-}^{2}+\kappa^{2}}{\omega_{-}^{2}-\kappa^{2}-i\,2\gamma\kappa+\dfrac{2\gamma}{\ell}\,e^{i\,\kappa\ell}}\biggr\}\,e^{-\frac{\lvert\kappa\rvert}{\Lambda}}\,,
\end{align}
Let us examine the sign of the generic expression
\begin{equation}
	\frac{1}{\kappa}\operatorname{Im}\frac{1}{\omega^{2}-\kappa^{2}-i\,2\gamma\kappa\pm\dfrac{2\gamma}{\ell}\,e^{i\,\kappa\ell}}\,,
\end{equation}
for all $\kappa$ because it will determine the sign of the heat capacity. Explicitly it takes the form
\begin{align}\label{E:sibfsers}
	\frac{2\gamma\ell^{2}\Bigl[1\mp\dfrac{\sin\kappa\ell}{\kappa\ell}\Bigr]}{\Bigl[(\kappa^{2}-\omega^{2})\ell\mp2\gamma\,\cos\kappa\ell\Bigr]^{2}+4\gamma^{2}\Bigl[\kappa\ell\mp\sin\kappa\ell\Bigr]^{2}}\,,
\end{align}
which is obviously positive for all $\kappa$. Note that the integrand in \eqref{E:dkbdkjse} is nowhere negative and the integral is well defined. Thus the heat capacity $C(\infty)$ in \eqref{E:dkbdkjse} is always positive for all nonzero temperatures even with the presence of bath-induced non-Markovian effects between the oscillators.  We observe that in the low temperature limit $\beta\to\infty$, the factor
\begin{align}\label{E:dfnksdf}
	\left(\frac{\dfrac{\beta\kappa}{2}}{\sinh\dfrac{\beta\kappa}{2}}\right)^{2}
\end{align}
goes to zero. Thus it implies that the heat capacity must vanish at zero temperature since the integral in \eqref{E:dkbdkjse} is regular. This argument may be too simplistic. In fact, when $\beta\to\infty$, the major contributions to the integral in \eqref{E:dkbdkjse} comes from the interval $\lvert\kappa\rvert<\mathcal{O}(\beta^{-1})$, within which the rest of the integrand is slowly varying. Thus we may pull the slowly changing component out of the integral and write \eqref{E:dkbdkjse} as
\begin{align}\label{E:gfkgbrsda}
	\lim_{\beta\to\infty}C(\infty)\simeq\frac{4\gamma\omega_{+}^{2}}{(\omega_{+}^{2}-\dfrac{2\gamma}{\ell})^{2}}\int_{-\beta^{-1}}^{\beta^{-1}}\!\frac{d\kappa}{2\pi}\;\left(\frac{\dfrac{\beta\kappa}{2}}{\sinh\dfrac{\beta\kappa}{2}}\right)^{2}=K\,\beta^{-1}\,,
\end{align}
where $K$ is a finite positive constant independent of $\beta$. Thus indeed the heat capacity vanishes algebraically fast as the temperature approaches zero.  Most important of all, the heat capacity vanishes for all permissible choices of oscillator separation, inter-oscillator coupling, and oscillator-bath interaction strength.

The analytical expression of the heat capacity for the two-oscillator system cannot be given without resort to approximation due to the retarded non-Markovian effect. However its low-temperature behavior is expected to be more complicated than its one oscillator counterpart because of additional scale $\ell$ dependence. A numerical example is given in Fig.~\ref{Fi:twosc}. In particular in Fig.~\ref{Fi:twosc}-(c), we compare the temperature dependence of heat capacity of this system between the conventional thermodynamics and the present nonequilibrium approaches.

\begin{figure}
\centering
    \scalebox{0.6}{\includegraphics{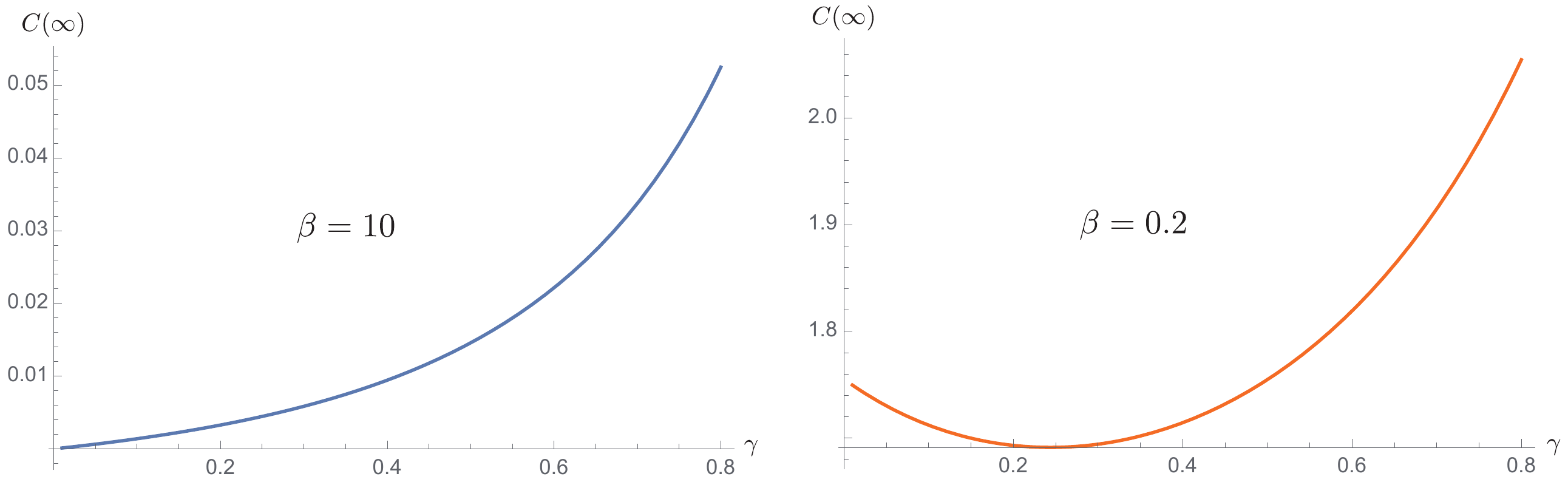}}
    \caption{Non-Markovian effect of heat capacity of system at short inter-oscillator separations. The non-monotonic behavior of the heat capacity with respect to the damping constant is most pronounced in the high-temperature limit $\beta\to0$. The plots are drawn with the choices of $\omega_{p}=5$, $\sigma=10$, $\ell=0.08$ and $\Lambda^{-1}=0.01$. Note that the heat capacity in the right plot may take values greater than 2. This is related to strong fluctuations of the bath at short distance and will be discussed later.}\label{Fi:non-Mak1}
\end{figure}

An interesting feature appears when the interaction strength between the oscillators and the bath is sufficiently strong. The heat capacity in the two-oscillator system does not increase monotonically as a function of $\gamma$ as is the case in the one-oscillator system. In fact, as shown in Fig.~\ref{Fi:non-Mak1}, when  the damping constant is greater than a critical value $\gamma_{c}$, the heat capacity will increase monotonically afterwards with increasing $\gamma$. This non-monotonic behavior is most easily seen at high temperatures. In this limit the critical value of $\gamma_{c}$ is approximately given by $\sigma\ell/4$ but slightly falls off with lower temperature. This explains why the non-monotonic behavior of the heat capacity is not seen at low temperatures.

The non-monotonicity in the heat capacity results from the presence of the term $\frac{2\gamma}{\ell}\,e^{i\,\kappa\ell}$ in \eqref{E:dkbdkjse}, which in turn corresponds to the retarded terms in the equations of motion \eqref{E:dfkdsdaq} and \eqref{E:dfkdsdaq1}. That is, the non-monotonicity in the heat capacity is due to the strong nonlocal effects mediated by the bath. The critical value $\gamma_{c}=\sigma\ell/4=\gamma\varsigma/2$ has a special physical significance, namely, it indicates the relative significance between the direct coupling and the indirect bath-induced causal influence among the oscillators.

\begin{figure}
\centering
    \scalebox{0.6}{\includegraphics{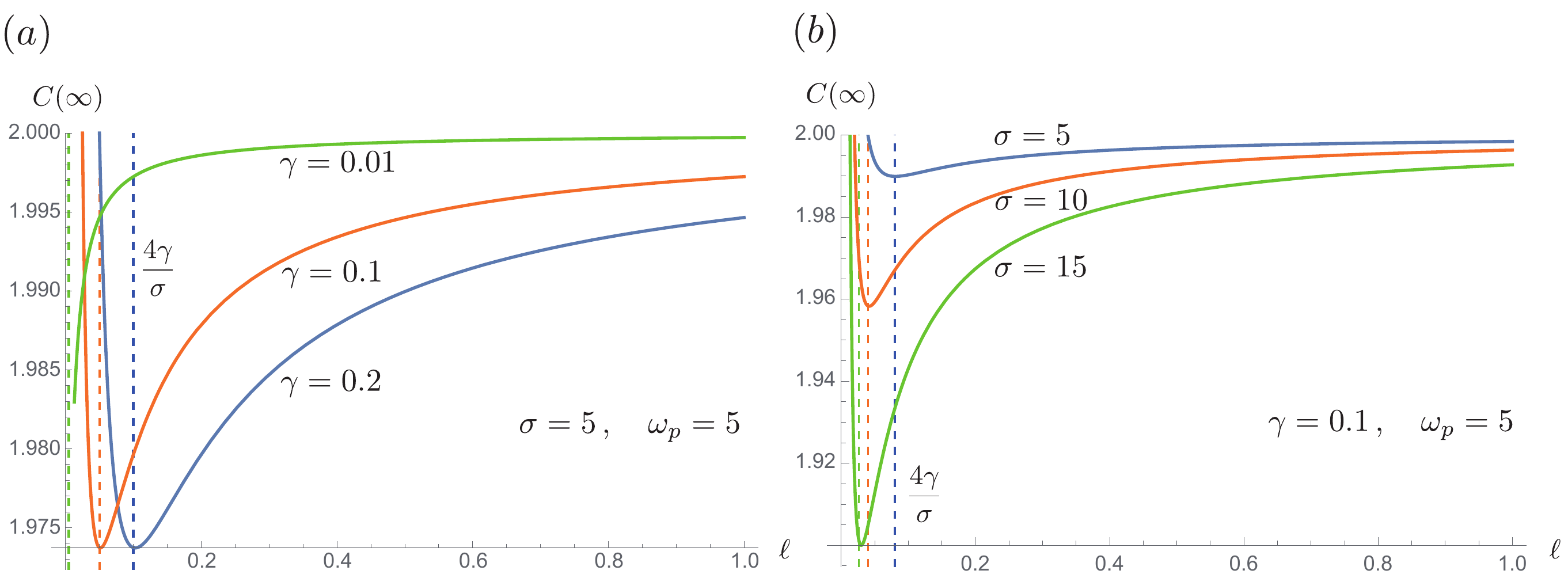}}
    \caption{Here we show that at high temperature $T\to\infty$, how the heat capacity at the final equilibrium state $C(\infty)$ varies with the oscillator separation $\ell$. In (a) we fix the inter-oscillator coupling $\sigma$ and in (b) the oscillator-bath interaction strength $\gamma$ is fixed. Generically we see the heat capacity is not monotonically decreasing when we shorten the separation. It is related to the large bath fluctuations at short distance.}\label{Fi:nonMark2}
\end{figure}

When $\gamma>\gamma_{c}$, i.e., $\varsigma<\mathcal{O}(1)$, the bath-induced effect dominates, whereas when $\gamma<\gamma_{c}$, i.e., $\varsigma>\mathcal{O}(1)$, the direct inter-oscillator coupling wins over. This can be further seen from Fig.~\ref{Fi:nonMark2}. In Fig.~\ref{Fi:nonMark2}--(a) the curves of heat capacity at high temperature $\beta\to0$ are plotted for different choices of damping constants $\gamma$ with respect to the oscillator separation $\ell$, while in (b) they are drawn for various selections of direct coupling strengths $\sigma$. They all show a monotonic behavior once $\ell$ is smaller than the critical separation $\ell_{c}=4\gamma/\sigma$. Similar behaviors also appear in thermal entanglement~\cite{HHPLB15}.

Mathematically it is not difficult to understand why this non-monotonicity in heat capacity is readily seen at high temperatures. From \eqref{E:dkbdkjse}, we immediately see that
\begin{equation}
	\lim_{\beta\to0}\left(\frac{\dfrac{\beta\kappa}{2}}{\sinh\dfrac{\beta\kappa}{2}}\right)^{2}=1\,,
\end{equation}
and since the remaining terms in the integrand does not depend on temperature $\beta^{-1}$, it implies that the temperature will not have any effect on the contributions from the causal influence $\frac{2\gamma}{\ell}\,e^{i\,\kappa\ell}$. Physically the rise of heat capacity with increasing $\gamma$ at short inter-oscillator separation  can be understood as a consequence of increasing thermal fluctuations of the scalar-field bath, since heat capacity is a measure of the variance of internal energy at least in conventional equilibrium thermodynamics~\footnote{In fact it is used as a criterion for the validity of canonical ensemble and the thermodynamic stability of the system.}

\begin{figure}
\centering
    \scalebox{0.62}{(a)\includegraphics{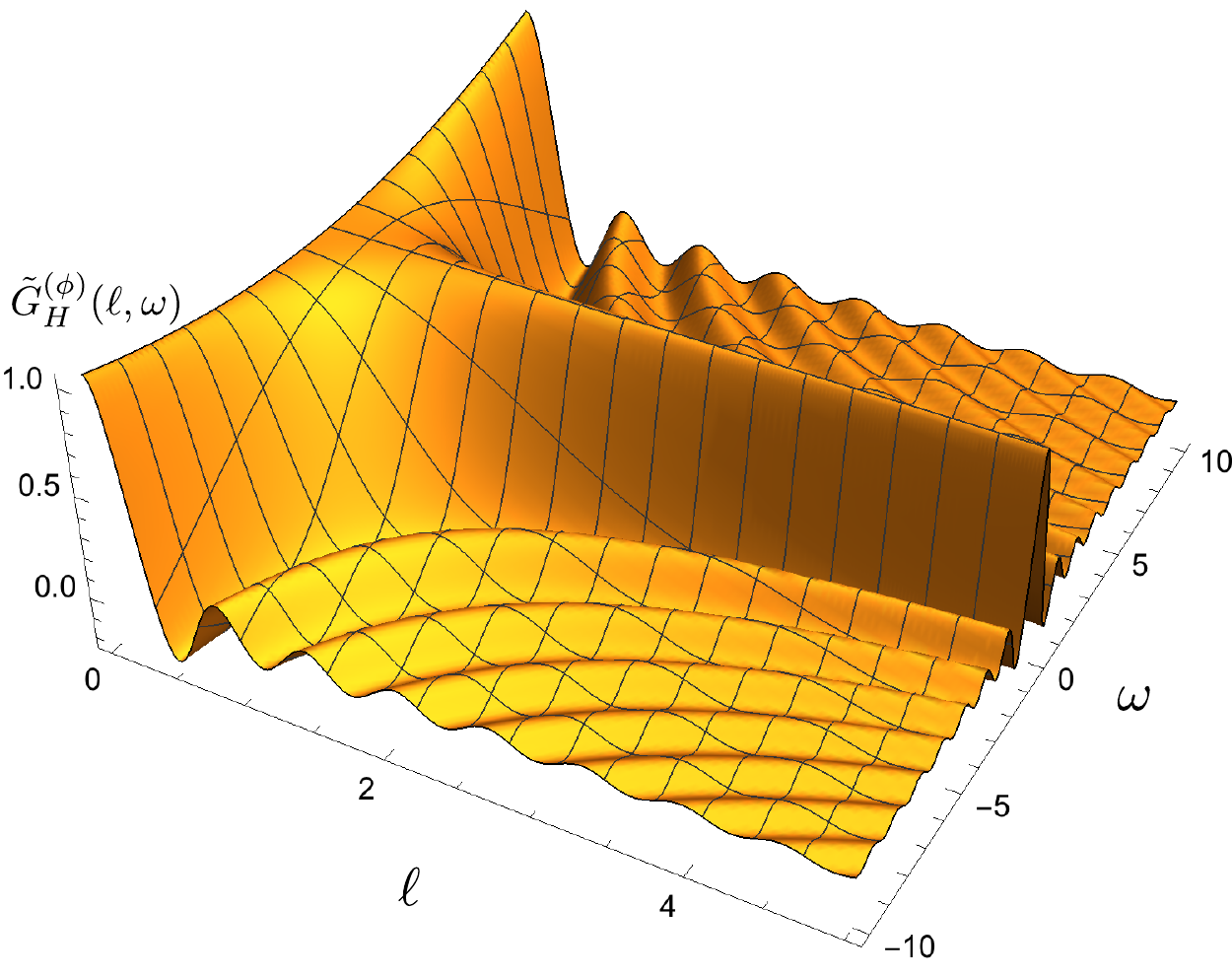}(b)\includegraphics{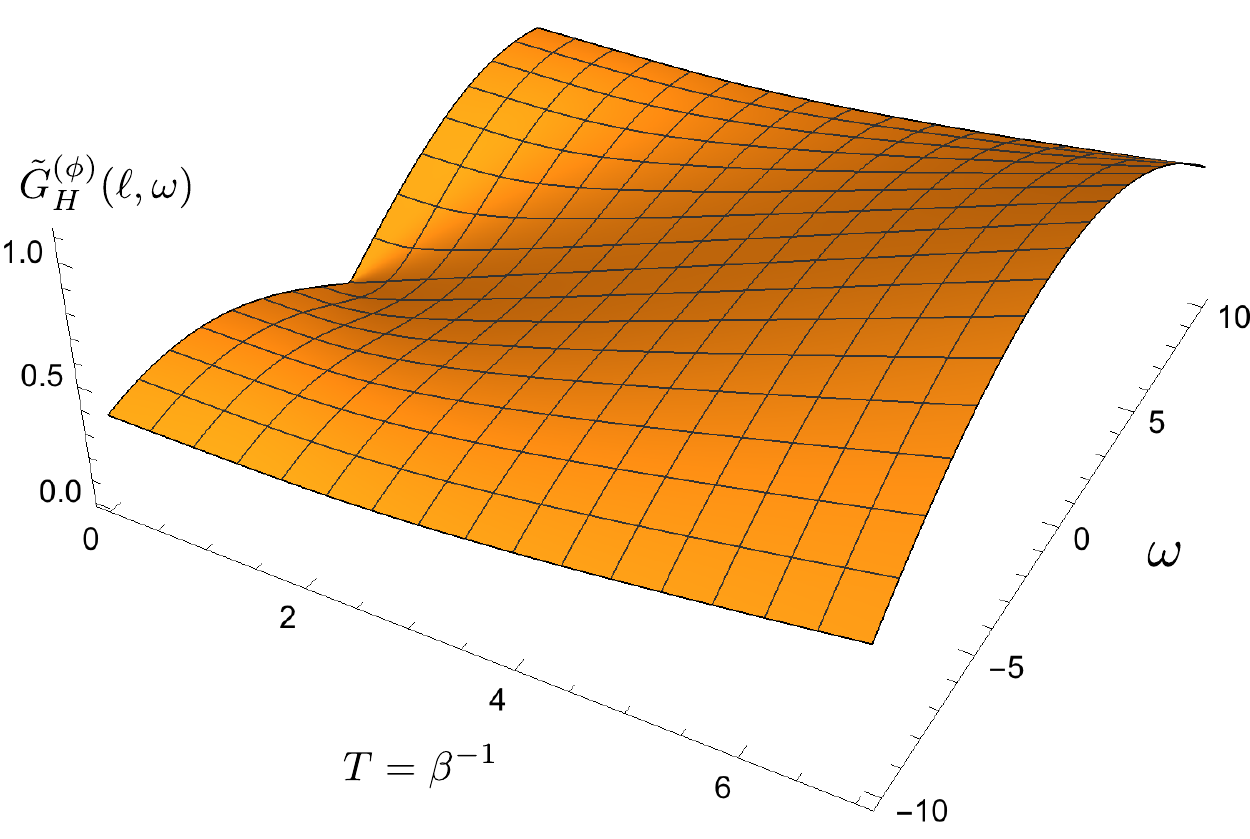}}
    \caption{To illustrate the effects of the large bath fluctuations, we draw the power spectrum of the scalar field bath with respect to oscillator separation $\ell$ in (a) and temperature $T$ in (b). Following the gridlines, we observe the intensity of the power spectrum is greater for shorter separation and hotter temperature.}\label{Fi:spectra}
\end{figure}

After the system equilibrates its dynamics is governed by the noise force, thus large quantum or thermal fluctuations from the bath can in principle induce large fluctuations in the energy of the system, which shows up as large heat capacity. To be specific, the Hadamard function of the scalar field \eqref{E:ejfdhwe1} depicts the correlation of the noise forces in \eqref{E:dfkdsdaq} and \eqref{E:dfkdsdaq1}, and Fig.~\ref{Fi:spectra} shows the corresponding power spectrum, which is the temporal Fourier transform of the Hadamard function. In general the spectrum of the scalar-field bath shows larger values for shorter inter-oscillator separation $\ell$ and higher bath temperature $\beta^{-1}$. They reflect the mere facts that when we probe into a smaller spatial region, we see larger quantum field fluctuations (simply seen on account of the uncertainty principle), and at higher temperatures thermal fluctuations become more pronounced. These large fluctuations will then produce large variances in the variables of the system via, for example, \eqref{E:uirysbf}.

This non-monotonic behavior of heat capacity at short distance may not be easy to observe because the transition occurs on a scale smaller than $\ell_{c}$ at moderate temperatures. This scale may be shorter than the minimal separation at which the motion of the system remains stable or fall below the physical size of realistic systems modeled by oscillators.

\section{Difference between ONEq and the CGTs Setups}\label{S:fkbser}
Here we use a simple calculation based on the ubiquitous quantum Brownian motion to illustrate the difference between ONEq and the CGTs setups.

Consider a system of an harmonic oscillator bilinearly coupled to an environment consisting of harmonic oscillators, also referred to as the quantum Brownian motion (QBM) model. The Hamiltonian of the combined system is:
\begin{align}
    H_C &= H_S + H_E + H_I\notag \\
      &= \biggl(\frac{P^2}{2M_{0}} + \frac{1}{2}M_{0}\Omega^2 Q^2\biggr) + 
      \biggl(\sum_n \frac{p_n^2}{2m_n}+\frac{1}{2}m_n \omega_n^2 q_n^2 \biggr) + 
      \sum_n g_n q_n Q\,,  
\end{align}
where the variables in capital are associated with the system, and $M_{0}$, $\Omega$ are the mass and the oscillating frequency of the system oscillator. The parameter $g_{n}$ is the coupling strength between each bath mode $q_{n}$ and the system oscillator $Q$, while $p_{n}$ and $P$ are their respective conjugate momenta. Such a combined system is sufficiently simple to serve our current purpose. At some initial time $t_i$ the system and environment are assumed to be uncorrelated. 
Moreover, the environment is in the thermal state with respect to its isolated 
Hamiltonian $H_E$:
\begin{align}\label{E:dkbersfgs}
    \rho(t_i) &= \rho_S(t_i) \otimes \frac{e^{-\beta H_E}}{Z_E}\,.
\end{align}
In the above equation $Z_E = \text{Tr}_E\left\{ e^{-\beta H_E} \right\}$. 
If the environment is very large and the system-environment coupling is such 
that the system relaxes to a unique steady (or equilibrium) state irrespective of the initial 
state, we call the environment a thermal reservoir or a heat bath. 
Under these assumptions of equilibration it has been shown that \cite{SFTH} the 
unique steady (or equilibrium) state $\rho_S^{eq}$ has the form:
\begin{align}\label{eq:eq}
    \rho_S^{eq} &\equiv\lim_{t\rightarrow \infty}\rho_S(t) = \operatorname{Tr}_E
  \left [ \frac{e^{-\beta\left( H_S + H_E + H_I \right)}}{Z_C} \right]\,,
\end{align}
where $Z_C$ is defined analogous to $Z_E$ above, and $\rho_{S}(t)$ is the reduced state of the system, evolved out of the initially uncorrelated global state \eqref{E:dkbersfgs}. This state is referred to as the equilibrium state to distinguish it from the thermal state given by the Boltzmann-Gibbs distribution with respect to the  system Hamiltonian alone.

It is important to note that although the system relaxes to the equilibrium 
state, the global state of the system+environment is not in equilibrium. 
In particular,
\begin{align}\label{E:erbdf}
    \lim_{t\rightarrow \infty}\rho_C(t) \ne \frac{e^{-\beta(H_S+H_E+H_I)}}{Z_C} \, .
\end{align}
The density matrix $\rho_{C}$ here is the state of the combined system and environment, evolved from the initial state \eqref{E:dkbersfgs}; thus $\rho_{S}(t)=\operatorname{Tr}_{E}\rho_{C}(t)$. Eq.~\eqref{E:erbdf} is true because the thermal state of the combined system+environment is a  stationary state of the full Hamiltonian and thus cannot be reached from a  nonstationary state under Hamiltonian dynamics.  What \eqref{eq:eq} says is that the reduced system state is consistent with a global thermal state and that no further information about the nonequilibrium state of the combined system+bath can be obtained if one has access to the system only. In other words the information on the non-equilibrium state of the combined system+environment is not stored in the system but rather in the environment and in the correlations between the system and the environment.

To demonstrate this point, we focus our attention to 
a single bath mode. 
The solutions to the equation of motion are:
\begin{align}
    q_k(t) &= q_k(t_i) \cos[\omega_k(t-t_i)] + \frac{p_k(t_i)}{m_k \omega_k}\, \sin[\omega_k(t-t_i)] + \int_{t_i}^t ds\; \frac{\sin[\omega_k(t-s)]}{m_k \omega_k}\, g_k\, Q(s)\,, \\
    Q(t) &= Q(t_i)\, M_{0} \dot{G}(t-t_i) + P(t_i)\, G(t-t_i) + \int_{t_i}^t ds\; G(t-s)\,\xi(s)\,, \\
    \xi(s) &= \sum_n g_n \left\{ q_n(t_i) \cos[\omega_n(t-t_i)] + p_n(t_i) \,\frac{\sin[\omega_n(t-t_i)]}{m_n \omega_n} \right\}\,.
\end{align} 
Here $G(t)$ is the Green's function of the system oscillator, the exact form of which is not important for this discussion (see Ref. \cite{SFTH} for details). Under the assumptions of equilibration $G(t) \rightarrow 0$ as $t\rightarrow \infty$. This is the reason why the system cannot keep any of the memory of its initial state. However, since the closed system dynamics is unitary, that information is not lost. Rather, it is distributed over the bath modes. The first two terms in the expression for $q_k(t)$ do keep the memory of the initial thermal state of the bath. The last term is the only place where the memory of the initial state of the system survives in the bath. 
\begin{align}
    \int_{t_i}^t ds\; \frac{\sin(\omega_k(t-s))}{m_k \omega_k}\, &g_k \left[ Q(t_i) M_{0} \dot{G}(s-t_i) + P(t_i) G(s-t_i) \right] \\
    \propto M_{0}Q(t_i)& \int_{t_i}^t ds\; \sin[\omega_k(t-s)]  \dot{G}(s-t_i) + P(t_i) \int_{t_i}^t ds \sin[\omega_k(t-s)] G(s-t_i)\,. \notag
\end{align}
The important observation is that although $G(t)$ and $\dot{G}(t)$ are decaying functions, the above integrals are oscillatory. (This can be seen explicitly by choosing $G(t) \propto e^{-\gamma t}$ (Ohmic case) and doing the integrals explicitly.) As a result, the information on the initial state of the system survives forever in the bath degrees of freedom. Similarly, information on initial state also survives in the correlations between the system and bath modes as can be seen by studying the late-time limit of $\langle Q(t) q_k(t) \rangle$. However, the coupling to each mode is extremely small under the assumption of equilibration, and this makes it practically impossible to extract this information in reality.

Under the assumption of equilibration, the reduced system relaxes to the equilibrium state at late times, irrespective of the initial state of the system. As a result any thermodynamic formulation that only relies on the reduced state of the system for its definitions, will be independent of the initial state of the system. Moreover, all the information about the reduced state of the system can be obtained by assuming a global thermal state without introducing any errors.

In the weak coupling limit, $|H_I|\rightarrow 0$, the equilibrium state approaches the thermal state.  But in general, the equilibrium state differs from the thermal state. It is common to define the so called ``Hamiltonian of mean force'' to quantify this difference. In essence, Hamiltonian of mean force is the operator with respect to which the equilibrium state has the Boltzmann-Gibbs form, where the temperature is that of the bath.

In conclusion, a) for strong coupling between the system and bath, if the system can  approach the equilibrium state, then the reduced density matrix of the open system is the same as the reduced density matrix in the CGTs framework upon integrating out the bath. b) However, the final global states are different, despite the fact in both cases the dynamics is generated by the same Hamiltonian ($H_C=H_S+H_I+H_B$). This is because two global systems start out with different initial states, and the unitary evolution does not change the distinguishability between the states. c) Note that in the CGTs set up, even though the closed system is assumed to be in a global thermal state, the system is not necessarily in a thermal state.  One needs extra assumptions, such as the system is very weakly coupled to the bath. It is under the same condition that  the open system can reach thermalization. One should be mindful of these presumptions when comparing the  thermal states in both setups. We mention these facts as a cautionary note to remind ourselves and for those who want to compare quantities calculated in these three different setups.

\section{Difference between Equilibrium States and Thermal States in Classical Thermodynamics}\label{S:Cthrt}

To help understand the important differences between equilibrium states and thermal states, we give a short yet general derivation below where one can see how their differences can be quantified.

Consider a classical system ($S$) described with position and momentum variables $x=(r,p)$ in the phase space. The system might consist of many particles in arbitrarily many dimensions. In that case $x$, $r$, $p$ are vectors. The thermal state of this system is described by the following probability distribution on the phase space 
\begin{equation}
	p^{th}_S(x) = \frac{e^{-\beta H_S(x)}}{Z_S}\,.
\end{equation}
where $H_S(x)$ is the system Hamiltonian. Now assume the system is coupled to an environment whose position and momenta we denote by $y=(r',p')$. The Hamiltonian of the environment is $H_E(y)$ and the interaction Hamiltonian is $H_I(x,y)$. Consider the thermal state of the combined system plus environment. The probability distribution on the combined phase space is given by:
\begin{equation}
	p^{th}_{S+E}(x,y) = \frac{e^{-\beta[H_S(x) + H_I(x,y) + H_E(y)]}}{Z_{S+E}}\,.
\end{equation}
If we are only interested in the state of the system ($S$), we integrate this distribution over the environment variables $y$. The result is the \textit{equilibrium} distribution for the system alone: 
\begin{equation}
	p^{eq}_{S}(x) = e^{-\beta H_S(x)}\;\frac{\displaystyle\int dy\;e^{-\beta (H_I(x,y) +H_E(y))}}{Z_{S+E}}\,.
\end{equation}
It is conventional to lump the second term on the righthand side
into the exponent by defining the \textit{potential of mean force} as
\begin{equation}
	H^*(x) = H_{S}(x) - \beta^{-1} \ln\,\frac{\displaystyle\int dy\;\exp\Bigl\{-\beta \bigl[H_I(x,y)+H_E(y)\bigr]\Bigr\}}{\displaystyle\int dy\;\exp\Bigl\{-\beta H_E(y)\Bigr\}}\,.
\end{equation}
where upon the probability density of the system becomes 
\begin{align*}
	p_S^{eq}(x) &=\frac{e^{-\beta H^*(x)}}{Z^*}\,,&Z^{*}&=\int\!dx\;e^{-\beta H^*(x)}\,.
\end{align*}
It is clear that $H(x)\ne H^*(x)$ in general. As a result, the thermal state differs from the equilibrium state classically as well as quantum mechanically.

A more rigorous way to quantify the difference between equilibrium and thermal distributions for classical systems is to use the Kullback--Leibler divergence which is a measure of distinguishability for probability distributions. Its quantum analog is the quantum relative entropy.

\section{Causality Issues with the $M$-oscillator Bath}\label{S:rbjdhe}

Here we add some cautionary comments on the differences between an oscillator bath which has been used extensively in Brownian motion studies,  e.g., \cite{FeyVer,CalLeg,GSI88,HPZ} and the scalar-field bath, which also has been used by many, e.g, \cite{LebSpo77,UnrZur89,LinHu09,HHPRD}.  We show, with the assumption of two or more system oscillators placed in the same spatial location, e.g., in \cite{CHY,PazRec} which is often assumed in the $M$-oscillator bath situation, there is a causality issue. No such issues exist with system oscillators in spatially separate locations in a scalar-field bath.

Recall how one proceeds from a harmonic oscillator representation to field theory~\cite{HM94}. Consider the plane-wave expansion of a massless scalar quantum  field 
\begin{equation}\label{E:bhskjdf}
	\phi(x)=\frac{1}{\sqrt{V}}\sum_{\mathbf{k}}\frac{1}{\sqrt{2\omega}}\Bigl [a_{\mathbf{k}}(t)\,e^{i\,\mathbf{k}\cdot\mathbf{x}}+\text{H.c.}\Bigr]\,.
\end{equation}
The creation and annihilation operators $a^{\dagger}_{\mathbf{k}}(t)$, $a^{\vphantom{\dagger}}_{\mathbf{k}}(t)$ satisfy an equation of motion similar to that of the ladder operators for the quantum harmonic oscillators
\begin{equation}\label{E:bhrssf}
	\ddot{a}_{\mathbf{k}}(t)+\omega^{2}a_{\mathbf{k}}(t)=0\,,
\end{equation}
with $\omega=\lvert\mathbf{k}\rvert$. In particular when $\mathbf{x}=0$, Eqs.~\eqref{E:bhskjdf} and \eqref{E:bhrssf} together represent a collection of quantum harmonic oscillators. The Hamiltonian of the massless scalar field in the plane-wave expansion \eqref{E:bhskjdf}
\begin{align}
	H=\int\!d^{3}\mathbf{x}\;\biggl\{\frac{1}{2}\,\bigl(\partial_{t}\phi\bigr)^{2}+\frac{1}{2}\,\bigl(\pmb{\nabla}\phi\bigr)^{2}\Bigr\}&=\sum_{\mathbf{k}}\Bigl [a_{\mathbf{k}}^{\dagger}a_{\mathbf{k}}^{\vphantom{\dagger}}+\frac{1}{2}\Bigr]\,\omega\,,
\end{align}
is to be compared to the Hamiltonian of a collection of quantum harmonic oscillators,  
\begin{equation}
	H_{\textsc{ho}}=\sum_{i=1}^{M}\frac{m}{2}\,\dot{\chi}_{i}^{2}+\frac{m\omega_{i}}{2}\,\chi_{i}^{2}=\sum_{i}\Bigl[a_{i}^{\dagger}a_{i}+\frac{1}{2}\Bigr]\,\omega_{i}\,,
\end{equation}
where $m$ is the mass and $\chi_{i}$ is the displacement of $i^{\text{th}}$ oscillator, whose natural frequency is given by $\omega_{i}=i\,\Delta$. The parameter $\Delta$ is some fundamental mode frequency. The ladder operators $a^{\vphantom{\dagger}}_{i}$, $a^{\dagger}_{i}$ in the latter case are defined by
\begin{equation}
	a_{i}=\sqrt{\frac{m\omega_{i}}{2}}\,\Bigl(\chi_{i}+\frac{1}{m\omega_{i}}\,p_{i}\Bigr)\,,
\end{equation}
where $p_{i}$ is the canonical momentum conjugate to $\chi_{i}$. For this and other reasons most people would not make a distinction between two bath models.  For example, this is so if one does not ask where each of the system oscillators is located. If they are stacked up at one particular spatial location or when the dipole approximation $e^{i\,\mathbf{k}\cdot\mathbf{x}}\simeq1$ is applicable, then there is no difference whether the bath is described by a bunch of oscillators or a massless quantum scalar field,  since one can arbitrarily shift the location of the system such that $\mathbf{x}=\mathbf{z}_{i}=0$. However, there are still fundamental differences between these two models.

In quantum field theory, the creation and annihilation operators $a_{\mathbf{k}}^{\vphantom{\dagger}}$, $a_{\mathbf{k}}^{\dagger}$ depend on the mode functions we use to expand the quantum field. In the plane wave expansion case, the mode function takes a simple form $e^{i\,\mathbf{k}\cdot\mathbf{x}}$ so the vacuum annihilated by the corresponding $a_{\mathbf{k}}^{\vphantom{\dagger}}$ and the associated number states have a definite three-momentum $\mathbf{k}$. Thus, even merely using the uncertainty principle argument, one can see that these quanta are highly nonlocalized, distributed over the whole configuration space. By contrast for harmonic oscillators, the ground state and the excited states of each bath oscillator are essentially confined by the corresponding harmonic potential $m\omega^{i}\chi_{i}^{2}/2$. Thus the higher the value of the natural frequency $\omega_{i}$ is, the more localized the corresponding mode.

The difference between a field and a collection of harmonic oscillators shows up, for example, when there is a spatial boundary present. The boundary will modify the mode functions of the field and alter the two-point functions of the bath to recognize the effects of the boundary.  It is not obvious how this boundary-induced spatial dependence can be naturally implemented in the bath-oscillator model.

Additional complexity emerges when the system contains more than one oscillator which are spatially separated from one another, since the $M$-oscillator bath model does not have the dynamical degrees of freedom to register the locations of the spatially-separated constituent oscillators in the system unless extra input of the spatial information of the system constituents and how it enters in their interaction with the bath is provided. More often than not, one simply ignores the spatial distances between the system constituents by assuming that this separation is so small that we can essentially view them as being situated at the same spatial location or by taking the dipole approximation. This creates a causality problem. For relativistic quantum fields, the influence of an object at one spacetime point on another object at another spacetime point is affected by the former's imprint on the field, which propagates causally to the latter when it begins to exert its influence. The dynamics of the latter will in turn trigger a new disturbance, on top of the previous one it received, in the field, which propagates at finite time and exerts its influence on the motion of all the other components. This field-induced interaction among the spatially separated components of the system depends on the history of its constituents and is thus fundamentally non-Markovian in nature (see, the plots in e.g., \cite{LinHu09}). Notably this causal propagation feature is mostly lacking in the oscillator-bath mode if the spatial information of the system constituents is not properly accounted for~\footnote{There may exist a nonlocal dynamics due to the appropriate choices of the system-bath coupling constants, but the motion is at best temporally nonlocal, unlike the field-bath where the dynamics is in general spatio-temporally nonlocal.}. The field-bath model being relativistic naturally incorporates the spatial correlation and respects the causality.


\begin{thebibliography}{99}

\bibitem{Mahler} J. Gemmer, M. Michel and G. Mahler, ``Quantum Thermodynamics - Emergence of Thermodynamic Behavior within Composite Quantum Systems'', 2nd Edition (Springer Verlag, Berlin, 2004).

\bibitem{ktln2} See, e.g., the conference series ktlog2: \url{http://seneca.fis.ucm.es/parr/ktlog2_12/Home.html}  

\bibitem{NEqTDbooks} E.g., S. R. de Groot and P. Mazur, ``Non-Equilibrium Thermodynamics'', (North Holland Publishing Co, 1962) 

\bibitem{Kirkwood} J. G. Kirkwood, ``Statistical mechanics of fluid mixtures'', J. Chem. Phys. \textbf{3}, 300 (1935).

\bibitem{GE16} C. Gogolin and J. Eisert, ``Equilibration, thermalization, and the emergence of statistical mechanics in closed quantum systems'', Rep. Prog. Phys. \textbf{79}, 056001 (2016).

\bibitem{stochTD} E.g., K. Sekimoto, ``Langevin equation and thermodynamics'', Prog. Theor. Phys. Suppl. \textbf{130}, 17 (1998); U. Seifert, ``Stochastic thermodynamics, principles and perspectives'', Eur. Phys. J. B, \textbf{64}, 423 (2008); ``Stochastic thermodynamics, fluctuation theorems and molecular machines'', Rep. Prog. Phys, \textbf{75}, 126001 (2012);  C. G. Bulnes, A. Engel and M. Esposito, ``Stochastic thermodynamics of rapidly driven quantum systems'', New J. Phys. \textbf{17}, 055002 (2015); C. Van den Broeck, S.-I. Sasa and U. Seifert (eds), Focus issue on stochastic thermodynamics, New J. Phys. \textbf{18}, 020401 (2016).


\bibitem{FeyVer} R. P. Feynman and F. L. Vernon, ``The theory of a general quantum system interacting with a linear dissipative system'', Ann. Phys. (N.Y.) \textbf{24}, 118 (1963).

\bibitem{CalLeg} A. O. Caldeira and A. J. Leggett, ``Path integral approach to quantum Brownian motion'', Physica (Amsterdam) \textbf{121A}, 587 (1983).

\bibitem{HPZ} B. L. Hu, J. P. Paz, and Y. Zhang, ``Quantum Brownian motion in a general environment: exact master equation with nonlocal dissipation and colored noise'', Phys. Rev. D \textbf{45}, 2843 (1992); ``Quantum Brownian motion in a general environment. II. nonlinear coupling and perturbative approach'', \textbf{47}, 1576 (1993).

\bibitem{Kosloff} R. Kosloff, ``Quantum thermodynamics: a dynamical viewpoint'', Entropy \textbf{15}, 2100 (2013). 

\bibitem{GSI88} H. Grabert, P. Schramm, G.L. Ingold, ``Quantum Brownian motion: the functional integral approach'', Phys. Rep. \textbf{168}, 115 (1988).

\bibitem{GelTho} M. F. Gelin and M. Thoss, ``Thermodynamics of a subensemble of a canonical ensemble'', Phys. Rev. E \textbf{79}, 051121 (2009).

\bibitem{KimMah10} I. Kim and G. Mahler, ``Clausius inequality beyond the weak-coupling limit: the quantum Brownian oscillator'', Phys. Rev. E \textbf{81}, 011101 (2010).

\bibitem{HTL11} S. Hilt, B. Thomas, and E. Lutz, ``Hamiltonian of mean force for damped quantum systems'', Phys. Rev. E \textbf{84}, 031110 (2011).

\bibitem{Seifert16} U. Seifert, ``First and second law of thermodynamics at strong coupling'', Phys. Rev. Lett. \textbf{116}, 020601 (2016).

\bibitem{PhiAnd} T. G. Philbin and J. Anders, ``Thermal energies of classical and quantum damped oscillators coupled to reservoirs", J. Phys. A \textbf{49}, 215303 (2016).

\bibitem{HIT08} P. H\"anggi, G.-L. Ingold and P. Talkner, ``Finite quantum dissipation: the challenge of obtaining specific heat'', New J. Phys. \textbf{10}, 115008 (2008).

\bibitem{SIW15} B. Spreng, G.-L. Ingold and U. Weiss, ``Anomalies in the specific heat of a free damped particle: the role of the cutoff in the spectral density of the coupling'',  Phys. Scr. T \textbf{165}, 014028 (2015).

\bibitem{Deutsch91} J. M. Deutsch, ``Quantum statistical mechanics in a closed system'', Phys. Rev. A \textbf{43}, 2046 (1991).

\bibitem{Srednicki94} M. Srednicki, ``Chaos and quantum thermalization'', Phys. Rev. E \textbf{50}, 888 (1994).

\bibitem{GLTZ06} S. Goldstein, J. L. Lebowitz, R. Tumulka and N. Zangh￬, ``Canonical Typicality'', Phys. Rev. Lett. \textbf{96}, 050403 (2006).

\bibitem{PSW06} S. Popescu, A. J. Short and A. Winter, ``Entanglement and the foundations of statistical mechanics'', Nat. Phys. \textbf{2}, 754 (2006).

\bibitem{LPSW09} N. Linden, S. Popescu, A. J. Short and A. Winter, ``Quantum mechanical evolution towards thermal equilibrium'', Phys. Rev. E \textbf{79}, 061103 (2009).

\bibitem{SF12} A. J. Short and T. C. Farrelly, ``Quantum equilibration in finite time'', New J. Phys. \textbf{14}, 013063 (2012).
 
\bibitem{Reimann08} P. Reimann, ``Foundation of statistical mechanics under experimentally realistic conditions'', Phys. Rev. Lett. \textbf{101}, 190403 (2008).
 
\bibitem{Rigol09} M. Rigol, ``Breakdown of thermalization in finite one-dimensional systems'', Phys. Rev. Lett. \textbf{103}, 100403 (2009).

\bibitem{PSSV11} A. Polkovnikov, K. Sengupta, A. Silva and M. Vengalattore, ``Nonequilibrium dynamics of closed interacting quantum systems'', Rev. Mod. Phys. \textbf{83}, 863 (2011).
 
\bibitem{CR10} M. A. Cazalilla and M. Rigol, `` Focus on dynamics and thermalization in isolated quantum many-body systems'', New J. Phys. \textbf{12}, 055006 (2010).

\bibitem{2LawSaga} E.g., A. E. Allahverdyan and T. M. Nieuwenhuizen, ``Extraction of work from a single thermal bath in the quantum regime'', Phys. Rev. Lett. \textbf{85}, 1799 (2000);  G. W. Ford and R. F. O'Connell, ``A quantum violation of the second law?'', Phys. Rev. Lett. \textbf{96}, 020402 (2006);  S. Deffner and E. Lutz, ``Generalized Clausius inequality for nonequilibrium quantum processes'', Phys. Rev. Lett. \textbf{105}, 170402 (2010).

\bibitem{GTOS}
J.-T. Hsiang and B. L. Hu, ``Thermodynamics of quantum systems strongly coupled to a heat bath I. Operator thermodynamic functions and relations'', arXiv:1710.03882.

\bibitem{DuaCal} O. S. Duarte and A. O. Caldeira, ``Effective quantum dynamics of two Brownian particles'', Phys. Rev. A \textbf{80}, 032110 (2009).

\bibitem{CarrWeiss15} M. Carrega, P. Solinas, A. Braggio, M. Sassetti and U. Weiss, ``Functional integral approach to time-dependent heat exchange in open quantum systems: general method and applications'' New J. Phys. \textbf{17}, 045030 (2015). 

\bibitem{EspositoPRL15} M. Esposito, M. A. Ochoa and M. Galperin, ``Quantum thermodynamics: a nonequilibrium Green's function approach'', Phys. Rev. Lett. \textbf{114}, 080602 (2015).

\bibitem{HasegawaJMP} H. Hasegawa, ``Specific heat anomalies of small quantum systems subjected to finite baths'', J. Math. Phys. \textbf{52}, 123301 (2011).

\bibitem{HanIng06} P. H\"anggi and G.-L. Ingold, ``Quantum Brownian motion and the third law of thermodynamics'', Acta. Phys. Pol. B \textbf{37}, 1537 (2006). 

\bibitem{IHT09} G.-L. Ingold, P. H\"anggi and P. Talkner, ``Specific heat anomalies of open quantum systems'', Phys. Rev. E \textbf{79}, 061105 (2009).

\bibitem{AIW14} R. Adamietz, G.-L. Ingold and U. Weiss, ``Thermodynamic anomalies in the presence of dissipation: from the free particle to the harmonic oscillator'', Eur. Phys. J. B \textbf{87}, 90 (2014).

\bibitem{FordOC}  R. F. O'Connell, ``Does the third law of thermodynamics hold in the quantum regime'', J. Stat. Phys. \textbf{124}, 15 (2006).

\bibitem{Vedral} M. Wie\'sniak, V. Vedral and \u{C}. Brukner, ``Heat capacity as an indicator of entanglement'', Phys. Rev. B \textbf{78}, 064108  (2008). 

\bibitem{Vedral1} E. Rieper, J. Anders and V. Vedral,  ``Entanglement at the quantum phase transition in a harmonic lattice'' New J. Phys. \textbf{12}, 025017 (2010).

\bibitem{HHPRD} J.-T. Hsiang and B. L. Hu, ``Distance and coupling dependence of entanglement in the presence of a quantum field'', Phys. Rev. D \textbf{92}, 125026 (2015). 

\bibitem{QTD2}
J.-T. Hsiang, C. H. Chou, Y. Subasi and B. L. Hu, ``Quantum thermodynamics from the nonequilibrium dynamics of open systems: II. entropy, entanglement and the Second Law'', in preparation.

\bibitem{EspositoJSM} L. Pucci, M. Esposito and L. Peliti, ``Entropy production in quantum Brownian motion'', J. of Stat. Mech. P04005 (2013).

\bibitem{CE09} M. Cramer and J. Eisert, ``A quantum central limit theorem for non-equilibrium systems: exact local relaxation of correlated states'', New J. Phys. \textbf{12}, 055020 (2009).

\bibitem{CEF11} P. Calabrese, F. H. L. Essler and M. Fagotti, ``Quantum quench in the transverse field Ising chain'', Phys. Rev. Lett. \textbf{106} 227203 (2011).

\bibitem{Agarwal}  G. S. Agarwal, ``Entropy, the Wigner distribution function, and the approach to equilibrium of a system of coupled harmonic oscillators" Phys. Rev. A \textbf{3}, 828 (1971).


\bibitem{CTP} J. S. Schwinger, ``Brownian motion of a quantum oscillator'', J. Math. Phys. \textbf{2}, 407  (1961); L. Keldysh, ``Diagram technique for nonequilibrium processes'', Zh. Eksp. Teor. Fiz. \textbf{47}, 1515 (1964); K.-C. Chou, Z.-B. Su, B.-L. Hao, and L. Yu, ``Equilibrium and nonequilibrium formalisms made unified'', Phys. Rept. \textbf{118}, 1 (1985).

\bibitem{HH15} J.-T. Hsiang and B. L. Hu, ``Nonequilibrium steady state in open quantum systems: influence action, stochastic equation and power balance'', Ann. Phys. \textbf{362}, 139 (2015).

\bibitem{ASH06} C. Anastopoulos,  S. Shresta and B. L. Hu,  ``Quantum entanglement under non-Markovian dynamics of two qubits interacting with a common electromagnetic field'',  arXiv:quant-ph/0610007.

\bibitem{LinHu09} S. Y. Lin and B. L. Hu, ``Temporal and spatial dependence of quantum entanglement from field theory perspective'', Phys. Rev. D 	\textbf{79}, 085020 (2009). 

\bibitem{CHY} C. H. Chou, T. Yu and B. L. Hu, ``Exact master equation and quantum decoherence for two harmonic oscillators in a general environment'',  Phys. Rev. E \textbf{77}, 011112 (2008). 

\bibitem{PazRec} J. P. Paz and A. J. Roncaglia, ``Dynamics of the entanglement between two oscillators in the same environment'', Phys. Rev. Lett. \textbf{100}, 220401 (2008). 

\bibitem{SFTH} 
Y. Suba\c{s}\i, C. H. Fleming, J. M. Taylor and B. L. Hu, ``Equilibrium states of open quantum systems in the strong coupling regime'', Phys. Rev. E \textbf{86}, 061132 (2012).

\bibitem{WHL12}
T.-H. Wu, J.-T. Hsiang, and D.-S. Lee, ``Subvacuum effects of the quantum field on the dynamics of a test particle'', Ann. Phys. \textbf{327}, 522 (2012).

\bibitem{HJ85}
R. A. Horn and C. R. Johnson, ``Matrix Analysis'', (Cambridge University Press, Cambridge, 1985).

\bibitem{AH95}
A. Hurwitz, ``On the conditions under which an equation has only roots with negative real parts'', Math. Ann. (Leipzig) \textbf{46}(2), 273 (1895).

\bibitem{PD85}
A. B. Pippard and R. H. Dicke, ``Response and Stability'', (Cambridge University Press, Cambridge, 1985).

\bibitem{VE11}
K. Veseli\'c, ``Damped Oscillations of Linear Systems - A Mathematical Introduction'', Lecture Notes in Mathematics, Vol. 2023 (Springer Verlag, Berlin/Heidelberg, 2011). 

\bibitem{EisPle} J. Eisert and M. B. Plenio, ``Quantum correlations in Brownian motion'', Phys. Rev. Lett. \textbf{89}, 137902 (2002).

\bibitem{HB05} C.  H\"orhammer and H. B\"uttner, ``Thermodynamics of quantum Brownian motion with internal degrees of freedom: the role of entanglement in the strong-coupling quantum regime'', J. Phys. A \textbf{38}, 7325 (2005).

\bibitem{HBJSP08} C.  H\"orhammer and H. B\"uttner, ``Information and entropy in quantum brownian motion - thermodynamic entropy versus von Neumann entropy'', J. Stat. Phys. \textbf{133}, 1161 (2008).

\bibitem{HiltLutz} S. Hilt and E. Lutz, ``System-bath entanglement in quantum thermodynamics'', Phys. Rev. A \textbf{79}, 010101 (R) (2009).

\bibitem{ELvdB} M. Esposito, K. Lindenberg and C. Van den Broeck, ``Entropy production as correlation between system and reservoir'', New J. Phys. \textbf{12}, 013013 (2010). 

\bibitem{DeffLutz} S. Deffner and E. Lutz, ``Nonequilibrium entropy production for open quantum systems'', Phys. Rev. Lett. \textbf{107}, 140404 (2011).

\bibitem{Kim12} I. Kim, ``Non-equilibrium dynamics in the quantum Brownian oscillator and the second law of thermodynamics'', J. Stat. Phys. \textbf{146}, 217 (2012).

\bibitem{AurEic15} E. Aurell and R. Eichhorn, ``On the von Neumann entropy of a bath linearly coupled to a driven quantum system'', New J. Phys. \textbf{17} 065007 (2015).

\bibitem{AnkPek14} J. Ankerhold and J. P. Pekola, ``Heat due to system-reservoir correlations in thermal equilibrium'',  Phys. Rev. B \textbf{90} 075421 (2014).

\bibitem{EspositoPRB15} M. Esposito, M. A. Ochoa and M. Galperin, ``Nature of heat in strongly coupled open quantum systems'', Phys. Rev. B \textbf{92} 235440 (2015)

\bibitem{CarrWeiss16} 
M. Carrega, P. Solinas, M. Sassetti and U. Weiss, ``Energy exchange in driven open quantum systems at strong coupling'', Phys. Rev. Lett. \textbf{116}, 240403 (2016).


\bibitem{CGEA} B. L. Hu,  Lectures at the Seventh International Latin-American Symposium on General  Relativity (SILARG VII). Proceeding appeared as {\it Relativity and Gravitation: Classical and Quantum}, edited by J. D'Olivo et al (World Scientific, Singapore, 1991); Yuhong Zhang, Ph.D thesis (University of Maryland, 1990); E. Calzetta, B. L. Hu, and F. D. Mazzitelli, ``Coarse-grained effective action and renormalization group theory in semiclassical gravity and cosmology'', Phys. Rep. {\bf 352}, 459 (2001).

\bibitem{JH1} P. R. Johnson and B. L. Hu, ``Stochastic theory of relativistic particles moving in a quantum field: scalar Abraham-Lorentz-Dirac-Langevin equation, radiation reaction, and vacuum fluctuations'', Phys. Rev. D {\bf 65}, 065015 (2002).


\bibitem{PazRom} L. D. Romero and J. P. Paz, ``Decoherence and initial correlations in quantum Brownian motion'', Phys. Rev. A, {\bf 55}, 4070 (1997).

\bibitem{FRH} C. H. Fleming, A. Roura and B. L. Hu, ``Initial state preparation with dynamically generated system-environment correlations'', Phys. Rev. E {\bf 84}, 021106 (2011).

\bibitem{DC06} O. S. Duarte and A. O. Caldeira, ``Effective coupling between two Brownian particles'', Phys. Rev. Lett. 97, 250601 (2006).

\bibitem{HZ95}
A. Hanke and W. Zwerger, ``Density of states of a damped quantum harmonic oscillator'', Phys. Rev. E \textbf{52}, 6875 (1995).

\bibitem{HWL08} J.-T. Hsiang, T.-H. Wu, and D.-S. Lee, ``Stochastic Lorentz forces on a point charge moving near the conducting plate'', Phy. Rev. D \textbf{77}, 105021 (2008).
 
\bibitem{BC63} R. Bellman and K. L. Cooks, ``Differential-Difference Equations'' (Academic, New York, 1963).

\bibitem{roh}
F. Rohrlich, ``Classical Charged Particles'' (Addison-Wesley, Reading, Mass., 1965); A. D. Yaghjian, ``Relativistic Dynamics of a Charged Sphere - Updating the Lorentz-Abraham Model'', Lecture Notes in Physics, Vol. 686, (Springer-Verlag, New York, 2006).

\bibitem{HHPLB15}
J.-T. Hsiang and B. L. Hu, ```Hot entanglement'? - A nonequilibrium quantum field theory scrutiny'', Phy. Lett. B \textbf{750}, 396 (2015).

\bibitem{LebSpo77} H. Spohn, J. L. Lebowitz, ``Stationary non-equilibrium states of infinite harmonic systems'', Comm. Math. Phys. \textbf{54}, 97 (1977).

\bibitem{UnrZur89} W. G. Unruh and W. H. Zurek, ``Reduction of a wave packet in quantum Brownian motion'', Phys. Rev. D \textbf{40}, 1071 (1989).

\bibitem{HM94} B. L. Hu and A. Matacz, ``Quantum Brownian motion in a bath of parametric oscillators: a model for system-field interactions'', Phys. Rev. D \textbf{49}, 6612 (1994).


\end{thebibliography}
\end{document}